\documentclass[10pt,a4paper]{article}
\usepackage{jheparxiv}
\usepackage[utf8]{inputenc}
\usepackage{amsmath}
\usepackage{amsmath,bm}
\usepackage{sujai}
\usepackage{amsfonts}
\usepackage{float}
\usepackage{amssymb}
\usepackage{latexsym}
\usepackage{mathrsfs}
\usepackage{braket}		
\usepackage{graphicx}
\usepackage{color}
\usepackage{xcolor}
\usepackage{slashed}
\usepackage{mathtools}
\usepackage{breqn}
\usepackage[all]{xy}
\usepackage{subcaption}
\newcommand{\RNum}[1]{\uppercase\expandafter{\romannumeral #1\relax}}

\subheader{\hfill \texttt{}}

\definecolor{bg2}{rgb}{0.00,0.52,0.79}

\newcommand{\cB}{{\cal B}}

\newcommand{\beq}{\begin{equation}}
\newcommand{\eeq}{\end{equation}}
\newcommand{\bea}{\begin{eqnarray}}
\newcommand{\eea}{\end{eqnarray}}
\def\be{\begin{equation}}
\def\ee{\end{equation}}

\renewcommand{\[}{\left[}
\renewcommand{\]}{\right]}

\def\be{\begin{equation}}
\def\ee{\end{equation}}
\def\ba{\begin{eqnarray}}
\def\ea{\end{eqnarray}}

\def\p{\partial}
\def\d{\mathrm{d}}

\def\({\left(}
\def\){\right)}

\def\nn{\nonumber}

\def\d{\mathrm{d}}

\usepackage[normalem]{ulem}

\def\ba{\begin{eqnarray}}
\def\ea{\end{eqnarray}}

\begin{document}

\title{Constraints on Regge behaviour from IR physics}

\author[a,b]{Claudia de Rham,}
\author[a]{Sumer Jaitly,}
\author[a,b]{Andrew J. Tolley}
\affiliation[a]{Theoretical Physics, Blackett Laboratory, Imperial College, London, SW7 2AZ, U.K.}
\affiliation[b]{Perimeter Institute for Theoretical Physics,
31 Caroline St N, Waterloo, Ontario, N2L 6B9, Canada}

\emailAdd{c.de-rham@imperial.ac.uk}
\emailAdd{sumer.jaitly14@imperial.ac.uk}
\emailAdd{a.tolley@imperial.ac.uk}

\abstract{We consider positivity constraints applicable to the Effective Field Theory (EFT) of gravity in arbitrary dimensions. By considering scattering of indefinite initial and final states, we highlight the existence of a gravitational scattering amplitude for which full crossing symmetry is manifest and utilize the recently developed crossing symmetric dispersion relations to derive compact non-linear bounds. We show that the null constraints built into these dispersion relations lead to a finite energy sum rule for gravity which may be extended to a one-parameter family of continuous moment sum rules. These sum rules enforce a UV-IR relation which imposes constraints on both the Regge trajectory and residue. We also highlight a situation where the Regge trajectory is uniquely determined in terms of the sub-Regge scale amplitude. Generically the Regge behaviour may be split into an IR sensitive part calculable within a given EFT which mainly depends on the lightest fields in Nature, and an IR independent part which are subject to universal positivity constraints following from unitarity and analyticity.}

\maketitle


\section{Introduction}
The physics of low energy degrees of freedom of a system of fields may be described via an effective field theory (EFT) built out of (in general) all operators that respect the symmetries of the system, along with their coupling constants. The Standard Model and the Lagrangian formulation of general relativity are understood best as effective field theories, with a cutoff scale beyond which new degrees of freedom are needed to extend their applicability to higher energies \cite{Donoghue:1994dn, Donoghue:1995cz, Burgess:2003jk, Donoghue:2012zc, Weinberg:2018apv}. It is now well understood that for the UV completion to satisfy the familiar properties of causality, unitarity, locality\footnote{For non-gravitational theories locality imposes a Froissart-like bound. The role of locality in quantum gravity is expected to be softer, however one would expect the main input from locality, namely a bound in the growth of scattering amplitudes to remain true \cite{Haring:2022cyf},  and this will be sufficient to justify the positivity bounds.} and Lorentz invariance there must be non-trivial constraints on its low energy description \cite{deRham:2022hpx}.
In our analysis these constraints take the form of the well-established positivity bounds that restrict the values of the low energy effective action coupling constants by way of inequalities \cite{Pham:1985cr,Ananthanarayan:1994hf,Adams:2006sv}. \\

The most familiar positivity bounds are applied to the amplitude in the forward limit $t\xr0$ where the optical theorem (unitarity) immediately grants positivity to the imaginary part (or rather the absorptive part in the general spin case) of the amplitude. However it has long been understood that these positivity properties may be extended to the finite fixed $t>0$ amplitude within a region of analyticity in the complex $t$-plane \cite{Pennington:1994kc, Vecchi:2007na, Manohar:2008tc, Nicolis:2009qm} and this plays an important role in more recent developments \cite{deRham:2017avq,deRham:2017zjm,deRham:2018qqo,Alberte:2019xfh,Alberte:2019zhd,Arkani-Hamed:2020blm,Bellazzini:2020cot,Tolley:2020gtv,Caron-Huot:2020cmc,deRham:2022sdl}. The application of these methods to gravitational effective theories is however problematic due to a number of issues related to the divergences at $t=0$ that prevent a direct continuation of the usual dispersion relation from the physical region $t<0$ to $t>0$. One straightforward way to deal with this is to focus on weakly coupled theories for which graviton loops (and hence the problematic $t=0$ branch points) are absent. Familiar positivity bounds may then be applied to scattering amplitudes with one more subtraction than usual for which the $t$-channel pole drops out. This is the approach taken for example in \cite{Bellazzini:2015cra,Bern:2021ppb}, and we shall follow a similar procedure here for some of the bounds, while folding-in information about the UV to connect with double subtracted ones. A second approach is the smeared amplitude method of \cite{Caron-Huot:2021rmr} which constructs positive quantities out of integrals over $t<0$. More recently this has been applied successfully in four dimensions \cite{Caron-Huot:2022ugt} and higher-dimensional gravitational EFTs \cite{Caron-Huot:2022jli}.\\

Causality constraints on higher curvature operators in an EFT of gravity have been considered in the past.  There is long standing theoretical evidence to suggest constraints on the two quartic curvature couplings, $ (R_{\mu\nu\rho\sigma}R^{\mu\nu\rho\sigma})^2$ and $(R^{\alpha\beta\mu\nu}\epsilon_{\mu\nu\rho\sigma}R^{\rho\sigma}_{\phantom{\rho\sigma}\alpha\beta}/2)^2$. In \cite{Gruzinov:2006ie} it is shown that positive signs for both quartic curvature corrections ensures sub-luminal group velocity of UV perturbations. Positivity of the quartic coefficients has been shown via analyticity and unitarity arguments in \cite{Bellazzini:2015cra} and \cite{Bern:2021ppb} where in the latter, a tighter bound is derived involving the size of the cubic curvature coupling. Recently further considerations of `infrared causality' have been shown to corroborate these previous bounds whilst also producing lower a bound on the cutoff of the gravitational EFT \cite{deRham:2021bll}. The quartic curvature terms are of particular interest as they are known to arise in the low energy action of compactified string theories \cite{Gross:1986iv} and potential implications for gravitational wave observations were considered in \cite{Endlich:2017tqa, Cardoso:2018ptl, Sennett:2019bpc}. Cubic curvature couplings are also known to arise from non-supersymmetric string UV completions \cite{Metsaev:1986yb}, with effects on observables considered in \cite{Brandhuber:2019qpg,AccettulliHuber:2020oou,AccettulliHuber:2020dal,deRham:2020ejn}. Recently approaches \cite{Caron-Huot:2022ugt,Caron-Huot:2022jli} have sharpened the constraints on some of the operators, in particular make use of the null constraints considered for scalar theories in \cite{Tolley:2020gtv,Caron-Huot:2020cmc} to strengthen the power of the positivity bounds.\\

In the present work, we consider the positivity bounds that apply to effective field theories of gravity in any dimension, including four. For simplicity we shall focus on the case of EFTs where the only light state is the graviton, although additional massless or light degrees of freedom such as the photon or dilaton may easily be incorporated. Our central object then is the graviton four-point scattering amplitude, and we explicitly compute the scattering amplitude up to and including dimension-12 EFT operators ($(\nabla R)^4$). We show that in any dimension, there is a judicious choice of indefinite incoming polarization states for which we can construct a one parameter family of manifestly crossing symmetric amplitudes with the same analytic structure of a scalar theory \eqref{triplecrossingamp}. It is straightforward to obtain positivity bounds for gravitational theories in any dimension both from semi-analytic arguments and numerical optimization procedures leading rapidly to results comparable with previous approaches \cite{Bern:2021ppb,Caron-Huot:2022ugt,Caron-Huot:2022jli}. In particular we make use of the elegant crossing symmetric dispersion relations utilized in \cite{Sinha:2020win,Haldar:2021rri,Raman:2021pkf,Chowdhury:2021ynh}. \\

One of the virtues of the crossing symmetric approach is that it allows a trivial derivation of the null constraints and sum rules which follow from them. In particular, it is straightforward to see that in the scattering amplitude for gravitons\footnote{This results also holds in higher than four dimensions, at the very least when focusing on the scattering of the four-dimensional subset of polarizations.} the leading $s^2$ (or its triple crossing symmetric version, $x=st+tu+us$)  always identically vanishes in any consistent  gravitational amplitude (be it following from string theory or from a local and covariant EFT).
It is of course precisely the presence of this coefficient and its positivity which is crucial in the arguments of \cite{Pham:1985cr,Adams:2006sv}, and its absence for example for Galileon theories is what rules them our from having such a conventional completion \cite{Tolley:2020gtv}. In gravitational EFTs, it is the presence of the $t$-channel graviton pole which also scales as $s^2$ which prevents us making any analogue conclusion directly. We show that the vanishing of this coefficient imposes a sum rule\footnote{Related sum rule are obtained by slightly different means in \cite{Caron-Huot:2022ugt,Caron-Huot:2022jli}.} which for a specific choice of indefinite polarizations contains no contribution from any low energy coefficient other than the Planck mass. This sum rule is closely analogous to the superconvergence relations which were influential in the development of Dolen-Horn-Schmid duality \cite{Dolen:1967jr} which was the motivation for the Veneziano amplitude. In particular, following a similar argument to \cite{Dolen:1967jr}, considered more recently in \cite{Tokuda:2020mlf,Alberte:2021dnj}, we show that in a gravitational effective theory in any dimension, there is a {\it finite energy sum rule} which can be used to relate the Regge behaviour of the would-be UV completion to the IR expansion coefficients. Remarkably this sum rule allows to infer how IR physics, contributes to the Regge behaviour through a particular combination of the Regge trajectory and residue, without relying on ad hoc resummations of perturbative diagrams. We further extend this to a one-parameter family of `continuous moment sum-rules' which can in principle be used to give separate information on the Regge trajectory and residue.
\\

The rest of this work is organized as follows: We start with providing a one-parameter family of manifestly triple crossing symmetric gravitational amplitudes in Sec.~\ref{sec:xssingPositivity}. With this formalism at hand, we can then derive null constraints for gravity and infer a set of analytic bounds on the coefficients of the gravitational amplitude in Sec.~\ref{sec:GraPositivity}. We then apply these constraints to the EFT of gravity in Sec.~\ref{sec:EFT}, involving up to dimension-12 operators providing constraints in agreement with previously found analytic and numerical ones as well as new higher order constraints. Sec.~\ref{sec:Regge}  focused on improved versions of the finite energy sum rule and on the implications for the trajectory and residue of the Regge behaviour, emphasizing how IR physics is inbuilt in that behaviour. We provide various outlooks in Sec.~\ref{sec:conclusion}. In order to gain a better insight of the crossing symmetric dispersion relations used in this work, we compare them with the Mandelstam double spectral representation in Appendix~\ref{app:mandel} for a manifestly crossing symmetric amplitude.
We also provide explicit examples from (partial) UV completions or string theory in Appendix~\ref{app:coeff}, while details on how to use the Hilbert series to infer the number of independent dimension-12 operators are summarized in Appendix~\ref{app:Hilbert}.

\section{Kinematics, crossing and positivity}
\label{sec:xssingPositivity}

\subsection{Factorizable polarization}
\label{sec:factorizable}

If we consider the scattering of gravitons with helicity $h$, we can  denote the scattering amplitude for the process with in-state of helicity $|h_1 , h_2 \rangle$ and out-state of helicity $|-h_3,-h_4 \rangle$ as
\be
\cA_{h_1 h_2 \rightarrow -h_3 -h_4}(s,t,u)  = \cA_{h_1 h_2  h_3 h_4}(s,t,u)\, .
\ee
The form $ \cA_{h_1 h_2  h_3 h_4}(s,t,u)$ describes all helicities as ingoing. As the external states are massless and bosonic, $s$-$u$ crossing symmetry is simple and is encoded in
\be
\cA_{h_1 h_2  h_3 h_4}(s,t,u) = \cA_{h_1 h_4  h_3 h_2}(u,t,s)\,.
\ee
Similarly $s$-$t$ crossing is encoded in the statement
\be
\cA_{h_1 h_2  h_3 h_4}(s,t,u) =\cA_{h_1 h_3  h_2 h_4}(t,s,u) \, ,
\ee
and $u$-$t$ crossing in
\be
\cA_{h_1 h_2  h_3 h_4}(s,t,u) =\cA_{h_1 h_2  h_4 h_3}(s,u,t) \, .
\ee
For the elastic scattering of an indefinite helicity state $\ket{\rm in}=\sum_{h_1, h_2}a_{h_1 h_2}\ket{h_1,h_2}$, the corresponding $u$-channel amplitude defined via $s$-$u$ crossing  as
\ba
\cA_u(s,t,u) = \cA_s(u,t,s) &= & \sum_{h_i=\pm 2} a_{h_1h_2} a^*_{-h_3-h_4} \cA_{h_1 h_2  h_3 h_4}(u,t,s)  \nn \\
&= &  \sum_{h_i=\pm 2} a_{h_1h_2} a^*_{-h_3-h_4} \cA_{h_1 h_4  h_3 h_2}(s,t,u)\,,
\ea
does not in general have a positive discontinuity since the final expression does not correspond to an elastic amplitude. However if we focus on factorizable indefinite helicity combinations for which $a_{h_1 h_2 }=\alpha_{h_1} \beta_{h_2}$ then
\ba
\cA_u(s,t,u) &=& \sum_{h_i=\pm 2} \alpha_{h_1} \beta_{h_2}\alpha_{-h_3}^* \beta_{-h_4}^*\cA_{h_1 h_4  h_3 h_2}(s,t,u) \nn \\
&=& \sum_{h_i=\pm 2} \alpha_{h_1} \beta_{-h_2}^* \alpha_{-h_3}^* \beta_{h_4} \cA_{h_1 h_2  h_3 h_4}(s,t,u) \, ,
\ea
and so $\cA_u(s,t,u)$ can be interpreted as an elastic scattering amplitude with initial state $|{\rm in}  \rangle = \sum_{h_{1,2}=\pm 2} \alpha_{h_1} \beta_{-h_2}^* |h_1 h_2 \rangle$.
The scattering amplitude will exhibit manifest $s$-$u$ crossing symmetry if
\be\label{crossing1}
\beta_{h}=\beta_{-h}^* e^{i \gamma_1}\, .
\ee

\subsection{Manifest $s\leftrightarrow t$ crossing symmetry}

In principle, we could apply positivity bounds to any factorizable indefinite polarizations. However, in order to obtain more precise bounds, it is useful to input the consequences of $s \leftrightarrow  t$ crossing symmetry which are largely hidden in the standard fixed $t$ dispersion relations. Indeed, it was shown in \cite{Tolley:2020gtv,Caron-Huot:2020cmc} that the imposition of $s \leftrightarrow  t$ crossing symmetry imposes a set of `null-constraints' which can be used to infer upper and lower bounds on Wilson coefficients.

The $t$-channel amplitude defined via $s$-$t$ crossing for factorizable states is given by
\ba
\cA_t(s,t,u) = \cA_s(t,s,u) &= & \sum_{h_1,h_2,h_3,h_4} \alpha_{h_1} \beta_{h_2} \alpha_{-h_3}^* \beta_{-h_4}^* \cA_{h_1 h_2  h_3 h_4}(t,s,u)   \\
&= &  \sum_{h_i=\pm 2}  \alpha_{h_1} \alpha_{-h_3}^*   \beta_{h_2} \beta_{-h_4}^* \cA_{h_1 h_3  h_2 h_4}(s,t,u) \\
&= &  \sum_{h_i=\pm 2}  \alpha_{h_1} \alpha_{-h_2}^*   \beta_{h_3} \beta_{-h_4}^* \cA_{h_1 h_2  h_3 h_4}(s,t,u)\,,
\ea
where in the last step we interchanged labels. It is therefore sufficient to impose
\be\label{crossing2}
\beta_{h}=\alpha_{-h}^* e^{i \psi}\, ,
\ee
to ensure that the scattering amplitude is manifestly $s\leftrightarrow t$ symmetric
\be
\cA_t(s,t,u) = \cA_s(s,t,u) \, .
\ee
Up to an overall irrelevant phase, there is a two-parameter family of such in-states
\ba
\label{eq:instate}
| {\rm in} \rangle = \sum_{h_{1,2}=\pm 2} \alpha_{h_1} \beta_{h_2} |h_1 h_2 \rangle\,,
\ea
for which $\alpha_{+2} = \cos \chi$ and $\alpha_{-2} =\sin \chi e^{i \phi}$ so that
\be
|{\rm in} \rangle = e^{i \psi}\(\cos \chi \sin \chi e^{-i \phi} | + + \rangle + \cos^2\chi | + - \rangle +  \sin^2 \chi  | - + \rangle  + \cos \chi  \sin \chi  e^{i \phi} | -- \rangle\) \, .\nn
\ee

\subsection{Manifest triple crossing symmetry}

By combining \eqref{crossing1} and \eqref{crossing2} we can construct elastic scattering amplitudes which are manifestly crossing symmetric under all three interchanges $s\leftrightarrow t$, $s\leftrightarrow u$ and $t\leftrightarrow u$,
\be
\cA_u(s,t,u) =\cA_t(s,t,u)= \cA_s(s,t,u) \, .
\ee
Up to an overall irrelevant phase, there remains a one parameter family of in-states \eqref{eq:instate} which respect \eqref{crossing1} and \eqref{crossing2}. Full crossing symmetry is manifest upon setting the phases $ \psi=\phi$, and $\chi=\pi/4$, so that
\begin{equation}
\label{eq:triplecrossing}
    \alpha_+=\frac{1}{\sqrt{2}}\,,\quad\alpha_-=\frac{e^{i\phi}}{\sqrt{2}}\,,\quad\beta_+=\frac{1}{\sqrt{2}}\,,
    \quad\beta_-=\frac{e^{i\phi}}{\sqrt{2}}\,,   \quad    \phi \in \mathbb{R}\,.
\end{equation}
For a general in-coming state as defined in \eqref{eq:instate},
using the shorthand notation $a_1=\alpha_+\beta_+$, $a_2=\alpha_+\beta_-$, $a_3=\alpha_-\beta_+$, $a_4=\alpha_-\beta_-$ the indefinite helicity elastic amplitude (simplified using the crossing symmetric properties of our amplitudes) is expressed as,
\ba
\label{expansindef}
\bra{\rm in}\hat T\ket{\rm in}
&=&\cA_{11}(|a_1|^2+|a_4|^2)+\cA_{22}(|a_2|^2+|a_3|^2)+2\cA_{23}{\rm Re}(a_2a_3^*)
+2\cA_{14}{\rm Re}(a_1a_4^*)\\
&+&2\cA_{21}\({\rm Re}(a_1a_2^*)+{\rm Re}(a_1a_3^*)+{\rm Re}(a_2a_4^*)+{\rm Re}(a_3a_4^*)\)\,,\nn
\ea
where the finite helicity amplitudes are $\cA_{ij}\coloneqq\bra{j}\hat T\ket{i}$, and we have used the shorthand notation
\begin{equation}
\label{eq:Tij}
    \ket{1}=\ket{++}\,,\quad\ket{2}=\ket{+-}\,,\quad\ket{3}=\ket{-+}\,,\quad\ket{4}=\ket{--}\,.
\end{equation}
With the triple crossing symmetric parametrisation \eqref{eq:triplecrossing} we therefore get
\ba
\label{expans}
\bra{\rm in}\hat T\ket{\rm in}
=\frac 12 \(\cA_{11}+\cA_{22}+\cA_{23}\)+2\cA_{21}\cos \phi
+\frac 12 \cA_{14}\cos 2\phi \,.
\ea

In the low-energy EFT of gravity, the manifestly crossing symmetric amplitude is given in what follows in \eqref{triplecrossingamp}, but we note that this formalism is applicable for the scattering of particles of any spin, including photons and higher spins.

\subsection{Partial Wave Expansion}

For convenience we start by focusing on the scattering of two massless spin-2 particles in four dimensions and the extension to arbitrary dimensions is carried out in the next subsection~\ref{eq:secHigherD}. We also emphasize once again that the procedure is directly applicable to scattering of particles of other spin.
In four dimensions, the partial wave expansion in the standard $s$-channel is given in\footnote{This differs by a factor of 2 from the result in \cite{deRham:2017zjm} due to the inclusion of a factor of $\sqrt{2}$ in the normalization of identical scattering states as in \cite{Hebbar:2020ukp}. This convention difference will not be important in what follows.} \cite{deRham:2017zjm,Hebbar:2020ukp}
\be \label{eq:partialwave}
\cA_{h_1 h_2  h_3 h_4}(s,\theta) = 32 \pi \sum_J (2J+1) d_{\lambda \mu}^J(\theta) T^J_{h_1 h_2 h_3 h_4}(s)\,,
\ee
where
\be
\lambda = h_1-h_2 \,, \qquad \mu = h_4-h_3 \, ,
\ee
and
\be
 T^J_{h_1 h_2 h_3 h_4}(s) = \langle p_f J M -h_3 -h_4 | \hat T | p_u J M h_1 h_2 \rangle \, .
\ee
The Wigner matrices satisfy $d_{\lambda \mu}(-\theta) =(-1)^{\lambda-\mu}  d_{\lambda \mu}(\theta) $. As long as the external states are spin-2 (or spin-0) then $(-1)^{\lambda-\mu}  =1$ so that the partial wave expansion may be written as
\be
\cA_{h_1 h_2  h_3 h_4}(s,\theta) = 16 \pi \sum_J (2J+1) \(  d_{\lambda \mu}^J(\theta) +d_{\lambda \mu}^J(-\theta) \) T^J_{h_1 h_2 h_3 h_4}(s)\,,
\ee
which is to say that all amplitudes will be even in $\theta$. Further using  the following relation for the Wigner matrices \cite{deRham:2017zjm}
\be
 d_{\lambda \mu}^J(\theta) +d_{\lambda \mu}^J(-\theta) = 2 e^{i \frac{\pi}{2} (\lambda-\mu)} \sum_{\nu=-J}^J d_{\lambda \nu}^J\(\frac{\pi}{2} \)d_{\mu \nu}^J\(\frac{\pi}{2} \) \cos(\nu \theta) \, ,
\ee
and considering an arbitrary indefinite helicity elastic scattering amplitude associated with a scattering in-state $| {\rm in} \rangle = \sum_{h_1,h_2} a_{h_1 h_2} |h_1 h_2 \rangle$,
\be
\cA_s(s,t,u) = \sum_{h_1,h_2,h_3,h_4} a_{h_1h_2} a^*_{-h_3-h_4} \cA_{h_1 h_2  h_3 h_4}(s,t,u)\,,
\ee
it is straightforward to see that
\ba
{\rm Disc}_s \cA(s,\theta)  =\sum_J (2J+1) \sum_{\nu=-J}^{J} \cos(\nu \theta) F^{\nu,J}(s)
= \sum_{\nu=0}^{\infty}\cos(\nu \theta) \tilde F^{\nu,J}(s)\, , \label{disc1}
\ea
where the function $F^{\nu,J}$ is defined as
\ba
F^{\nu,J}(s) &=& \sum_{h_i=\pm 2} {C_{-h_3 -h_4}^{\nu}}^*{\rm Disc}_s \( T^J_{h_1 h_2 h_3 h_4}(s)\) C_{h_1 h_2}^{\nu} \\
&=&\sum_{h_i=\pm 2} {C_{h_3 h_4}^{\nu}}^*{\rm Disc}_s \( T^J_{h_1 h_2 \rightarrow -h_3 -h_4}(s)\) C_{h_1 h_2}^{\nu}
 \ge 0 \, ,
\ea
with
\be
C_{h_1 h_2}^{\nu} = e^{i \frac{\pi}{2} (h_1-h_2)} d_{(h_1-h_2) \nu}^J\(\frac{\pi}{2}\) a_{h_1 h_2} \, ,
\ee
which in turn implies positivity of the spectral densities $ \tilde F^{\nu,J}(s) \ge 0$.
Using the positivity of
\be
\frac{\p^n \cos(\nu \theta)}{\p \cos^n \theta} \ge 0 \, ,
\ee
and given $t = - \frac{1}{2} s (1 - \cos \theta)$, we infer that
\be\label{eq:positivity}
\left.\frac{\partial^n }{\partial t^n} {\rm Disc}_s \cA_s(s,t) \right|_{t=0}\ge 0 \, ,  \text{ for } s>0 \, ,
\ee
for all integer $n \ge 0$. This is true as long as these derivatives exist which assumes that ${\rm Disc}_s \cA(s,t)$ is analytic in $t$ at $t=0$. In general loops of massless particles destroy this, so we can only use this property for amplitudes computed at tree level in the low energy effective theory in which massless loops are not included. In reality for spin $S \le 1$ particles we can always introduce an IR regulating mass. As for the graviton loops, they are always suppressed by additional powers of the Planck scale.

 With the factorizable choice of polarization made in Sec.~\ref{sec:factorizable}, we can show that we also have positivity along the left hand branch cut
\be
\left.\frac{\partial^n }{\partial t^n} {\rm Disc}_s \cA_u(s,0) \right|_{t=0} \ge 0 \, ,  \text{ for } s>0 \, .
\ee
This is sufficient to ensure a set of positivity bounds for factorizable indefinite states. Positivity bounds for such indefinite states have been considered at length in the literature \cite{Bellazzini:2015cra,Cheung:2016yqr,deRham:2017imi,deRham:2018qqo,Alberte:2019xfh,Alberte:2019zhd,Alberte:2021dnj}.

\subsection{Higher Dimensions}
\label{eq:secHigherD}

In four dimensions scattering amplitudes of massless particles suffer from IR divergences related to the infinite range nature of their interactions and Coulomb asymptotics. This problem can partly be avoided by only computing tree level scattering amplitudes. At tree level, the only remaining IR divergence is that attributed to the $t$-channel spin-2 pole which drops out of higher subtracted positivity bounds. In more than four dimensions, the gravitational potential falls off sufficiently fast at large distances so that scattering amplitudes are IR finite. Thus it is of great interest to consider positivity bounds for higher dimensional theories. The full set of such bounds becomes increasingly complicated due to the increased number of polarizations. However, there is a rather simple truncation which is already sufficient to determine significant bounds. In a higher dimensional theory, we can always focus on scattering states which have only momenta and polarization dependence in four of the $D$ dimensions. This has the significant advantage that the scattering amplitude in $D \ge 5$ dimensions for such external states can still be expressed via a partial wave expansion of the form \eqref{eq:partialwave}, with positivity properties for the discontinuity, with the only difference being the precise implementation of unitarity. \\

To illustrate this, consider the simpler problem of the scattering of scalars in $D$ dimensions where the natural expansion variables are Gegenbauer polynomials $C^{(D-3)/2}_{\ell}(\cos \theta)$ where $\ell$ is the multipole and $\theta$ the scattering angle. It is an elementary result that the Gegenbauer polynomials in $D$ dimensions admit an expansion in terms of $D=4$ Legendre polynomials with positive coefficients, or indeed in terms of $D=3$ Chebyshev polynomials of the first kind with positive coefficients. To be precise the expansion of $D$ dimensional Gegenbauer polynomials in terms of $\tilde D<D$ Gegenbauer's is \cite{ismail2005classical}
\be
\hspace{-0.3cm}C^{(D-3)/2}_{\ell}(\cos \theta) = \sum_{k=0}^{[\ell/2]} \frac{\(\frac{1}{2}(D-\tilde D)\)_k \(\frac{1}{2}(D-3)\)_{\ell -k}}{k! \(\frac{1}{2}(D-2)\)_{\ell-k}} \( \frac{(\tilde D-3)/2+\ell-2 k}{(\tilde D-3)/2} \) C^{(\tilde D-3)/2}_{\ell-2k}(\cos \theta) \, ,
\ee
where for the case $\tilde D=3$ we should recognize the limit in terms of Chebyshev polynomials
\ba
&& \lim_{\tilde D \rightarrow 3}  \( \frac{(\tilde D-3)/2+n}{(\tilde D-3)/2} \) C^{(\tilde D-3)/2}_{n}(\cos \theta) =
\begin{cases}
  2 \cos( n \theta) \, , & \mbox{if } n \ge 1  \\
  1, & \mbox{for } n =0.
\end{cases}
\ea
Here $(x)_k$ are the Pochammer symbols which are all manifestly positive for $x>0$ and $k \ge 0$. Since these expansion coefficients are positive, all of the positivity arguments we have made remain intact.

\paragraph{Polarizations and momenta:}
To be specific, in four dimensions the scattering amplitudes are computed with the following momenta and vector polarizations,
\begin{equation}
	\begin{split}
		&k_1^\mu = k (1,0,0,1)\\
		&k_2^\mu = k (1,0,0,-1)\\
		&k_3^\mu = -k(1,\sin\theta,0,\cos\theta)\\
		&k_4^\mu = -k(1,-\sin\theta,0,-\cos\theta)\,
	\end{split}
	\quad\quad
	\begin{split}
		&\vep_1^\mu = (0,1,i h_1,0)/\sqrt2\\
		&\vep_2^\mu = (0,-1,i h_2,0)/\sqrt2\\
		&\vep_3^\mu = (0,\cos\theta,i h_3,-\sin\theta)/\sqrt2\\
		&\vep_4^\mu = (0,-\cos\theta,i h_4,\sin\theta)/\sqrt2\,,
	\end{split}
\end{equation}
where $k=\sqrt{s}/2$ and $h_i=\pm1$ are the helicities of each vector polarization. The graviton polarizations are simply given by the direct product of vector polarizations, i.e. $\vep_i^{\mu\nu}=\vep_i^\mu\vep_i^\nu$ which describe states with helicity $2h_i$. For a $D$-dimensional spacetime, we use the momenta $K_i^A = (k_i^\mu,0)$ and vector polarizations $E_i^A = (\vep_i^\mu,0)$, with $A=0,\ldots,D-1$. In $D$ dimensions, the graviton has $(D-4)(D+1)/2$ additional polarization states, however we are free to focus on scattering between the polarizations described above, which again have helicities $\pm2$. The additional polarizations enter internal lines in a dimension-dependent way, however positivity of the appropriately subtracted amplitudes is ensured  from unitarity in any dimensions.

\section{Gravitational Positivity Bounds}
\label{sec:GraPositivity}

The most familiar positivity bounds focus on the forward and positive $t$ region of the scattering amplitude where the discontinuity is known to satisfy positivity properties \eqref{eq:positivity}.
The presence of the spin-2 exchange pole at $t=0$ as well as the branch cut associated to graviton loops obstructs continuation of the partial wave expansion from the physical region $t<0$, to positive values of $t$. In addition, the pole itself has a residue growing faster with $|s|$ than what the Froissart/Jin-Martin  bound allows \cite{Froissart:1961ux, Jin:1964zz} or the equivalent Regge bound for massless amplitudes \cite{Haring:2022cyf}, and thus cannot be subtracted without leading to the resulting pole-subtracted amplitude violating the bound. This prevents us from deriving the usual positivity bounds for theories of gravity.
There are essentially 3 ways to deal with this:
\begin{enumerate}
\item Consider further subtractions to remove the $t$-channel pole. This is the approach taken for example in \cite{Bellazzini:2015cra,Bern:2021ppb}. This approach only works when the $t$-channel branch cut may be ignored as in a weakly coupled UV completion for which the branch cut may be assumed to start at a finite value $s=\Lambda^2$ \footnote{Typically $\Lambda^2$ is either $4m^2$ where $m$ is the mass of the lightest state with spin $S<2$ or $\Lambda^2=M_*^2$ where $M_*$ is the mass of the lightest spin $ S \ge 2$ state (usually $S=4$).}.
\item Work exclusively in the negative $t$ region and obtain independent positivity statements. This is the approach taken in \cite{Caron-Huot:2021rmr,Caron-Huot:2022ugt,Caron-Huot:2022jli} as well as in other related bootstrap programs \cite{Guerrieri:2021ivu}. To be functional it also relies on the assumption of weak coupling in order to consider $t$ as negative as $ t \sim - \Lambda^2$ to get useful bounds.
\item Fold in an assumed UV behaviour into the dispersion relation to allow us to analytically continue from $t<0$ to $t \ge 0$. This is the approach considered in \cite{Tokuda:2020mlf,Aoki:2021ckh,Noumi:2021uuv,Alberte:2021dnj,Noumi:2022zht,Herrero-Valea:2022lfd} and will be discussed further below.
\end{enumerate}
All three approaches assume the existence of a dispersion relation with two subtractions for $t<0$. The first and third approach seek to leverage information at $t>0$ whereas the second uses only $t<0$. All three approaches rely on assuming some level of weak coupling or neglect of loops of massless states so that the branch cut starts at finite values $\Lambda^2$. In particular the strength of the second approach comes from considering amplitudes for $0>t\sim - \Lambda^2$.

\subsection{Crossing Symmetric Dispersion Relation}

For scattering amplitudes which exhibit manifest triple crossing symmetry, as in the case of our one parameter family of elastic indefinite polarizations, we can use the crossing symmetric dispersion relation reintroduced in \cite{Sinha:2020win,Haldar:2021rri,Raman:2021pkf,Chowdhury:2021ynh}, based on earlier work \cite{Auberson:1972prg,Mahoux:1974ej}.
For this we introduce the standard crossing symmetric variables
\be
\label{eq:xandy}
x=st+tu+us=-\frac 12\(s^2+t^2+u^2\)\,, \quad {\rm and} \quad y=stu\,,
\ee
and further defined $a=y/x$. It is apparent that as $|s| \rightarrow \infty$, $a \rightarrow t$. Thus $a$ may be regarded as the crossing symmetric version of $t$. With this in mind, it is natural to ask if there is a dispersion relation in the variable $x \sim -s^2$ defined at fixed $a$. For large $x$, this will be equivalent to the dispersion relation in $s$ at fixed $t$, which in turn means we expect the same overall number of subtractions, namely that for small $a<0$ two subtractions in $s$ which is equivalent to one subtraction in $x$ as $x\sim - s^2$, should be sufficient. Putting this together, we obtain a crossing symmetric dispersion relation
\ba\label{crossamp1}
\cA(s,t,u)&=& c(a)+\frac{1}{\pi}\int_{\Lambda^2}^{\infty}\d \mu \,  \disc_s\cA\(\mu,\tau(\mu,a) \) \( \frac{s^2}{\mu^{2}(\mu-s)}+\frac{t^2}{\mu^{2}(\mu-t)}+\frac{u^2}{\mu^{2}(\mu-u)} \) \,, \\
&=& c(a)+\frac{x}{\pi}\int_{\Lambda^2}^{\infty}\d \mu \,  \frac{\disc_s\cA\(\mu,\tau(\mu,a) \)}{\mu^3} \( \frac{(2\mu-3 a)\mu^2}{x (a-\mu) -\mu^3} \) \,, \\
&=& c(a)+x\int_{\Lambda^2}^{\infty}\d \mu \, \sum_{\nu=0}^{\infty} \rho_{\nu}(\mu) T_\nu\(1+\frac{2 \tau(\mu,a)}{\mu}\) \( \frac{(2\mu-3 a)\mu^2}{x (a-\mu) -\mu^3} \) \, ,
\ea
where $c(a)$ is a subtraction function and we have defined the positive distribution
\be
\rho_\nu(\mu)\equiv \frac{\tilde F^{\nu}(\mu)}{\pi\mu^3}>0\, ,
\ee
associated with each Chebyshev polynomial $T_\nu(\cos \theta) = \cos(\nu \theta)$. $t\leftrightarrow u$ crossing symmetry guarantees that $\rho_\nu(\mu)=0$ for $\nu$ odd, thus the partial waves only include even $3D$ multipoles. \\

The dispersion relation \eqref{crossamp1} may be put in a simpler form by performing a redefinition of the spectral variable to
\be
\omega = \frac{\mu^3}{\mu-a}\,,
\ee
so that
\be\label{crossamp10}
\cA(x,a)= c(a)-\frac{x}{\pi}\int_{\frac{\Lambda^6}{(\Lambda^2-a)}}^{\infty}\d \omega \,  \disc_s\cA\(\mu,\tau(\mu,a) \) \frac{1}{\omega(\omega+x)}\,,
\ee
at the price of a dependence on $a$ in the lower limit and a less transparent $\omega$ dependence of the discontinuity through $\mu=\mu(\omega,a)$. \\

In practice, locality imposes the subtraction function to be of the form
\be
c(a) = c_0 + \frac{c_1}{a} \,,
\ee
with $c_1$ associated with any massless spin-0 pole since other powers of $a$ would give unphysical $1/x$ singularities. Unless otherwise stated, in what follows we shall  only be interested in theories in which only the graviton is massless for which $c_1=0$, but it is easily reintroduced when necessary. The variable $\tau$ is defined via
\be
\tau(\mu,a)=-\frac{\mu}{2} \( 1-\sqrt{\frac{\mu+3a }{\mu-a}} \) \, ,
\ee
which is simply how $t$ is expressed as a function of $a$ at fixed $s=\mu$, that is $\tau(s,a)=t$. Indeed, if we compute the discontinuity of \eqref{crossamp1} at fixed $t<0$ on the right hand branch cut (where $a$ remains $<0$ throughout) we obtain
\be
\disc_s \cA(s,t,u) =\disc_s\cA\(\mu,\tau(\mu,a) \) \Big|_{\mu=s}=\disc_s\cA\(s,t\)\, .
\ee
The validity of this crossing symmetric dispersion relation superficially requires $-\frac{1}{3} \Lambda^2 \le a \le \Lambda^2$ in order that $\tau$ is real, and so necessitates a gap $\mu \ge \Lambda^2$. Thus \eqref{crossamp1} can only be applied to amplitudes for which massless loops are not included. Outside of this region we expect double discontinuities in the manner of the Mandelstam double spectral representation \cite{Mandelstam:1958xc}. From this (as we will argue in Appendix \ref{app:mandel}) we can see that \eqref{crossamp1} acquires a branch cut beginning at $a=2 \Lambda^2/3$. \\

Unlike the standard fixed $t$ dispersion relation, expanding the dispersive integrands in powers of Mandelstam variables $s,t,u$ will give rise to an infinite number of terms with inverse powers of $x$, coming from powers of $a=y/x$, i.e. terms of the form $y^m/x^n$ with $m\ge 0, n>0$ positive. Since the scattering amplitude is local, all such terms must vanish, and this gives an independent derivation of the null constraints utilized in \cite{Tolley:2020gtv} and \cite{Caron-Huot:2020cmc} which are constraints on the discontinuity.\\

In the case of scattering in a gravitational theory with a massless spin-2 state, \eqref{crossamp1} is only expected to hold for $a<0$ due to the same $t$-channel pole which is realized here in crossing symmetric form
\be
{\cal A} \sim  \frac{x}{\mpl^{D-2}a}\,.
\ee
Consequently, to run the traditional positivity arguments it is necessary to perform another subtraction in the dispersion relation, which in analogy with the fixed $t$ case is given simply by
\ba\label{twicesubtracteddispersion}
 \cA(s,t,u) &=&\frac{x^2}{\mpl^{D-2} y}+c(a)+x \, b(a)+\frac{x^2}{\pi}\int_{\Lambda^2}^{\infty}\rd\mu\,\frac{\disc_s\cA(\mu,\tau(\mu,a))}{\mu^6} \( \frac{(\mu-a)(2\mu-3 a)\mu^2}{( \mu^3-x (a-\mu))} \)\,, \nn \\
&=&\frac{x}{\mpl^{D-2} a}+c(a)+x \, b(a)+x^2\int_{\Lambda^2}^{\infty}\rd\mu\,\sum_{\nu=0}^{\infty}\rho_\nu(\mu) T_\nu\(\sqrt{\frac{\mu+3a}{\mu-a}} \)\( \frac{(\mu-a)(2\mu-3 a)\mu^2}{\mu^3( \mu^3-x (a-\mu))} \) \, . \nn
\ea
Here $b(a)$ is an unknown subtraction function which itself admits an expansion in $a$ at low energies. Indeed locality demands that
\be
b(a) = b_0+b_1 a \, ,
\ee
since any higher powers of $a$ would lead to spurious $1/x$ dependence in the amplitude.

\subsection{Leading positivity bounds}
The pole subtracted amplitude
\be
\hat\cA(x,y)=\cA(x,a)-\frac{x}{\mpl^{D-2} a} \, ,
\ee
admits the expansion\footnote{This assumes there are no other massless poles. If spin-0 massless poles are present then these should also be subtracted.}
\ba
\label{eq:exp1}
	\hat\cA(x,y)=\sum_{i\ge 0}\sum_{j\ge0}a_{i,j}x^i y^j=\sum_{i\ge 0}\sum_{j\ge0}a_{i,j}x^{i+j} a^j\,.
\ea
Due to the number of subtractions, the traditional positivity bounds can only give us useful information on the coefficients $a_{i,j}$ for $i+j\ge 2$.
As in the case of the fixed $t$ dispersion relation, we may easily derive an infinite number of nonlinear positivity bounds. For example, defining an $N\times N$ Hankel matrix $H(a)$ whose $nm$ elements with $n,m =0\dots N-1$  are specified by
\ba
H_{nm}(a) &=& (-1)^{2+n+m} \frac{1}{(2+n+m)!}\del_x^{2+n+m}\hat\cA(0,a) \\
&=&\int_{\Lambda^2}^{\infty}\rd\mu\, \disc_s\cA(\mu,\tau(\mu,a)) \frac{(\mu-a)^{1+n+m}(2\mu-3 a)}{\mu^{7+3n+3m}}  \, ,
\ea
then we have
\be
{\rm Det}[H(a)] > 0 \, , \quad 0 \le  a< \frac{2}{3}\Lambda^2 \,,
\ee
together with positivity of the determinant of any minors of this matrix. Bounds of this type have been considered for fixed $t$ dispersion relations in \cite{Arkani-Hamed:2020blm,Bellazzini:2020cot,Chiang:2021ziz,Chiang:2022ltp}. These bounds constrain higher orders in the EFT expansion when those can be computed explicitly, however in what follows we only work with EFT expansions including the  first few orders. Thus the real question we want to ask is given the first few terms in the EFT expansion, what are the strongest bounds that can be made and how this can be folded back into statements on the UV behaviour. Below we explicitly calculate the graviton scattering amplitudes to order $(s,t,u)^6$ in Mandelstam variables, (or up to cubic in $x$, quadratic in $y$). \\

In the forward limit, positivity of the following discontinuities
\ba
\frac{(-1)^n}{2} \p_x^{2n}\hat{A}(0,0)=\frac {n!} \pi \int_{\Lambda^2}^\infty \rd \mu\frac{\disc_s\cA(\mu,0)}{\mu^{2n+1}}>0 \,,
\ea
directly implies
\ba
(-1)^n a_{n,0}>0\quad  \forall \quad n\ge 2\,.
\ea
Away from the forward limit, the leading bounds come from the positivity of
\be
\frac{1}{2} \partial_x^2 \hat\cA(0,a) = \frac{1}{\pi}\int_{\Lambda^2}^{\infty}\rd\mu\,\disc_s\cA(\mu,\tau(\mu,a)) \frac{(\mu-a)(2\mu-3 a)}{\mu^7} > 0\, , \quad 0 \le  a< \frac{2}{3}\Lambda^2 \, .
\ee
In terms of the above parameterization this is
\be
a_{2,0} + a_{1,1} a+ a_{0,2} a^2 >0 , \quad {\rm for }\quad 0 \le  a< \frac{2}{3}\Lambda^2 \,,
\ee
where we obviously recover $a_{2,0}>0$ from $a=0$ and sliding $a$ to its maximum at $a=(2/3)\Lambda^2$, we infer
\be
a_{2,0} + \frac{2 }{3}a_{1,1}\Lambda^2 + \frac{4}{9} a_{0,2}\Lambda^4  >0 \, . \label{cond2}
\ee
Similarly, differentiating\footnote{By taking further derivatives of the dispersion relation and considering its forward limit, we can also obtain
$19 +10 \hat a_{1,1}+4\hat a_{0,2}>0$, however by considering the lower bound on $\hat a_{11}$, this bound is not as strong as the one obtained from allowing $a$ to take its maximum value in the expansion of $\del_x^2\hat\cA(0,a)$.} with respect to $a$ and evaluating at $a=0$ we obtain
\be
\frac{1}{2} \partial_a \partial_x^2 \hat\cA(0,0) ={\frac{1}{\pi}} \int_{\Lambda^2}^{\infty}\rd\mu\,  \( \frac{2 \partial_{\tau}\disc_s\cA(\mu,0)}{\mu^5}-\frac{5\disc_s\cA(\mu,0)}{\mu^6} \) \, ,
\ee
from which we easily infer
\be
 \partial_a \partial_x^2 \hat\cA(0,0)+ \frac{5}{2 \Lambda^2} \partial_x^2 \hat\cA(0,0) >0\,,
\ee
or equivalently
\be
a_{1,1}+\frac{5}{2\Lambda^2}a_{2,0}>0\,. \label{cond1}
\ee
Safe in the knowledge that the leading coefficient $a_{2,0}>0$, it will be useful in what follows to define the dimensionless coefficient ratios
\be
\hat a_{n,m}= \frac{a_{n,m}}{a_{2,0}} \Lambda^{4(n-2)+6 m} \, ,
\ee
with $\hat \mu = \mu/\Lambda^2$. The above linear bounds can then be stated as
\be
\hat a_{0,2}+\frac{3}{2} \hat a_{1,1} >- \frac{9}{4} \, , \quad {\rm and }\quad \hat a_{1,1}>-\frac{5}{2} \, .
\ee

\subsection{Two-sided nonlinear bounds}

To further sharpen these statements we can use the explicit form of the partial wave expansion in terms of Chebyshev polynomials from which we find
\be
a_{2,0} = \biggl[ \frac{2}{\mu^2} \biggr] \, , \quad a_{1,1}= \biggl[ \frac{4 \nu^2-5}{\mu^3}  \biggr]\, , \quad a_{0,2} = \biggl[ \frac{4 \nu^4-34 \nu^2+9}{3\mu^4}  \biggr]\, ,
\ee
together with an infinite set of null constraints from the vanishing of $\partial_a^{m} \partial_x^2\hat\cA(x,0)$ with $m\ge 3$ whose leading two expressions are
\be
\biggl[ \frac{ \nu^2(316-95 \nu^2+4 \nu^4)}{\mu^5} \ \biggr]=0 \, , \quad {\rm and}\quad  \biggl[ \frac{  \nu^2 (-3292 + 1183 \nu^2 - 98 \nu^4 + 2 \nu^6) }{ \mu^6} \ \biggr]=0 \, ,
\ee
and we have defined the integral/sum over $\mu$ and $\nu$ via
\be
\biggl[ X(\mu,\nu^2)  \biggr] =\int_{\Lambda^2}^{\infty}\rd\mu\,\sum_{\nu=0}^{\infty}\rho_\nu(\mu)  X(\mu,\nu^2) \, .
\ee
To the order that we calculate the amplitudes in the next section we can also make use of the coefficients of $x^3$ which is the same order as the coefficient $a_{0,2} $ of $y^2$,
\be
a_{3,0} = -\biggl[ \frac{2}{\mu^4} \biggr] \, .
\ee
An obvious bound is then $0<-a_{3,0} < a_{2,0}/\Lambda^4$. \\

Defining the following normalized moments
\be
\biggl<X(\mu,\nu^2)  \biggr>_{\nu} =\frac{\int_{\Lambda^2}^{\infty} \d\mu\, \rho_\nu(\mu)  \frac{1}{\mu^{2}} X(\mu,\nu^2) }{\int_{\Lambda^2}^{\infty}\rd\mu'\,\sum_{\nu'=0}^{\infty}\rho_{\nu'}(\mu') \frac{1}{(\mu')^{2}}} \, ,
\ee
so that
\be
\biggl<X(\mu,\nu^2)  \biggr> = \sum_{\nu=0}^{\infty }\biggl<X(\mu,\nu^2)  \biggr>_{\nu} \, ,
\ee
we have
\be
\hat a_{1,1} = \biggl<  \frac{2 \nu^2-\frac{5}{2}}{\hat \mu}   \biggr>\, , \quad \hat a_{0,2}  =  \biggl< \frac{2 \nu^4-17 \nu^2+\frac{9}{2}}{3 \hat \mu^2}  \biggr>  \, ,  \quad \hat a_{3,0}  =  -\biggl< \frac{1}{ \hat \mu^2}  \biggr> \, .
\ee
The leading two null constraints are
\be
\biggl< \frac{ \nu^2(316-95 \nu^2+4 \nu^4)}{\hat \mu^3} \ \biggr>=0 \, , \ {\rm and }\quad \biggl< \frac{ \nu^2 (-3292 + 1183 \nu^2 - 98 \nu^4 + 2 \nu^6) }{\hat \mu^4} \ \biggr>=0 \, , {\rm etc}\dots \, . \label{null1}
\ee
As first noted in \cite{Tolley:2020gtv,Caron-Huot:2020cmc}, the null constraints automatically impose two sided bounds on the amplitude coefficients. There are a number of ways to see this, but this is most explicit from recognizing that the only negative contribution from these first two null constraints comes from the $\nu=4$ multipole, (remembering that only even $\nu$ contribute by virtue of crossing symmetry). Explicitly we have (the notation on the RHS indicating summation over $\nu\geq 6$),
\ba
\label{eq:nullcubic}
\biggl< \frac{ 1}{\hat \mu^3}  \biggr>_4&=& \frac{1}{2880}\biggl< \frac{ \nu^2(316-95 \nu^2+4 \nu^4)}{\hat \mu^3} \biggr>_{\nu \ge 6} \, , \\
\label{eq:null}
 \biggl< \frac{ 1}{\hat \mu^4}  \biggr>_4&=& \frac{1}{20160 }\biggl< \frac{ \nu^2 (-3292 + 1183 \nu^2 - 98 \nu^4 + 2 \nu^6) }{\hat \mu^4}  \biggr>_{\nu \ge 6}  \, ,
\ea
where the terms in brackets on the RHS of these two equations are positive for all even $\nu \ge 6$. Pragmatically this implies that all higher multipoles with $\nu \ge 6$ are bounded in terms of $\nu=4$. From this observation alone it is straightforward to derive semi-analytic bounds on the Wilson coefficients.

\subsection{Semi-analytic Bounds}

To demonstrate this, let us consider separately the contribution from $\nu \ge 6$ and define
\be
\hat b_{1,1} \equiv \biggl<  \frac{2 \nu^2-\frac{5}{2}}{\hat \mu}   \biggr>_{\nu \ge 6}\, , \quad \hat b_{0,2}  \equiv  \biggl< \frac{2 \nu^4-17 \nu^2+\frac{9}{2}}{3 \hat \mu^2}  \biggr>_{\nu \ge 6}  \, ,   \quad \hat b_{3,0}  \equiv  -\biggl< \frac{1}{ \hat \mu^2}  \biggr>_{\nu \ge 6} \, .
\ee
In the prescribed range the coefficient of each term in brackets is positive. From a straightforward application of Cauchy-Schwarz $\langle X \rangle^2 \le \langle X^2 \rangle \langle 1 \rangle$ we have
\ba
\hat b_{1,1}^2 &\le&\beta\biggl< \frac{(2 \nu^2-\frac{5}{2})^2}{\hat \mu^2}   \biggr>_{\nu \ge 6}=\beta \(6 \hat b_{0,2} +\biggl<  \frac{(24 \nu^2-11/4)}{\hat \mu^2}   \biggr>_{\nu \ge 6}\) \, ,\\
&\le& \beta \(  6 \hat b_{0,2} +12 \biggl<  \frac{(2 \nu^2-\frac{5}{2})}{\hat \mu^2}   \biggr>_{\nu \ge 6} - \frac{109}{4} \hat b_{3,0} \)\, , \\
&\leq& \beta \( 6 \hat b_{0,2} +12 \hat b_{1,1}-\frac{109}{4} \hat b_{3,0} \)  \, ,\label{constraint1}
\ea
where
\be
\beta = \biggl< 1 \biggr>_{\nu \ge 6 } = \( 1- \sum_{\nu=0}^4  \biggl< 1 \biggr>_{\nu }  \) \, .
\ee
Following a similar reasoning using the second null constraint we have
\be\label{eq:b02}
\hat b_{0,2}^2 \leq  \frac{1}{9}\beta\biggl<  \frac{(2 \nu^4-17 \nu^2+9/2)^2}{\hat \mu^4}   \biggr>_{\nu \ge 6} \, .
\ee
Given that the null constraint \eqref{eq:null} imposes a bound on each multipole with $\nu \ge 6$ in terms of $\nu=4$ we may perform a straightforward maximization of the term in brackets in \eqref{eq:b02} in terms of the $ \nu=4$ multipole, which leads to
\be
\hat b_{0,2}^2 \le  43758\beta \, \biggl< \frac{ 1}{\hat \mu^4}  \biggr>_4 \, . \label{constraint2}
\ee
Given that each term in $b_{1,1}$ is positive we have as an extension of Cauchy-Schwarz
\be
b_{1,1}^3 \le \beta^2 \biggl<  \frac{(2 \nu^2-\frac{5}{2})^3}{\hat \mu^3}   \biggr>_{\nu \ge 6} \, .
\ee
and similarly maximizing the RHS subject to the constraint \eqref{eq:nullcubic} in terms of the $\nu=4$ multipole gives
\be
b_{1,1}^3 \le 12911 \beta^2 \biggl< \frac{ 1}{\hat \mu^3}  \biggr>_4 \, . \label{constraint3}
\ee
Together with the fact that $0 \le -\hat b_{0,3} \le \beta$ and $0 \le \beta \le 1$, the constraints \eqref{constraint1}, \eqref{constraint2} and \eqref{constraint3} impose compact bounds on $\hat b_{0,2}$, $\hat b_{1,1}$ and $\hat b_{0,3}$ in terms of the $\nu=4$ multipole. The actual amplitude coefficients are related to these by
\ba\label{a11}
&& \hat a_{1,1} = \hat b_{1,1} - \frac{5}{2}  \biggl< \frac{ 1}{\hat \mu}  \biggr>_0+\frac{11}{2} \biggl< \frac{ 1}{\hat \mu}  \biggr>_2+\frac{59}{2} \biggl< \frac{ 1}{\hat \mu}  \biggr>_4 \,, \\ \label{a02}
&& \hat a_{0,2} =     \hat b_{0,2}+ \frac{3}{2} \biggl< \frac{ 1}{\hat \mu^2}  \biggr>_0- \frac{21}{2}\biggl< \frac{ 1}{\hat \mu^2}  \biggr>_2+ \frac{163}{2}\biggl< \frac{ 1}{\hat \mu^2}  \biggr>_4   \, ,  \\ \label{a30}
&& \hat a_{3,0} = \hat b_{3,0}  -  \biggl< \frac{ 1}{\hat \mu^2}  \biggr>_0-\biggl< \frac{ 1}{\hat \mu^2}  \biggr>_2-\biggl< \frac{ 1}{\hat \mu^2}  \biggr>_4 \, .
\ea
Thus to infer bounds on the actual amplitudes it is sufficient to further extremize over the $\nu=0,2,4$ multipoles. Given each of these multipoles is independent (unrelated by the null constraints) it is sufficient to impose the positivity of the Hankel determinant of moments for $\nu=0,2,4$
\be
{\rm Det}_{nm } \left[ \biggl< \frac{ 1}{\hat \mu^{n+m}}  \biggr>_{\nu} \right]  > 0 \, ,
\ee
for $n+m=0,\dots, 4$, together with positivity of the minor determinants. A straightforward extremization over the $\nu=0,2,4$ moments yields the bounds
\be
\label{eq:compactBounds}
-\frac{5}{2} \le \hat a_{1,1}  \le  31.4 \, \quad -\frac{21}{2} \le \hat a_{0,2} \le  153 , \quad - 1 \le  \hat a_{3,0} \le  0\,.
\ee
Although not optimal, these bounds which follow from extremizing the first 3 moments are already powerful and apply equally in any dimension. \\

\subsection{Numerical Bounds}

A simple numerical linear optimization which treats all multipoles equally and approximates the scattering amplitude by a finite number of multipoles and imposes the null-constraints by hand over a range of $a$ (including negative $a$ $-\Lambda^2/3<a<(2/3) \Lambda^2$) results in
\be
\label{eq:numericalcompactBounds}
-\frac{5}{2} \le \hat a_{1,1} \le  31 \, \quad -\frac{21}{2} \le  \hat a_{0,2} \le 101 , \quad - 1\le \hat a_{3,0} \le 0\,.
\ee
The lower bounds remain the same because they are determined by the $\nu=0$ and $\nu=2$ multipoles from \eqref{a11}, \eqref{a02}, \eqref{a30}. The upper bound for $a_{1,1}$ matches well the simple semi-analytic approach \eqref{eq:compactBounds} whereas the bound on $a_{0,2}$ is strengthened. Note however that this makes use of the simplest simple linear optimization technique and implementing the bounds  (including non-linear ones) within a fully numerical setup has the potential to significantly tighten these bounds.

\subsection{Preferred expansion variables}

It is clear from the above considerations that the form of the bounds is largely determined by the $\nu=0,2,4$ multipoles. Given this, it is natural to work with a combination of the variables $a_{1,1}$, $a_{0,2}$ and $a_{3,0}$ which is dominated by each multipole specifically.
The more appropriate choice of variables is\footnote{For example, from \eqref{a11} if the amplitude consists solely of a single mass $\mu=\Lambda^2$, $\nu=0$ multipole, then $\alpha_2=\alpha_4=0$.}
\begin{subequations}\label{eq:alpha}
\begin{align}
\label{eq:alpha0}
\alpha_0 &= \frac{1}{512} (12 a_{0,2} - 46 a_{1,1} - 379 a_{3,0}) \, ,\\
\label{eq:alpha2}
\alpha_2 &=\frac{1}{128} (-4 a_{0,2} + 10 a_{1,1} - 31 a_{3,0}) \, ,\\
\label{eq:alpha4}
\alpha_4 &=\frac{1}{512} (4 a_{0,2} + 6 a_{1,1} - 9 a_{3,0}) \, .
\end{align}
\end{subequations}
The semi-analytic bounds on these coefficients are
\begin{subequations}\label{eq:alphaa}
\begin{align}
\label{eq:alpha0a}
& -0.95 \le \alpha_0 \le 2.85 \, ,  \\
\label{eq:alpha2a}
& -3.27 \le \alpha_2 \le 1.10 \, ,\\
\label{eq:alpha3a}
& -0.007  \le \alpha_4 \le 1.56 \, .
\end{align}
\end{subequations}
A numerical linear optimization gives
\begin{subequations}\label{eq:alphan}
\begin{align}
\label{eq:alpha0n}
& -0.6 \le \alpha_0 \le 1 \, ,  \\
\label{eq:alpha2n}
& -0.5 \le \alpha_2 \le 1 \, ,\\
\label{eq:alpha4n}
& -0.007  \le \alpha_4 \le 1.17 \, .
\end{align}
\end{subequations}

\subsection{Smeared positivity bounds}

Although we will not directly follow this approach in what follows, it is straightforward to implement the analogue of the smeared `impact parameter' bounds introduced in \cite{Caron-Huot:2021rmr}. The crossing symmetric dispersion relation automatically implies a set of sum rules which follow from the null constraints/locality. These are obtained by expanding the amplitude in powers of $x$ at fixed $a$. The sum rule that follows from the term linear in $x$ for which the $t$-channel pole contributes will be considered in section \ref{crossingregge}.
Those from higher powers of $x$ take the form (for $n \ge 2$)
\be
\sum_{j=0}^n a_{n-j,j} a^j = (-1)^n \int_{\Lambda^2}^{\infty}\rd\mu\,\sum_{\nu=0}^{\infty}\rho_\nu(\mu) T_\nu\(\sqrt{\frac{\mu+3a}{\mu-a}} \)\( \frac{(\mu-a)^{n-1}(2\mu-3 a)\mu^2}{\mu^{3n}} \)\,. \
\ee
Following the spirit of  \cite{Caron-Huot:2021rmr}, at each order $n$ we may introduce a set of functions $f_n(p)$ defined with $a=-p^2$ with $0<p^2<\Lambda^2/3$ so that positivity of the following integrals holds for all even $\nu$,
\be
\int_0^{\Lambda/\sqrt{3}} \d p \, f_n(p) \int_{\Lambda^2}^{\infty}\rd\mu\,\sum_{\nu=0}^{\infty}\rho_\nu(\mu) T_\nu\(\sqrt{\frac{\mu-3p^2}{\mu+p^2}} \)\( \frac{(\mu+p^2)^{n-1}(2\mu+3 p^2)\mu^2}{\mu^{3n}} \)>0 \, \quad \forall \,  \nu \text{ even}\,.\nn
\ee
It then follows that
\be
\int_0^{\Lambda/\sqrt{3}} \d p \, f_n(p) (-1)^{n+j} \sum_{j=0}^n a_{n-j,j} p^{2j}>0 \, ,
\ee
setting up a linear optimization problem which may be used to constrain the set of coefficients $a_{n-j,j} $ for each $n$. The case $n=1$ can be treated similarly except that we must account for the additional contribution from the pole
\be
\frac{1}{\mpl^{D-2}} \frac{1}{a}+a_{1,0} +a_{0,1} a = - \int_{\Lambda^2}^{\infty}\rd\mu\,\sum_{\nu=0}^{\infty}\rho_\nu(\mu) T_\nu\(\sqrt{\frac{\mu+3a}{\mu-a}} \)\( \frac{(2\mu-3 a)\mu^2}{\mu^{3n}} \) \,,
\ee
so that
\be
\int_0^{\Lambda/\sqrt{3}} \d p \, f_1(p)  \(\frac{1}{\mpl^{D-2}} \frac{1}{p^2}-a_{1,0} +a_{0,1} p^2  \) >0 \, .
\ee
Crucially $a_{1,0}=0$ and $a_{0,1}$ may be chosen to vanish for a specific choice of polarizations as discussed later.

\section{Constraints on Low Energy Gravitational EFTs}
\label{sec:EFT}

\subsection{EFT of Gravity}
We now consider a low energy effective action of gravity in $D$-dimensions and focus on even-parity operators relevant for the 2--2 graviton scattering amplitude at tree level. Relevant to this amplitude, the EFT contains 11 parity-even operators of dimension 4 to 12 as well as 5 parity-odd operators.

A formalism using the Hilbert Series method for the derivation of a non-redundant operator basis for the EFT of gravity in any dimension is presented in \cite{Ruhdorfer:2019qmk} together with the explicit EFT of gravity in $D=4$ up to dimension-10 operators and includes the first line of \eqref{eq:EFT1}. We are here interested in the EFT at higher order and  the Hilbert series is discussed further in Appendix~\ref{app:Hilbert}.
Up to dimension-12, the EFT can be expressed in terms of the parity-even set of operators,
\ba
\label{eq:EFT1}
S=\mpl^{D-2}\int\rd^D x\sqrt{-g}\,\left\{\frac{R}{2}+ \frac{c_{\rm GB}}{\Lambda^2}\cG+\frac{c_3}{\Lambda^4} R^3+\frac{c_1}{\Lambda^6} \cC^2+\frac{c_2}{\Lambda^6}\Tilde{\cC}^2+\frac{e_1}{\Lambda^8}[\mathcal{F}]\cC+\frac{e_2}{\Lambda^8}[\tilde{\mathcal{F}}]\tilde{\cC}\right. \\
+\left. \frac{f_1}{\Lambda^{10}}[\mathcal{F}]^2+\frac{f_2}{\Lambda^{10}}[\tilde{\mathcal{F}}]^2+
\frac{g_1}{\Lambda^{10}}[\mathcal{F}^2]+\frac{g_2}{\Lambda^{10}}[\tilde{\mathcal{F}}^2]+\frac{j_1}{\Lambda^{10}} \mathcal J
\right\}\,,\nn
\ea
where square brackets represent the trace of tensors and we have defined,
\be
\cG\coloneqq R^2-4R_{\mu\nu}R^{\mu\nu}+R_{\mu\nu\rho\sigma}R^{\mu\nu\rho\sigma} \, ,
\ee
\be
	R^3\coloneqq R_{\alpha\beta\rho\sigma}R^{\rho\sigma}_{\phantom{\alpha\beta}\mu\nu}R^{\mu\nu\alpha\beta}\,,\quad\cC\coloneqq R_{\mu\nu\rho\sigma}R^{\mu\nu\rho\sigma}\,,\quad\Tilde{\cC}\coloneqq\frac12R^{\alpha\beta\mu\nu}\epsilon_{\mu\nu\rho\sigma}R^{\rho\sigma}_{\phantom{\rho\sigma}\alpha\beta}\,,
\ee
$$
	\mathcal{F}_{\alpha\beta}\coloneqq \nabla_{\alpha}R_{\mu\nu\rho\sigma}\nabla_{\beta}R^{\mu\nu\rho\sigma}\,,\quad\tilde{\mathcal{F}}_{\alpha\beta}\coloneqq\frac12 \nabla_{\alpha}R^{\gamma\delta \mu\nu}\nabla_\beta\left(\epsilon_{\mu\nu\rho\sigma}R^{\rho\sigma}_{\phantom{\rho\sigma}\gamma\delta}\right)\,,
$$
and
$$
\mathcal J = \nabla^\mu\nabla^\nu R^{\alpha\beta\gamma\delta}\nabla^\varepsilon R^{\chi}_{\phantom{\chi}\zeta\delta \gamma}\nabla^{\zeta}R_{\varepsilon\alpha \iota\beta}R^{\iota}_{\phantom{\iota}\nu\mu\chi}\,.
$$
In four dimensions, this includes all the even-parity operators up to dimension-12 which contribute at tree-level to the $2-2$ scattering amplitude.
Any other operator is hence either redundant (removable by field redefinition) or does not contribute to the graviton $2-2$ amplitude at tree-level.  Along with these parity-even operators, one could also include the 5 parity-odd operators $\cC \Tilde{\cC}, [\mathcal{F}]\Tilde{\cC}, [\mathcal{F}] [\Tilde{\mathcal{F}}], [\mathcal{F}\Tilde{\mathcal{F}}]$ and $\Tilde{\mathcal{J}}$. In what follows, we shall only focus on parity-even operators (aside when comparing with heterotic string theory as considered in Appendix~\ref{app:coeff}). \\

On computing the four-point graviton amplitudes at tree level we find the three independent helicity amplitudes\footnote{The remainder of the helicity amplitudes may be obtained using crossing symmetry and parity, e.g. $\cA_{11}(s,t,u)=\cA_{++--}(s,t,u)=\cA_{+-+-}(t,s,u)=\cA_{23}(t,s,u)$ and so on. Parity flips the helicities of all external gravitons, under which the amplitudes are invariant (e.g. $\cA_{++;-+}=\cA_{--;+-}$). We can see that as a consequence of crossing symmetry the last two lines are themselves totally symmetric in $s,t,u$.},
\be
\begin{aligned}
\mpl^{{D-2}}\cA_{11}&=\frac{s^3}{t u}-\frac{8(D-4)}{D-2}\frac{c_{\rm GB}^2}{\Lambda^4} s^3-18 c_{3}^2 s^3 \left(\frac{(D-4)}{(D-2)} s^2+2  s t+2  t^2\right)
\\&\phantom{=}+\frac{8}{\Lambda^6}c_+ s^4+\frac{4}{\Lambda^8} e_+s^5+\frac{2f_+ }{\Lambda^{10}}s^6+\frac{g_+}{\Lambda^{10}}s^4(u^2+t^2)
\,,\\
\mpl^{{D-2}}\cA_{14}&=\frac{12}{\Lambda^4}\(5c_3 -\frac{2 (D-4) }{(D-2)}c_{\rm GB}^2\)y-\frac{2}{\Lambda^8}\left(\frac{9(12-D)}{(D-2)}c_3^2+10e_-\right) x y+\frac{16}{\Lambda^6}c_-x^2\nn \\&\phantom{=}+\frac{2f_- }{\Lambda^{10}}  \left(3y^2-2x^3\right)
-\frac{g_-}{\Lambda^{10}}(3y^2+2x^3)-\frac{3}{4\Lambda^{10}}j_1 y^2
\,,\\
\mpl^{{D-2}}\cA_{13}&=\frac{6}{\Lambda^4}c_3 y-\frac{3}{16\Lambda^{10}}j_1 y^2\,,
\end{aligned}
\ee
where we have defined $c_\pm=c_1\pm c_2$, and similarly for  $e_\pm, f_\pm, g_\pm$ and where in the last two lines we have made use of the explicitly $s,t,u$ crossing symmetric variables $x$  and $y$ given in \eqref{eq:xandy}.
\\

In terms of the crossing symmetric variables $x$ and $y$,
the manifestly indefinite crossing-symmetric amplitude \eqref{expans} (now denoted $\cA\equiv\bra{\rm in}\hat T\ket{\rm in}$) is given by
\ba\label{triplecrossingamp}
\mpl^{D-2}\cA&=&\frac{x^2}{y}+\frac{6 }{ \Lambda ^4} \(2  c_3\cos \phi +5 c_3  \cos 2 \phi -4 c_{\rm GB}^2\cos ^2\phi  \frac{D-4}{D-2}\) y\\
&+& \frac{8}{\Lambda^6} \(  c_++   \cos 2 \phi c_-\) x^2
-\frac{10}{\Lambda^8} (e_++ \cos2 \phi\, e_-) x y+\frac{18 (D-12)  \cos ^2(\phi )}{\Lambda^8(D-2)}  c_3^2\,x y \nn\\
&+&
\frac{\left(f_+ +\cos2\phi f_-\right)}{\Lambda^{10}}\left(3y^2-2x^3\right)
- \frac{\left(g_+ +\cos2\phi g_-\right)}{2\Lambda^{10}}\left(3y^2+2x^3\right)-\frac{3(\cos\phi+\cos2\phi)j_1}{8\Lambda^{10}} y^2
\, .\nn
\ea
As already mentioned, a striking feature of this amplitude is the vanishing coefficient of $x$, i.e. $a_{1,0}$ in the low energy expansion whilst the coefficient of $y$, i.e. $a_{0,1}$ is generally non-zero. For scalar theories, positivity bounds on triple crossing symmetric amplitudes would not allow the coefficient of the $x$ term to vanish without resulting in both $a_{1,0}$ and $a_{0,1}$ vanishing, as $a_{0,1}$ is bounded from above and below by an amount proportional to $a_{1,0}$ \cite{Tolley:2020gtv}.

Interestingly, the  $e_\pm$ operators cannot be generically absorbed into that of $c_3^2$ and the coefficient of the $xy$ term needs not be generically positive definite. We shall therefore see that positivity bounds do imply an upper and lower bound on the coefficients $a_{1,1}$ (i.e. $a_{1,1}$ needs not be taken positive definite from the outset) which implies a genuine upper and lower bound on the dimension-10 operators captured by $e_\pm$.


\subsection{Constraints on Wilson coefficients}
\subsubsection{Indefinite helicity elastic amplitude}
Whilst our discussion has been thus far focussed around the triple crossing symmetric dispersion relation and scattering amplitude, we also have at our disposal the standard fixed $t$ positivity bounds, applicable to the \textit{elastic indefinite helicity} amplitude (given in Equation \ref{expansindef}). From the twice subtracted dispersion relation one may derive the following positive quantities,
\be
\begin{aligned}
	\del_s^4\cA(0,t)&>0\,,\quad\text{(a)}\\
	\del_s^6\cA(0,t)&>0\,,\quad\text{(b)}\\
	\del_s^5\cA(0,t)+\frac{5}{t+\Lambda^2}\del_s^4\cA(0,t)&>0\,,\quad\text{(c)}\\
	\del_s^4\del_t\cA(0,t)+\frac{5}{t+\Lambda^2}\del_s^4\cA(0,t)&>0\,,\quad\text{(d)}\\
	\del_s^4\del_t^2\cA(0,t)+\frac{10}{(t+\Lambda^2)}\left(\del_s^4\del_t\cA(0,t)+\frac{5}{t+\Lambda^2}\del_s^4\cA(0,t)\right)&>0\,,\quad\text{(e)}\\
	\del_s^5\del_t\cA(0,t)+\frac{5}{(t+\Lambda^2)}\left(\del_s^4\del_t\cA(0,t)+\frac{5}{t+\Lambda^2}\del_s^4\cA(0,t)\right)&>0\,,\quad\text{(f)}\,.\\
\end{aligned}
\ee
In the above inequalities the scale $\Lambda$ is the energy at which the branch cut of the exact scattering amplitude begins, which we have chosen to identify with the scale appearing in the EFT action. These inequalities when applied to the exact scattering amplitude are valid for positive real $t$ within the region of analyticity, however when applied to the EFT amplitude computed to finite order in powers of $s$ and $t$ have a reduced regime of validity in $t$. Due to this we evaluate the positive quantities above at $t=0$ to avoid any potential for inaccuracy. Applying these inequalities to the elastic indefinite helicity amplitude we find the following; equation (a) implies,
\begin{equation}
	c_{-} \sin (2 \theta ) \sin (2 \chi ) \cos (\psi +\phi )+c_{+}>0 \quad \implies\quad c_+>|c_-|\quad \iff\quad c_{1,2}>0\,.
\end{equation}
Equation (b) implies,
\begin{equation}
	(2 f_{-}+g_{-}) \sin (2 \theta ) \sin (2 \chi ) \cos (\psi +\phi )+2 f_+ +g_{+}>0\quad \implies\quad 2f_+ + g_+>|2f_- +g_-|\,.
\end{equation}
Bound (c) gives,
\begin{equation}
	\cos (2 \theta ) \cos (2 \chi ) \left(2 e_{+}-9 c_{3}^2 \frac{(D-4)}{(D-2)}\right)+4 (c_{-} \sin (2 \theta ) \sin (2 \chi ) \cos (\psi +\phi )+c_{+})>0\,.
\end{equation}
The second term is positive by virtue of bound (a) and so this inequality bounds a combination of $e_+$ and $c_3^2$ from above and below. For example taking $\cos(\phi+\psi)=0$ gives,
\begin{dmath}
	4c_+>\left| 2 e_{+}-9 c_{3}^2 \frac{(D-4)}{(D-2)}\right|\,,
\end{dmath}
which reduces in $D=4$ to an upper and lower bound solely on $e_+$,
\begin{dmath}
2c_+>	|e_+|\,.
\end{dmath}
Despite the fact that we made the particular choice of $\cos(\phi+\psi)=0$, this previous bound is in fact the strongest statement (when combined with $c_+>|c_-|$) that one can derive from the bound (c).
Bound (d) gives,
\begin{dmath}
	\frac{1}{D-2}\left(\sin (2 \theta ) \sin (2 \chi ) \cos (\psi +\phi ) \left(9 c_{3}^2 (D-12)+10 (D-2) (4 c_{-}-e_{-})\right)-5 \cos (2 \theta ) \cos (2 \chi ) \left(9 c_{3}^2 (D-4)-2 (D-2) e_{+}\right)+9 c_{3}^2 (D-12)+10 (D-2) (4 c_{+}-e_{+})\right)>0\,.
\end{dmath}
Again setting $\cos(\phi+\psi)=0$ as well as $\cos(2\theta)\cos(2\chi)=1$ simplifies this greatly to an upper and lower bound on the value of $c_3$, valid for any $D\geq3$,
\begin{equation}\label{c3bound}
	c_3^2< \frac{10 }{9 }c_+\,.
\end{equation}
Setting $\cos(\phi+\psi)=0$ and $\cos(2\theta)\cos(2\chi)=0$ and combining with bound (c) on the other hand gives,
\begin{equation}
	\frac{9 c_{3}^2 (D-12)}{10(D-2)}+4 c_{+}>e_{+}>-2c_++\frac{9(D-4)}{2(D-2)}c_3^2\,.
\end{equation}
We can compare this with bound (c) in the form,
\begin{equation}
2c_++\frac{9(D-4)}{2(D-2)}c_3^2>e_{+}>-2c_++\frac{9(D-4)}{2(D-2)}c_3^2\,,
\end{equation}
however we cannot conclude that either of these double sided bounds are stronger than the other as the difference between the two upper-bounding quantities is sign indefinite. For $D=4$ these two bounds read,
\begin{equation}
	4c_+-\frac{18}{5}c_3^2 > e_+ > - 2c_+ \,,\quad 2c_+ > e_+ > - 2c_+\,.
\end{equation}
Bound (e) gives,
\begin{dmath}
	\frac{1}{D-2}\left(\sin (2 \theta ) \sin (2 \chi ) \cos (\psi +\phi ) \left(360 c_{3}^2 (D-12)+(D-2) (1600 c_{-}-400 e_{-}+120 f_{-}+36 g_{-}-3 j_{1})\right)-20 \cos (2 \theta ) \cos (2 \chi ) \left(90 c_{3}^2 (D-4)-(D-2) (20 e_{+}-6 f_{+}-g_{+})\right)\right)+4 \left(\frac{90 c_{3}^2 (D-12)}{D-2}+400 c_{+}-100 e_{+}+30 f_{+}+9 g_{+}\right)-3 j_{1} (\sin (\theta ) \cos (\theta ) \cos (\phi )+\sin (\chi ) \cos (\chi ) \cos (\psi )) > 0\,,
\end{dmath}
which at $D=4$ reduces to,
\begin{dmath}
	\sin (2 \theta ) \sin (2 \chi ) \cos (\psi +\phi ) \left(-80 \left(18 c_{3}^2+5 e_{-}\right)+1600 c_{-} +3(40 f_{-}+12 g_{-}-j_{1})\right)-80 \left(18 c_{3}^2+5 e_{+}\right)+1600 c_{+}-20 \cos (2 \theta ) \cos (2 \chi ) \left((6 f_{+}+g_{+})-20 e_{+} \right)+12 (10 f_{+}+3 g_{+})-3 j_{1}  (\sin (\theta ) \cos (\theta ) \cos (\phi )+\sin (\chi ) \cos (\chi ) \cos (\psi ))>0\,.
\end{dmath}
One simple example case of this bound is at $\chi=\psi=0,\, \theta=\pi/4,\, \phi=\pi/2$ giving,
\begin{dmath}
400 c_{+}+30 f_{+}+9 g_{+} > 10\left(10e_+ +\frac{9 c_{3}^2 (12-D)}{D-2}\right)
\end{dmath}
Bound (f) gives,
\begin{dmath}
	\sin (2 \theta ) \sin (2 \chi ) \cos (\psi +\phi ) \left(9 c_{3}^2 (D-12)+(D-2) (40 c_{-}-10 e_{-}+6 f_{-}+3 g_{-})\right)+\cos (2 \theta ) \cos (2 \chi ) \left((D-2) (10 e_{+}-6 f_{+}-g_{+})-45 c_{3}^2 (D-4)\right)+9 c_{3}^2 (D-12)+(D-2) (40 c_{+}-10 e_{+}+6 f_{+}+3 g_{+})>0\,,
\end{dmath}
which in $D=4$ simplifies to,
\begin{dmath}
	\sin (2 \theta ) \sin (2 \chi ) \cos (\psi +\phi ) \left(-2\left(18 c_{3}^2+5 e_{-}\right)+40 c_{-} +3(2 f_{-}+g_{-})\right)-2 \left(18 c_{3}^2+5 e_{+}\right)+40 c_{+} -\cos (2 \theta ) \cos (2 \chi ) \left( (6 f_{+}+g_{+})-10 e_{+}\right)+3  (2 f_{+}+g_{+})>0\,.
\end{dmath}
From bound (f), taking $\chi=\theta=\psi=0$ and $\phi = \pi/2$ gives a bound on $g_+$ from below, valid for any $D\geq3$,
\begin{equation}
g_+>-2\left|9 c_{3}^2-10 c_{+} \right|\,.
\end{equation}
\subsubsection{Manifestly crossing symmetric amplitude}
\paragraph{Bounds on dim-8 operators, ($Riemann^4$):}
Having explicitly computed the amplitude for our one parameter family of polarizations it is now straightforward to read off the bounds. The leading bound is
\be
a_{2,0}= \frac{8}{\Lambda^2} (c_+ + \cos2 \phi c_-)>0\,,
\ee
which is clearly most informative at the extremes $\cos2 \phi=\pm 1$ where we learn that
\be
c_{1,2} >0 \, ,
\ee
consistent with \cite{Bellazzini:2015cra,Bern:2021ppb} as well as with pure infrared causality constraints \cite{Gruzinov:2006ie,deRham:2021bll}.

\paragraph{Bounds on $Riemann^3$ and dim-10 operators:}
Bounds on the Wilsonian coefficient of the $Riemann^3$ were already discussed in \cite{Bern:2021ppb} including dimension-10 operators in \cite{Caron-Huot:2022ugt} and for completeness we briefly discuss them in the context of our analytic compact bounds here.

Even though the amplitude is sensitive to the coefficient $c_3$ of the $Riemann^3$ operator, the subtleties associated with applying double-subtracted bounds imply that there is to date no positivity bounds on the coefficient $c_3$ directly, i.e. on its sign\footnote{Causality constraints can on the other provide direct bounds on the sign of $c_3$, which is required to be positive in any tree-level high energy completion \cite{deRham:2021bll}.}, and only $c_3^2$ can be bounded using higher order moments bounds. It enters the relevant coefficient $\hat a_{1,1}$ as follows
\ba
-\frac 5 2<\hat a_{1,1} &=&-\frac{1}{4} \frac{\(18  (1+\cos2 \phi)c_3^2+ 5 (e_++ \cos2 \phi\, e_-)\)}{\(  c_++   \cos 2 \phi  \,c_- \) }  < 31\,,
\ea
and evaluating again at the extremes $\cos 2\phi=\pm 1$ leads to the following two bounds
\ba
-124c_1< 18  c_3^2+ 5 e_1  < 10c_1 \quad {\rm and }\quad
-\frac{124}5c_2< e_2  <2c_2 \,. \label{eq:a11}
\ea
In addition, from the standard fixed $t$ positivity bounds given above we have the bound $c_3^2 < 10c_+ / 9$ which implies,
\begin{equation}
	-\frac{144}{5}c_1 -4 c_2 < e_1 < 2c_1\,,
\end{equation}
which are slightly weaker but similar in spirit to those obtained in \cite{Caron-Huot:2022ugt} at that order.
Any supersymmetric EFT would satisfy $c_3=0$ and hence would require
\ba
-\frac{124}5c_1<e_1^{\rm susy}<2c_1\,.
\ea

\paragraph{Bounds on dim-12 operators:}

The remaining coefficient ratios are
\ba
-\frac{21}2<\hat a_{0,2} &=& \frac{3}{8} \frac{\left(f_+ + f_-\cos2\phi -\frac{1}{2}(g_+ +g_-\cos 2\phi)-\frac{1}{8}j_1 (\cos\phi+\cos2\phi)\right)}{\(  c_++   \cos 2 \phi  \,c_- \) }<101 \\
 -1<\hat a_{3,0} &=&-\frac{1}{4} \frac{\left(f_+ + f_-\cos2\phi+\frac{1}{2}(g_+ +g_-\cos2\phi)\right)}{\(  c_++   \cos 2 \phi  \,c_- \) } <0 \, .
\ea
Evaluating them at the extremes, the relevant angles are at  $\phi =\pi/3, \pi/2$ and $\pi$, from where we infer the following five bounds that can be used as compact bounds for $f_{1,2}, g_{1,2}$ and $j_1$,
\ba
\label{eq:phipi}
{\rm at}\quad \phi&=&\pi, \quad-56c_1<2f_1- g_1 < \frac{1616}{3} c_1 \quad  {\rm and } \quad
0<  2f_1+ g_1 <8c_1\, ,\\
{\rm at}\quad \phi&=&\pi/2,\quad
-56 c_2< 2f_2- g_2+\frac18 j_1< \frac{1616}{3} c_2 \quad  {\rm and } \quad
 0<2f_2+g_2 <8c_2\, ,\quad \\
 {\rm at}\quad \phi&=&\pi/3,\quad
-56<\frac{2(f_1+2f_2)-(g_1+3g_2)}{c_1+3c_2}< \frac{1616}{3}\, .
\label{eq:phipi3}
\ea
These relations imply compact bounds on each individual Wilsonian coefficients in terms of $c_i$
\ba
& -14 c_1<f_1<\frac{410}{3} c_1 \quad {\rm and }\quad -\frac{808}{3} c_1<g_1<32c_1\\
&-14 c_2<f_2+\frac{1}{32}j_1<\frac{410}{3} c_2 \quad {\rm and }\quad -\frac{808}{3} c_2<g_2-\frac{1}{16}j_1<32c_2  \\
& |j_1|<1903 (c_++2 c_2)\,,
\ea
however the bounds (\ref{eq:phipi}--\ref{eq:phipi3}) are overall stronger and should remain satisfied.\\

In many of the positivity bounds above, we see the appearance of $c_3^2$ which, when considering a loop-level completion, is suppressed by factors of $\Lambda/\mpl$ compared to expressions linear in the Wilson coefficients, and so may be neglected in these cases. This is not the case for tree-level completions such as the string theories for which the Wilson coefficients are given in Appendix~\ref{app:coeff}. Interestingly though, in all supersymmetric completions, $c_3=0$, so only the case of bosonic string theory gives a concrete example where the contribution from that operator is relevant. \\

\begin{figure}
\centering
\includegraphics[width=0.3\textwidth]{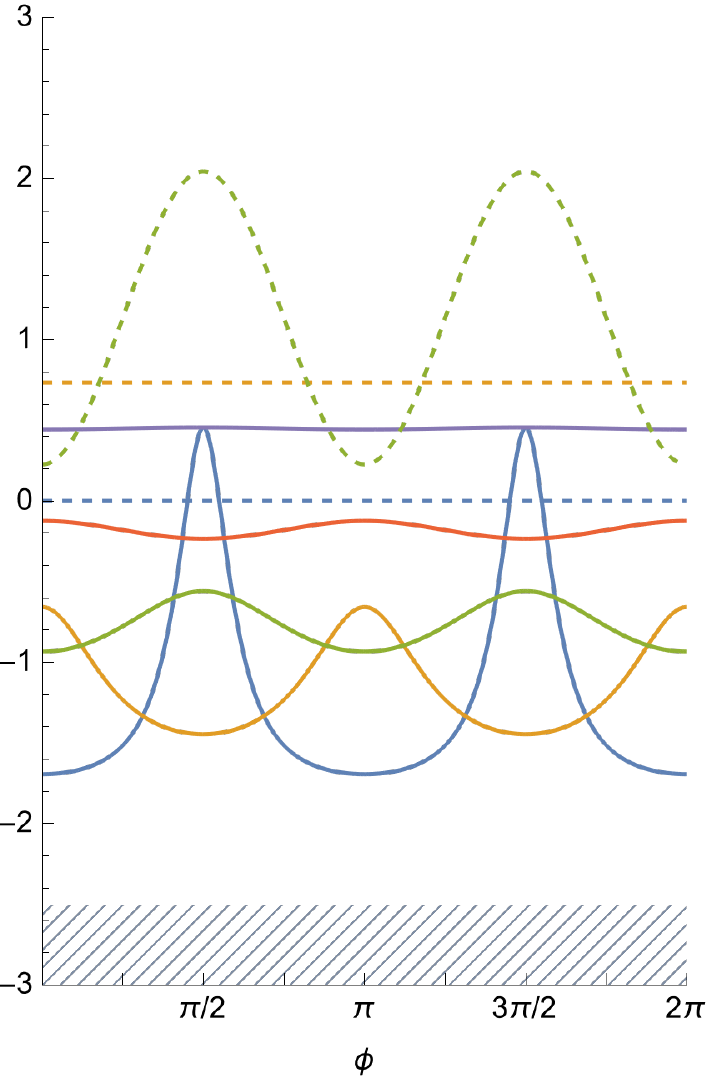}\ \
\includegraphics[width=0.3\textwidth]{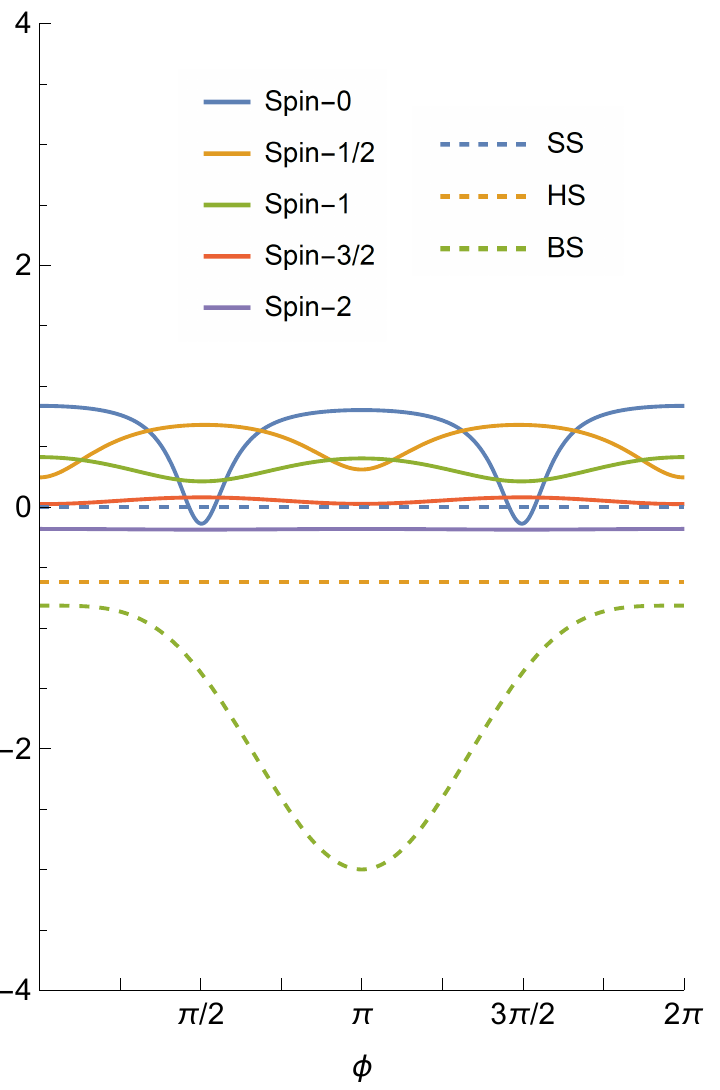}\ \
\includegraphics[width=0.3\textwidth]{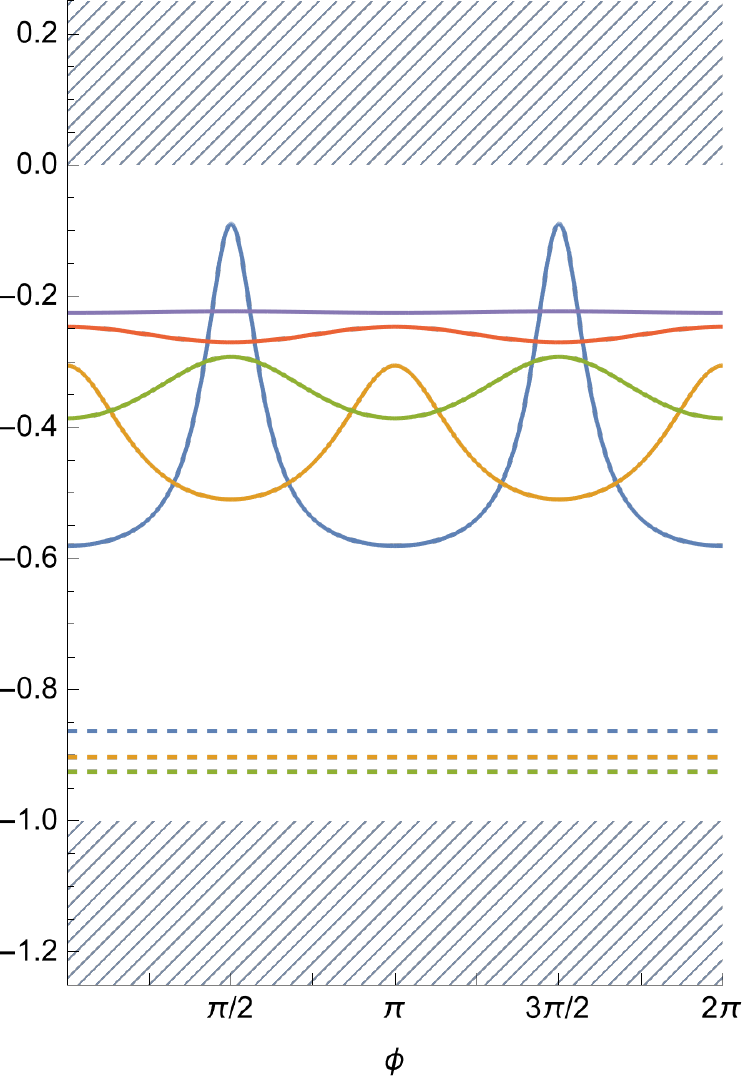}\\
\vspace{-7.2cm}
\hspace{0.5cm}$\hat a_{1,1}$ \hspace{4.5cm} $\hat a_{0,2}$ \hspace{4.1cm} $\hat a_{3,0}$ \vspace{6.7cm}
\caption{\label{fig:csbounds}
Coefficients arising from the string completions (dashed) and loops of lower spin (solid). As expected, for these (partial) UV completions,  all coefficients lie within the allowed range defined by positivity. Parameters lying within the shaded regions are deprived from ever enjoying a standard (Wilsonian) high energy completion. As indicated analytically in \eqref{eq:compactBounds} or semi-numerically in \eqref{eq:numericalcompactBounds}, the allowed region is compact for all three amplitude parameters $\hat{a}_{1,1}$, $\hat{a}_{0,2}$ and $\hat{a}_{3,0}$, however the upper or lower bounds are sometimes well away from the regions explored by BS, HS or SS string theory completions or by lower-spin contributions and hence not always represented in these figures. }
\end{figure}

We can visualise the allowed regions on different combinations of Wilson coefficients from both the fixed-$t$ positivity bounds and the crossing symmetric bounds and where different partial UV completions (arising from massive spin $\leq 2$ particles or string theories) lie in these regions. In all cases we take the spacetime dimension to be $D=4$. As expected, and can be seen explicitly in Fig.~\ref{fig:csbounds}, the double sided bounds from crossing symmetry are all satisfied by the coefficients arising from integrating out minimally coupled fields of spin $\leq2$. In addition, we also include the equivalent coefficients for the respective bosonic, heterotic and superstring tree-amplitudes (BS,HS and SS).\\

Using the explicit values of the coefficients given in  Appendix~\ref{app:coeff}, we can verify that the bounds are satisfied for loop-level partial UV completions of a single spin $\leq$ 2 particle of mass $m$ (for which $\Lambda = 2m$), as well as for the three string theory completions (BS,HS and SS). We can see that, especially for the lower spin loops, the choice of angle $\phi$ can lead to a relatively large change in the magnitude and also the sign of certain coefficients in the expansion of the amplitude and so may prove a useful tool for exploring the allowed parameter space.
\begin{figure}
\centering
\includegraphics[width=0.3\textwidth]{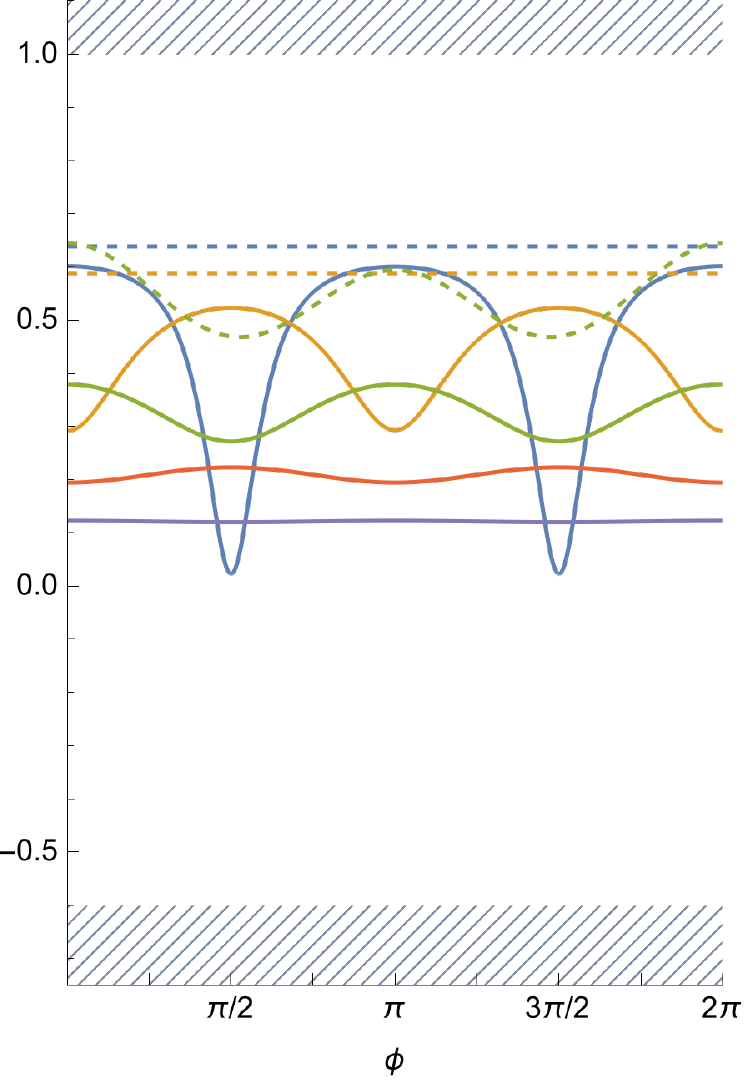}\ \
\includegraphics[width=0.3\textwidth]{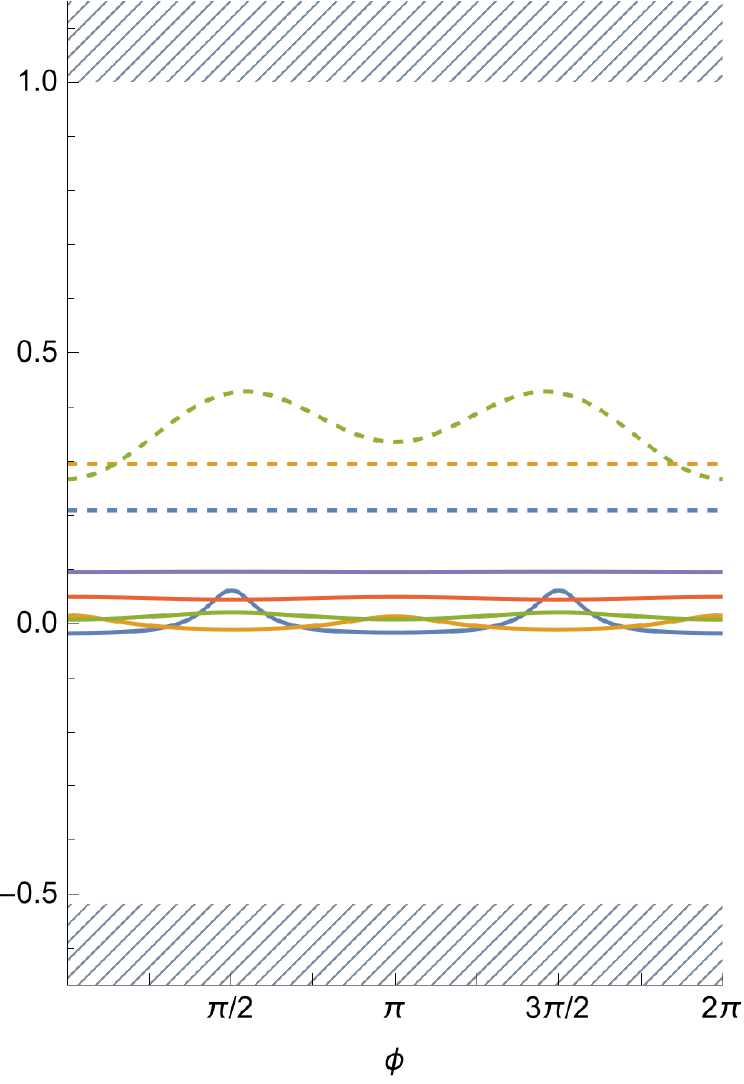}\ \
\includegraphics[width=0.3\textwidth]{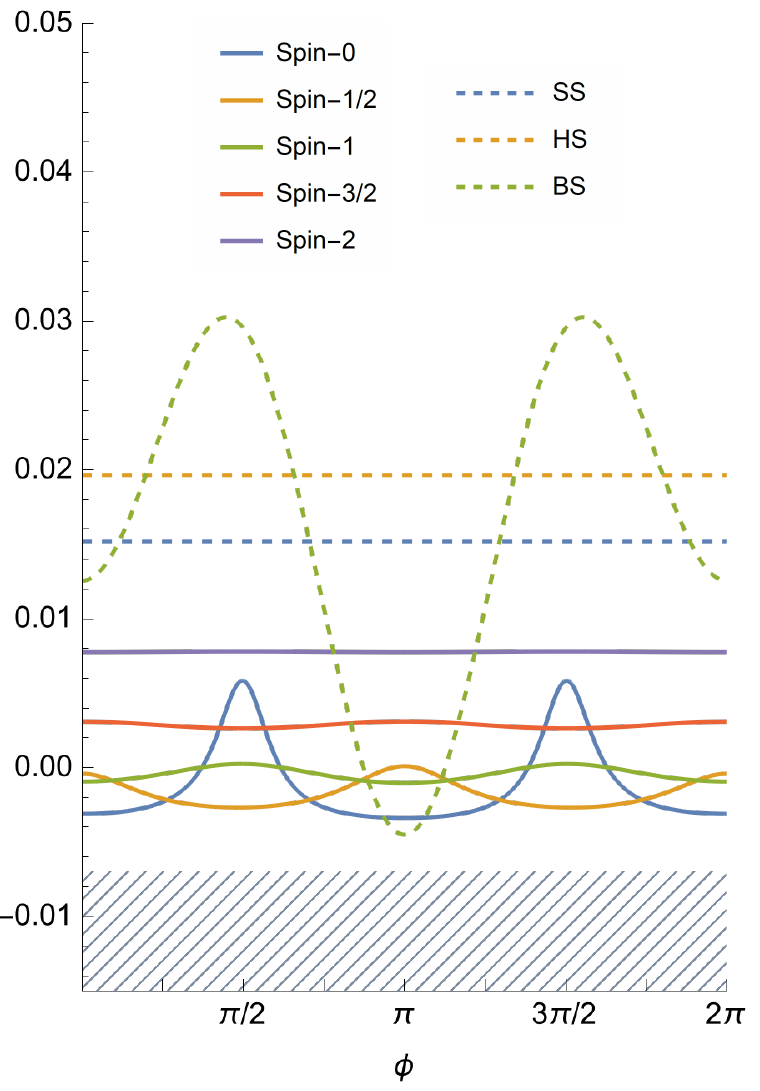}\\
\vspace{-7.2cm}
\hspace{0.5cm}$\alpha_0$ \hspace{4.5cm} $\alpha_2$ \hspace{4.4cm} $\alpha_4$ \vspace{6.7cm}
\caption{Amplitude contributions from the $\mu=0,2,4$ multipoles expressed as the $\alpha_{0,2,4}$ introduced in \eqref{eq:alpha},  from string completions (dashed) and loops of lower spin (solid). All coefficients lie within the allowed range defined by positivity bounds. The shaded regions denote  forbidden values from positivity bounds.
As indicated analytically in \eqref{eq:alphaa} or semi-numerically in \eqref{eq:alphan}, the contribution from all three multipoles are bounded, however the upper or lower bounds are sometimes well away from the regions explored by BS, HS or SS string theory completions or by lower-spin contributions and hence not always represented in these figures.
}
\end{figure}

\section{Constraints on UV Regge Behaviour from analyticity and positivity}
\label{sec:Regge}

 If we were dealing with a lower spin EFT (with light states of spin $<2$), the absence of a term linear in $x$ would by itself be sufficient to cast doubt on the existence of a {\it standard} high-energy completion, as is the case for instance in massless Galileon EFTs \cite{Adams:2006sv,deRham:2017imi,Tolley:2020gtv,Bellazzini:2020cot}. In the EFT of gravity, the presence of the $t$-channel pole prohibits the direct application of standard positivity bounds without folding in other assumptions \cite{Alberte:2020bdz,Alberte:2020jsk,Tokuda:2020mlf,Noumi:2021uuv,Aoki:2021ckh,Alberte:2021dnj,Herrero-Valea:2022lfd}. In what follows we shall therefore see that the absence of a term linear in $x$ implies constraints on the UV Regge behaviour\footnote{The existence of a Regge behaviour does not rely on the assumption of a string theory completion. Rather,  as discussed in \eqref{eq:largemu} and \eqref{CSRegge},  it is an unavoidable outcome of the graviton $t$-channel pole when assuming the existence of a dispersion relation with two subtractions for $t<0$ (or here $a<0$).}. In particular we will see that the Regge trajectory and residue are sensitive to the loops of the lightest massive particles and we are able to determine this contribution by utilizing analyticity. This is similar in spirit to what was proven in \cite{Alberte:2021dnj} in the context of graviton-photon scattering and \cite{Herrero-Valea:2022lfd} for photon-photon scattering. For example the fact that the Regge behaviour in the presence of $U(1)$ gauge fields was necessarily sensitive to the lightest {\it charged} particles was pointed in \cite{Alberte:2021dnj}. We will see that this behaviour appears to be ubiquitous to any gravitational EFT, involving in the UV a non-trivial dependence on the scale of the lightest massive particles, {\it charged or not}. This connection between the UV and IR was of course integral to the bootstrap approach of the historical finite energy sum rules \cite{Igi:1962zz,Logunov:1967dy,Igi:1967zzb,Igi:1967zza,Dolen:1967jr} and may be summarized in the following statement:
 \begin{itemize}
 \item Knowledge of the IR scattering amplitude for energies $<\Lambda_r$ is sufficient to determine the UV scattering amplitude for energies $>\Lambda_r$,
 \end{itemize}
 where $\Lambda_r$ is the energy scale at which the Regge behaviour kicks in. Our goal is to extend this observation to gravitational theories.

\subsection{Crossing symmetric Regge behaviour}
\label{crossingregge}

Following the procedure of \cite{Alberte:2021dnj}, we derive a fully crossing symmetric dispersion relation for the pole subtracted amplitude that is valid across the forward limit singularity. As already discussed, for a triple crossing symmetric configuration of polarizations the graviton scattering amplitude will contain exchange poles in the form $x^2/y=x/a$ (or $x/a$ where $a=y/x$) in the low energy expansion. Noting that this term does not explicitly appear in the once (in $x$) subtracted dispersion relation, it must be produced implicitly from a divergence of the integral over the discontinuity. In terms of the crossing symmetric variables, the dispersion relation reads,
\be\label{crossamp5}
\cA(x,a)= c(a)+\frac{x}{\pi}\int_{\Lambda^2}^{\infty}\d \mu \,  \frac{\disc_s\cA\(\mu,\tau(\mu,a) \)}{\mu^3} \( \frac{(2\mu-3 a)\mu^2}{x (a-\mu) -\mu^3} \)\,,\quad a < 0 \,.
\ee
Considering the amplitude at fixed $a$, we have
\ba
\left.\del_x\cA(x,a)\right|_{x=0} =\frac{1}{\mpl^{D-2}}\frac 1a + a_{1,0}+a_{0,1} a\,,
\ea
where {\it any} gravitational amplitude necessarily has $a_{1,0}=0$, in contradistinction to the more familiar scalar case. As for the coefficient $a_{0,1}$ it depends in principle on the Wilsonian coefficients of the EFT, so one could in principle be tempted to interpret the running of these coefficients with the scaling of $\Lambda$ in the dispersion relation. However this interpretation would be incorrect as can clearly be seen by choosing the choice of indefinite parameter $\phi=\phi_0$ for which the coefficient $a_{0,1}$ vanishes for any gravitational theory in the absence of gravitationally non-minimally coupled massless scalars (consistent with the realization of Nature we observe).
 For that specific choice of indefinite parameter $\phi_0$ (which in $D=4$ satisfies $\cos \phi_0=-5/2 \cos 2 \phi_0$), we see that the dispersion relation satisfies the universal relation (for negative $a$),
\ba
\label{eq:exactRelation}
\boxed{
-\frac{1}{\mpl^{D-2}}\frac 1a\equiv
\frac{1}{\pi}\int_{\Lambda^2}^{\infty}\d \mu \,  \frac{\disc_s\cA\(\mu,\tau(\mu,a),\phi_0 \)}{\mu^4} \( 2\mu -3 a \)\,,  \quad a<0\,,}
\ea
for any gravitational theory (in the absence of massless scalars), irrespectively of the scale $\Lambda$ and the details of IR or UV physics. This relation already proves the presence of non-trivial IR/UV mixing whereby IR contributions to the dispersion relation from the region $\int_{\Lambda^2}^{\Lambda_c^2}\d \mu$ ought to be entirely compensated by UV contributions in the region $\int_{\Lambda_c^2}^{\infty}\d \mu$, so that the combination of both always leads to the same $1/a$ outcome. \\

We can make the previous arguments more concrete by extending these relations to the region of positive $a$ where unitarity rules can be applied.
So far the dispersion relation is valid for negative values of $a$, and  diverges as $a\xr 0^-$ due to the graviton exchange pole. This implies that
\ba
\label{eq:largemu}
\lim_{\mu \rightarrow \infty}\disc_s\cA\(\mu,\tau(\mu,a) \)
\begin{cases}
<  \mu^2 , & \mbox{for } a<0 \\
>  \mu^2, & \mbox{for } a>0.
\end{cases}
\ea
In order to produce the necessary divergence as $a\xr 0^-$ we assume that at least in the vicinity of $a=0$, where this transition occurs, we have at high energies the crossing symmetric Regge behaviour\footnote{Subleading logarithmic corrections are considered in \cite{Herrero-Valea:2020wxz} in order to account for massless loops and a different ansatz is considered in \cite{Herrero-Valea:2022lfd}. Here we will not include the contribution of massless loops graviton loops since they prevent the dispersion relation being continued from $t<0$ to $t>0$ (equivalently here through $a=0$) and it is therefore unclear what if any positivity properties remain for $t>0$.}
\be\label{CSRegge}
	\lim_{\mu\xr\infty}\disc_s\cA\(\mu,\tau(\mu,a) \)=r(a)\Lambda_r^4\left(\frac{\mu}{\Lambda_r^2}\right)^{\alpha(a)}\,,
\ee
for some analytic $\alpha(a)$ and arbitrary scale $\Lambda_r$ for which $\alpha(0)=2$ and $\alpha'(0)>0$. We will not need to assume that this form is valid at larger $a$, and can simply work perturbatively about $a=0$, where $\alpha(a)=2+\alpha'(0)a+\ldots$. The ansatz above describes the asymptotic behaviour of the discontinuity at energies well above the scale, at which the `Reggeizing physics' becomes relevant, prompting the definition of the difference function,
\begin{dmath}
	R(\mu,a)\equiv \disc_s\cA\(\mu,\tau(\mu,a) \)-r(a)\Lambda_r^4\left(\frac{\mu}{\Lambda_r^2}\right)^{\alpha(a)}\,,
\end{dmath}
which vanishes as $\mu\xr\infty$. This allows us to perform the integral over energies above $\Lambda_r$ for $a<0$, subtract the pole and re-arrange the dispersion relation,
\ba\label{crossamp2}
\cA(x,a)-\frac{x}{\mpl^{D-2}a}&=& c(a)-\frac{2x}{\pi}\int_{\Lambda^2}^{\Lambda_r^2}\d \mu \,  \frac{\disc_s\cA\(\mu,\tau(\mu,a) \)}{\mu^3}-\frac{2x}{\pi}\int_{\Lambda_r^2}^{\infty}\d \mu \,  \frac{R(\mu,a)}{\mu^3}\\
&+&\frac{2xr(a)}{\pi(\alpha(a)-2)}-\frac{x}{\mpl^{D-2}a}  +\frac{x}{\pi}\int_{\Lambda^2}^{\infty}\d \mu \,  \frac{\disc_s\cA\(\mu,\tau(\mu,a) \)}{\mu^3} \( \frac{2x(a-\mu)-3a\mu^2}{x(a-\mu)-\mu^3} \) \,.\nn
\ea
In particular if we insert $\alpha(a)=2+\alpha'(0)a+\ldots$, then we produce the expected pole via the Regge behaviour,
\be\label{Reggecond1}
	\lim_{a\xr0^-}\frac{2xr(a)}{\pi(\alpha(a)-2)}\sim\frac{2r(0)}{\pi\alpha'(0)}\frac{x}{a}\implies \frac{r(0)}{\alpha'(0)}=\frac{\pi}{2} \frac{1}{\mpl^{D-2}}\,.
\ee
As the integrals on the right hand side of (\ref{crossamp2}) are now convergent for $0 \leq a < \Lambda^2$ we may continue the dispersion relation representation of $\cA(x,y)$ to $0 \leq a < \Lambda^2$.

Starting from this representation of the pole subtracted amplitude defined as
\ba
\hat{\cA}(x,a)\equiv\cA(x,a)-\frac{x}{\mpl^{D-2}a}\,,
\ea
we may now derive positivity bounds on the low-energy expansion coefficients, which involve the functions $r(a)$ and $\alpha(a)$ through the Regge function  $P(a)$ defined as
\be\label{PRalpha}
 P(a)\equiv\frac{2r(a)}{\pi(\alpha(a)-2)}-\frac{1}{\mpl^{D-2}a} \,.
\ee
Crucially, as defined, this Regge function is fully analytic in the vicinity of $a=0$. It contains no poles, $P$ and all its derivatives are finite and well-defined at $a=0$. In fact we will argue later that it is likely an analytic function of $a$ up to a right hand branch cut beginning at $a=2\Lambda^2/3$.

Using the expansion of the discontinuity in terms of Chebyshev polynomials $T_{n}(x)$,
\begin{equation} \disc_s{\cA}(\mu,\tau(\mu,a))=\sum_{\nu=0}^{\infty}T_{\nu}\left(1+\frac{2\tau}{\mu}\right)\tilde{F}^{\nu}(\mu)=\sum_{\nu=0}^{\infty}T_{\nu}\left(\sqrt{\frac{\mu+3a}{\mu-a}}\right)\tilde{F}^{\nu}(\mu)\,,
\end{equation}
the dispersion relation becomes
\ba\label{crossamp4}
	\hat{\cA}(x,a) = c(a)+x P(a)-\frac{2x}{\pi}\int_{\Lambda_r^2}^{\infty}\d \mu \,  \frac{R(\mu,a)}{\mu^3}-2x\int_{\Lambda^2}^{\Lambda_r^2}{\d \mu}   \sum_{\nu=0}^{\infty}\rho_\nu(\mu)\,T_{\nu}\left(\sqrt{\frac{\mu+3a}{\mu-a}}\right) \\ +x\int_{\Lambda^2}^{\infty}\d \mu   \sum_{\nu=-\infty}^{\infty}\rho_\nu(\mu)\,T_{\nu}\left(\sqrt{\frac{\mu+3a}{\mu-a}}\right)\( \frac{2x(a-\mu)-3a\mu^2}{x(a-\mu)-\mu^3} \) \,,\nn
\ea
which we can now use to infer finite sum rules at positive $a$.


\subsection{Finite Energy Sum Rule}

 Consider the low energy expansion of the crossing-symmetric amplitude in the variables $x$ and $a$, we have
\begin{equation}
	\hat\cA(x,a)
= \sum_{i\ge 0}\sum_{j=0}^i a_{i-j,j}\ x^i a^j \,.
\end{equation}
Terms with powers of $a$ greater than that of $x$ would lead to poles at $x=0$ which cannot be physically present. This  explains why the sum has to be truncated over finite powers of $a$, leading to an infinite number of  `null constraints'. Since we are working  in a framework where full crossing symmetry is manifest from the outset,  the existence of the null constraints is here trivial. In particular, at leading order in the $x$ expansion, this implies,
\begin{equation}
	\left.\del_x\hat\cA(x,a)\right|_{x=0} = a_{1,0}+a_{0,1} a\,.
\end{equation}
Applying this to the dispersion relation (\ref{crossamp4}) gives,
\begin{dmath}\label{dr}
a_{1,0}+a_{0,1} a = \frac{1}{\pi}\int_{\Lambda^2}^{\Lambda_r^2}\d \mu \,  \disc_s\cA\(\mu,\tau(\mu,a) \)\( \frac{3a-2\mu}{\mu^4} \) +P(a)-\frac{3a r(a)}{\pi\Lambda_r^2(\alpha(a)-3)}+\frac{1}{\pi}\int_{\Lambda_r^2}^{\infty}\rd\mu\,R(\mu,a)\( \frac{3a-2\mu}{\mu^4} \) = \frac{1}{\pi}\int_{\Lambda^2}^{\Lambda_r^2}\d \mu \,  \disc_s\cA\(\mu,\tau(\mu,a) \)\( \frac{3a-2\mu}{\mu^4} \) +\hat P(a) \,,
\end{dmath}
where in the final line we have absorbed the two sub-dominant terms into the function $\hat P(a)$. Henceforth we will assume that $\Lambda_r$ is taken sufficiently large that $\hat P(a) \approx P(a)$. \\

In our family of fully crossing symmetric amplitudes the coefficient $a_{1,0}$ is zero and $a_{0,1}$ continuously varies with the parameter $\phi$ and can take positive or negative values.
As discussed above \eqref{eq:exactRelation}, there is a single choice of parameter $\phi$ ($\phi=\phi_0$ with $\phi_0$ satisfying $\cos \phi_0=-5/2 \cos 2 \phi_0$) for which $a_{0,1}$ vanishes at {\it all} scales for {\it any} theory describing Nature, meaning the function $\hat P(a)$ encoding the information about the Regge behaviour is uniquely fixed in terms of the {\it finite energy sum rule} \cite{Igi:1962zz,Logunov:1967dy,Igi:1967zzb,Igi:1967zza,Dolen:1967jr}, now also valid for positive $a$,
\ba\label{sumrule}
\boxed{
\hat P(a)\equiv  \frac{1}{\pi}\int_{\Lambda^2}^{\Lambda_r^2}\d \mu \,  \disc_s\cA\(\mu,\tau(\mu,a),\phi_0 \)\( \frac{2\mu-3a}{\mu^4} \)\,.}
\ea
In particular this implies that for $a<2/3\Lambda^2$, IR physics encoded in the discontinuity at low-energy ought to be fundamentally encoded within the Regge behaviour $\hat P(a)$ as it cannot be absorbed in the remaining part of the positive dispersion relation. For arbitrary polarization $\phi$ we have
\ba\label{sumrule2}
\hat P(a)\equiv a_{0,1}a+ \frac{1}{\pi}\int_{\Lambda^2}^{\Lambda_r^2}\d \mu \,  \disc_s\cA\(\mu,\tau(\mu,a),\phi \)\( \frac{2\mu-3a}{\mu^4} \)\,.
\ea

\subsection{Dispersion Relation for Regge function}

For any perturbative low-energy EFT, the amplitude can be put in the Mandelstam double spectral representation form as discussed in Appendix~\ref{app:mandel}. Then comparing our crossing symmetric dispersion relation with this double spectral representation, we can infer that $\disc_s\cA\(\mu,\tau(\mu,a),\phi_0 \)$ is an analytic function of $a$ with a right hand branch cut starting at $a=\mu \Lambda^2(\Lambda^2+\mu)/(\mu^2+\mu \Lambda^2+\Lambda^4)$ and ending at $a=\mu^2$. In terms of  $\hat P(a)$, this implies that $\hat P(a)$ should be an analytic function of $a$ with a right hand branch cut starting at $a=2 \Lambda^2/3$ and ending at $a=\Lambda_r^2$, {\it i.e.}
\be\label{phatdisp}
\hat P(a) = \sum_{n=0}^{N_p-1} p_n a^n + \frac{a^{N_p}}{\pi} \int_{\frac{2\Lambda^2}{3}}^{\Lambda_r^2} \d \tilde a \frac{\disc_a \hat P(\tilde a)}{\tilde a^{N_p}(\tilde a-a)} \, ,
\ee
where we have allowed for $N_p$ subtractions. In fact it is reasonable to suppose $N_p=0$ given that we do not expected $\disc_a \hat P(\rho)$ to carry any singular behaviour near $\rho=\Lambda_r^2$ as this is an arbitrarily chosen point. The discontinuity of the Regge function is itself related to the double discontinuity of the amplitude via
\be
\disc_a \hat P(\tilde a) = \frac{1}{\pi} \int_{\Lambda^2}^{\Lambda_r^2} \d \mu \,  \rho(\mu,\tilde \mu) \frac{(2\mu-3 \tilde a)(\mu+\tilde \mu)}{\mu (\mu^2+\mu \tilde \mu+\tilde \mu^2)^2}\,,
\ee
where $\rho(\mu,\tilde \mu)$ is the Mandelstam double discontinuity (see Appendix \ref{app:mandel}) and $\tilde \mu=\tilde \mu(\tilde a,\mu)$ in the integrand is the solution of $\tilde a =\mu  \tilde \mu (\mu+\tilde \mu)/(\mu^2+\mu \tilde \mu+\tilde \mu^2)$.

\subsection{Trees versus Loops}

For weakly coupled UV completions, we can generally split the contributions to the amplitude into those from loops of low spin particles (spin $<4$), and those from tree level effects of higher spin particles (spin $\ge 4$). Explicitly splitting the Regge behaviour into each contribution
\ba
\hat P(a)=\hat P^{\rm (tree)}(a)+\hat P^{\rm (loop)}(a)\,.
\ea
Then on dimensional grounds, one would expect the following approximate scalings for the low energy expansion in $a$
\ba\label{treeloopexpansions}
\hat P^{\rm (tree)}(a)&\sim & \frac{1}{\mpl^{D-2} M_{*}^{2}}\sum_nc_n^{t}\frac{a^n}{M_*^{2n}} \label{tree}\\\label{treeloopexpansions2}
\hat P^{\rm (loop)}(a)&\sim & \frac{m^{D-4}}{\mpl^{2 (D-2)}} \sum_nc_n^{l}\frac{a^n}{m^{2n}} \label{loop}\,.
\ea
Here it is assumed that the dimensionless tree and loop coefficients $c_n^{t,l}$ are of order unity, $M_*$ is the scale at which the tower of higher spin states kicks in (i.e. typically the lightest spin-4 state) and $m$ is the mass if the lightest low spin state. In realistic examples we expect $m\ll M_*\lesssim \mpl$\footnote{For instance, current axionic models of dark matter consider masses as small as about $10^{-20}$eV, while models of dark energy consider massive fields with masses that could be as low as the Hubble scale today, $\sim H_0\sim10^{-32}$eV.}. For example the form \eqref{loop} is exactly what we obtain in the examples \eqref{loop1}, \eqref{loop2} and \eqref{loop3} considered below.

Owing to the overall additional $\mpl$ suppression in \eqref{loop} relative to \eqref{tree} that penalizes loops of lower spin particles, one might naively have expected tree level contributions to always dominate over loops. However the inverse scaling of the coefficients $\sim m^{-2n}$ in terms of the lightest mass $m$, implies that at higher orders in the $a$ expansion, loops from massive light fields will always dominate contributions to the Regge behaviour $\hat P(a)$. In particular, assuming that the low energy theory exhibits a single state of mass $m \ll M_*^2/\mpl$ in $D=4$ then loop contributions dominate all $a$ derivatives of $\hat P(a)$. Fundamentally this is because the branch cut from light loops starts at smaller values of $a$ and so the regime of convergence of the Taylor expansion is much smaller.
Indeed, following our previous argument that $P(a)$ is an analytic function of $a$ with branch cut beginning at $a=2\Lambda^2/3$ then expansions \eqref{treeloopexpansions}, \eqref{treeloopexpansions2} are expected to come from functions of the form
\ba
\hat P^{\rm (tree)}(a) &=& \frac{1}{\mpl^{D-2} M_{*}^{2}} \int_{2 M_*^2/3}^{\infty} \d \tilde a \frac{p_{\rm tree}(\tilde a)}{\tilde a-a} \, , \\
\hat P^{\rm (loop)}(a) &=& \frac{m^{D-4}}{\mpl^{2 (D-2)}} \int_{8 m^2/3}^{\infty} \d \tilde a \frac{p_{\rm loop}(\tilde a)}{\tilde a-a} \, ,
\ea
where $p_{\rm tree}(a)$ and $p_{\rm loop}(a)$ are order unity functions which is consistent with \eqref{treeloopexpansions}, \eqref{treeloopexpansions2}. In $D \ge 4$ for any finite $a$, loops are suppressed relative to tree level effects, but the Taylor expansion is clearly dominated by loops. Thus it is necessary to subtract out the light loop contribution by some means in order to make useful statements. We discuss how to achieve this below.

\subsection{Improved Regge Positivity Bounds}

The statement of unitarity, encoded through positivity of the low energy discontinuity for $a>0$ simply implies
\ba\label{eq:bound1}
\hat P(a)>0 \quad \forall \quad 0<a<2\Lambda^2/3\,.
\ea
By itself this condition does not go too far beyond what is already expected. Indeed in the case of tree level string theory, the Regge slope may be taken as linear $\alpha(a) = 2 + \alpha' a$. In this case
\be
P(a) = \frac{2}{\pi \alpha ' a}(r(a) -r(0)) \, .
\ee
On the other hand, unitarity of the UV part of the discontinuity requires that $r(a) \ge r(0)$ for $a \ge 0$, which already implies $ P(a)>0$. The bound \eqref{eq:bound1} is nevertheless independent since it does not require any linear assumption on the Regge slope.

As is often the case, statements derived from the dispersion relation may be strengthened by leveraging information regarding the low energy physics. Let us now assume that the low energy physics can be captured by an EFT with a cutoff $\Lambda_c$ that is parametrically below $\Lambda_r$ but above the beginning of the branch cut set by $\Lambda$. When this is the case, the part of the discontinuities from $\Lambda $ to $\Lambda_c$ is calculable. This is the situation when loops of light fields are included.

For an EFT at sufficiently low energy, the most natural and conservative perspective is that the leading EFT corrections are due to massive loops of a lower-spin($< 4$) particle, with mass $m$.  For concreteness  consider an EFT, valid well below $\Lambda_c$, that contains a  particle of mass $m$ (none of the following features will depend on the precise low-energy field content of the low-energy EFT, rather the sensitivity is on the mass of the lightest massive modes). Loops of the light field produce a discontinuity in the amplitude above $\mu=4m^2$. We could explicitly compute the discontinuity due to these loops and subtract their contribution (from the branch point at $\Lambda^2=4m^2$ up to a scale $\Lambda_c^2$ well-below the cutoff of the EFT that includes this massive particle). Following this approach, we  split $\hat P(a)$ into its  `low' and `high' energy contributions $\hat P^{\text{low}}$ and $\hat P^{\text{high}}$ as,
\begin{equation}\label{phatdef}
	\hat P(a) = \hat P^{\text{low}}(a)+\hat P^{\text{high}}(a)= \left(\int_{4m^2}^{\Lambda_c^2}+\int_{\Lambda_c^2}^{\Lambda_r^2}\right)\d \mu \,  \disc_s\cA\(\mu,\tau(\mu,a),\phi_0 \)\( \frac{2\mu-3a}{\pi\mu^4} \) \,.
\end{equation}
Performing a `small $a$' expansion for each part,
\be
\hat P(a) =\frac{1}{\mpl^{2(D-2)}}\sum_{n=0}^{\infty}d_n a^n\,,
\ee
and similarly for $\hat P^{\text{low}/{\rm high}}$, so that at each order
$d_n=d_n^{\rm low}+d_n^{\rm high}$.

Given that the branch cut from loops begins at $4m^2$, it is clear that the coefficients $d_n^{\rm low}$ will scale with increasing inverse powers of $m^2$ as $n$ increases reflecting the regime of convergence of the Taylor expansion. For concreteness and illustrative purposes, we focus here on the discontinuity from a single massive scalar loop in $D=4$ and establish its contribution to $\hat P(a)_{\text{low}}$, see Appendix~\ref{oneloopscalar} for details.  Specifically we find
\begin{dmath}\label{loop1}
	\hat P(a) = \mpl^{-4}\sum_{n=0}^{\infty}d_{n}^{\rm low} a^n + \int_{\Lambda_c^2}^{\Lambda_r^2}\d \mu \,  \disc_s\cA\(\mu,\tau(\mu,a),\phi_0 \)\( \frac{2\mu-3a}{\pi\mu^4} \)\,,
\end{dmath}
with
\begin{dmath}\label{loop2}
 d^{\rm low}_0 = \int_{4m^2}^{\Lambda_c^2}\rd\mu \,\disc_s\cA(\mu,0,\phi_0)\left(\frac{2}{\pi\mu^3}\right) = \frac{20 \log \frac{ \Lambda_c ^2}{m^2}+5 \cos (2 \phi_0) -68}{19200 \pi ^2}+\frac{m^2}{48 \pi ^2 \Lambda_c^2}-\frac{3 m^4}{32 \pi ^2 \Lambda_c^4}+\mathcal{O}\left(\frac{m^6}{\Lambda_c^6 }\right)\,,
\end{dmath}
and
\begin{dmath}\label{loop3}
 d^{\rm low}_1 = \int_{4m^2}^{\Lambda_c^2}\rd\mu \,\left(\frac{2\mu\, \del_\tau \disc_s\cA(\mu,0,\phi_0)-3\disc_s\cA(\mu,0,\phi_0)}{\pi\mu^4}\right) = \frac{6-5 \cos (2 \phi_0) }{80640 \pi ^2 m^2}-\frac{7}{1920 \pi ^2 \Lambda_c ^2}+\frac{5 m^2}{96 \pi ^2 \Lambda_c ^4}+\mathcal{O}\left(\frac{m^6}{\Lambda_c^6 }\right)\,,
\end{dmath}
where we recall that $\phi_0$ is the special value of the indefinite scattering for which the coefficient $a_{0,1}$ of the amplitude cancels at all scales and for all four-dimensional theories ($\cos \phi_0=-5/2 \cos 2 \phi_0$). To illustrate the scaling of these coefficients with the lightest mass, the leading terms of several higher order coefficients $d_n^{\rm low}$ are given,
\ba
	d_2^{\rm low} = \frac{1}{2!}\frac{5+3\cos\phi_0}{604800\pi^2 m^4 }+\ldots\,,&& \quad d_3^{\rm low} = \frac{1}{3!}\frac{5+4 \cos \phi_0 }{2217600 \pi ^2 m^6}+\ldots\,,\\
 d_4^{\rm low}=\frac{1}{4!}\frac{21+20 \cos \phi_0 }{20180160 \pi ^2 m^8}+\ldots\,, && \quad
	 d_5^{\rm low} = \frac{1}{5!}\frac{14+15 \cos \phi_0}{20180160 \pi ^2 m^{10}}+\ldots\,.
\ea
At higher order, the coefficients generically scale as $d_n^{\rm low} \sim m^{-2n}$, unless the high energy theory is precisely fine-tuned so as to allow for exact cancellations to occur.
Generically, higher terms in the expansion are hence dominated by the lightest mass states, and despite the additional $\mpl^2$ suppression, at higher orders in the expansion loops will dominate over tree level effects by virtue of the branch-cut.

\subsection{Linear and Nonlinear Regge Positivity Bounds}

In principle one could conceive of a situation where the contributions $d_n^{\rm high}$ arising from higher energy physics (below the Regge scale) could absorb the $d_n^{\rm  low}\sim m^{-2n}$  dependence on the low-energy physics. We shall prove below that such a situation can never occur due to positivity. At sufficiently high order in a small $a$ expansion beyond the forward limit, one cannot prevent the Regge behaviour to be dominated by the loops of the lightest massive states in Nature. To see this, we can start by focusing on the first five  coefficients $d_n^{\rm high}$, given by
\ba
&& d_0^{\rm high}=\biggl[ \frac{2}{\mu^3} \biggr]>0 \, , \quad d_1^{\rm high}= \biggl[ \frac{(-3 + 4 \nu^2) }{\mu^4}  \biggr]\, , \quad d_2^{\rm high}= \biggl[ \frac{2 \nu^2 (-11 + 2 \nu^2) }{3\mu^5}  \biggr]\, , \nn \\
&& d_3^{\rm high}= \biggl[ \frac{2 \nu^2 (151 - 65 \nu^2 + 4 \nu^4) }{45 \mu^6}  \biggr]\, ,  \quad d_4^{\rm high}= \biggl[ \frac{2 \nu^2 (-2235 + 728 \nu^2 - 70 \nu^4 + 2 \nu^6)}{315 \mu^7}  \biggr]\,  ,
\ea
with the definition
\be
\biggl[ X(\mu,\nu^2)  \biggr] =\mpl^{4}\int_{\Lambda_c^2}^{\Lambda_r^2}\rd\mu\,\sum_{\nu=0}^{\infty} \mu^3\rho_\nu(\mu) X(\mu,\nu^2) \, .
\ee
Defining the dimensionless ratios $b_n =\Lambda_c^{2n} \frac{d^{\rm high}_n}{d^{\rm high}_0}$ together with the moments
\be
\biggl< X(\mu,\nu^2)  \biggr> =\frac{\int_{\Lambda_c^2}^{\Lambda_r^2}\rd\mu\,\sum_{\nu=0}^{\infty}\rho_\nu(\mu) X(\mu,\nu^2) }{\int_{\Lambda_c^2}^{\Lambda_r^2}\rd\mu\,\sum_{\nu=0}^{\infty}\rho_\nu(\mu)} \, ,
\ee
we have
\ba
&&  b_{1}= \Lambda_c^2 \biggl< \frac{(-3 + 4 \nu^2) }{2\mu}  \biggr>\, , \quad b_{2}= \Lambda_c^4 \biggl< \frac{ \nu^2 (-11 + 2 \nu^2) }{3\mu^2}  \biggr>\, , \nn \\
&& b_{3}=\Lambda_c^6  \biggl< \frac{ \nu^2 (151 - 65 \nu^2 + 4 \nu^4) }{45 \mu^3}  \biggr>\, ,  \quad b_{4}= \Lambda_c^8  \biggl< \frac{ \nu^2 (-2235 + 728 \nu^2 - 70 \nu^4 + 2 \nu^6)}{315 \mu^4}  \biggr>\,  .
\ea
As usual we can take linear combinations which are manifestly positive
\ba
&& b_1+\frac{3}{2} >0 , \quad b_2+\frac{11}{6}  b_1+\frac{11}{4}>0  \, , \quad  b_3+\frac{13}{6} b_2 + \frac{102245}{43488} b_1 + \frac{102245}{28992}>0 \, , \nn \\
&& b_4 + \frac{5}{2}  b_3 + \frac{1625}{768}  b_2 + \frac{104261}{32256 }  b_1 + \frac{3}{2} >0 \,,\ {\rm etc}\ldots\,.\label{eq:b}
\ea
We may also use Cauchy-Schwarz to derive nonlinear bounds of the form
\ba
&& 0<\( b_1+\frac{3}{2} \)^2 < 6  b_2+11  b_1 +\frac{33}{2} \, , \label{nonlinear1}\\
&& 0<\( b_1+\frac{3}{2} \)^3 < 90  b_3+195  b_2+\frac{1403}{4} b_1+\frac{1059}{2} \, , \label{nonlinear2}\\
&& 0<\( b_2+\frac{11}{6}  b_1+\frac{11}{4}\)^2<70  b_4+175  b_3+359  b_2+\frac{10757}{12}  b_1+\frac{10757}{8} \, . \label{nonlinear3}
\ea
Unlike the low energy positivity bounds, we can no longer use crossing symmetry to obtain compact bounds since we have already used all of the information from the null constraints to obtain \eqref{sumrule}. We can rewrite \eqref{nonlinear1} as
\be
b_2>\frac{1}{6}(b_1-4)^2-\frac{121}{24} \, ,
\ee
which tells us that $b_2>-\frac{121}{24}$. Similarly we can rewrite  \eqref{nonlinear2} as
\be
b_3+\frac{13}{6} b_2>\frac{1}{90}\( b_1+\frac{3}{2} \)^3 -\frac{1403}{360} b_1-\frac{353}{60} \, .
\ee
By minimizing the RHS subject to the constraint that $b_1>-3/2$ we infer
\be
b_3+\frac{13}{6} b_2>-28.1 \,.
\ee

\subsection{Regge Imprint}

We are now in a position to see why the high-energy (meaning $\Lambda_c^2< \mu< \Lambda_r^2$) contribution to $P(a)$ cannot cancel the low energy ($\mu<\Lambda_c^2$). Even though the higher order coefficients of $\hat P^{\rm high}(a)$ can in principle be negative, their level of negativity is always limited from the above positivity bounds,
\ba
d_0^{\rm high}>0, \quad
d_1^{\rm high}>-\frac 32 \frac{d_0^{\rm high}}{\Lambda_c^{2}}, \quad
d_2^{\rm high}>-\frac {121}{24} \frac{d_0^{\rm high}}{\Lambda_c^{4}},\ {\rm etc}\ \cdots\,.
\ea
Consider if from the outset we had $d_0^{\rm high}\gtrsim \Lambda_c^2/m^2$, then the Regge behaviour would clearly be dominated by the scale $m$ of the lightest IR loops, $d_0>d_0^{\rm high}\gtrsim \Lambda_c^2/m^2$. So to remain conservative, it is safe to assume $0<d_0^{\rm high}\lesssim \Lambda_c^2/m^2$ in what follows. Then clearly, $d_2^{\rm high}> -\mathcal{O}(1)/\Lambda_c^2 m^2$ and $d_2=d_2^{\rm low}+d_2^{\rm high}>\mathcal{O}(1)/m^4$. Already at that level, we see that the finite energy sum rule implies that the coefficient $d_2$ of the Regge expansion be dominated by loops of IR fields. \\

At this order in the Regge expansion, this result is similar to that obtained in \cite{Alberte:2021dnj} in the context of graviton-photon scattering and in \cite{Alberte:2020bdz,Herrero-Valea:2022lfd} in the context of photon-photon scattering, however the result here is shown to be generic and can be carried out to all orders by virtue of the relation \eqref{eq:exactRelation}. At cubic order in the small $a$ expansion, our modest linear positivity bounds are already sufficient to imply $b_3+13b_2/6>-28.1$. There again, to remain  conservative it is safe to assume  $|b_2|\ll |b_3|$, in which case, the previous bound imposes $d_3^{\rm high}\gtrsim - \mathcal{O}(10) d_0^{\rm high}/\Lambda_c^6\gtrsim - \mathcal{O}(10)/(\Lambda_c^4 m^2)\gtrsim -\mathcal{O}(10) d_3^{\rm low} m^4/\Lambda_c^4 $, so unless $\Lambda_c\lesssim 2 m$ (in which case the mass gap is nonexistent), the coefficient $d_3$ of the Regge behaviour is also dominated by IR physics, $d_3\sim d_3^{\rm low}\sim m^{-6}$. If one were to relax the assumption $|b_2|\ll |b_3|$ one could in principle consider a situation where $d_3^{\rm high}\sim - d_3^{\rm low}$ so that $d_3$ is parametrically suppressed, however this comes at the cost of having $b_3= \Lambda_c^6 d_3^{\rm high}/d_0^{\rm high}\sim - \Lambda_c^6 d_3^{\rm low}/d_0^{\rm high}\sim - \Lambda_c^6 /(m^6 d_0^{\rm high})$ and hence $b_2\gtrsim |b_3|\gtrsim  \Lambda_c^6 /(m^6 d_0^{\rm high})$ leading to $d_2=d_2^{\rm low}+d_2^{\rm high}\gtrsim  \Lambda_c^2 /m^6 $, and hence an even stronger dependence of the Regge coefficients on IR physics.
As could already have been anticipated from the universal rule \eqref{eq:exactRelation}, this implies that the scale of IR physics has to be imprinted in the ultimate high energy completion in a non-trivial way. In other words, at each (non-zero) order in the small $a$ expansion, the coefficients of the Regge behaviour are dominated by loops of IR fields and carry insight on the lightest massive modes, $d_n\sim m^{-2n}$ which dominate over the contribution from higher spin trees.

\subsection{Separating UV physics from IR physics}

\label{separating}

As we have seen, the low energy contribution to the Regge function $\hat P^{\rm low}(a)$ is dominated by the lightest mass states, which cannot be compensated by $\hat P^{\rm high}(a)$ by virtue of positivity bounds. However, the situation is worse than that: $\hat P^{\rm high}(a)$ is also sensitive to the light fields, because it contains contributions from the discontinuities in $a$ as is apparent from the dispersion relation \eqref{phatdisp}. The decomposition in \eqref{phatdef} does not cleanly separate low energy and high energy contributions.  What we need is a better way to disentangle low-energy and high-energy physics, which requires removing all low-energy contributions to both single and double discontinuities.\\

As implied by the Mandelstam double spectral representation, we start by considering that the single discontinuity $\disc_s\cA\(\mu,\tau(\mu,a)\)$ is itself an analytic function with the prescribed double discontinuity and a fixed number $N_s$ of subtractions at least for $\mu <\Lambda_c^2$.
\ba
\disc_s\cA\(\mu,\tau(\mu,a)\) &=&\sum_{n=0}^{N_s-1} \kappa_n(\mu) (a+a_*)^n  \\
&+&\frac{(a+a_*)^{N_s}}{\pi} \int_{\Lambda^2}^{\infty}  \d \tilde \mu \frac{\rho(\mu,\tilde \mu)}{(b(\mu ,\tilde \mu)+a_*)^{N_s}} \frac{(\mu+2 \tilde \mu) (\mu-b(\mu,\tilde \mu))}{\mu \tilde \mu (\mu+\tilde \mu)-a(\mu^2+\mu \tilde \mu + \tilde \mu^2)} \, , \nn
\ea
where $b(\mu,\tilde \mu)$ is the value of $a$ when the denominator vanishes,
\ba
b(\mu,\tilde \mu)=\frac{\mu\tilde \mu (\mu+\tilde \mu)}{\mu^2+\mu\tilde \mu+\tilde \mu^2}\,,
\ea
and with subtraction coefficients
\be
 \kappa_n(\mu) =\frac{1}{n!} \partial_a^n\disc_s\cA\(\mu,\tau(\mu,-a_*)\)\, .
\ee
Here we have chosen an arbitrary subtraction point $a=-a_*$ with $a_*>0$ to be clear of the RH branch cut. Furthermore we can take $a_* \sim {\varepsilon} \Lambda_c$ with $\varepsilon$ small enough to trust the low energy EFT calculations, but with ${\varepsilon} \Lambda_c\gg \Lambda^2= 4m^2$ to ensure that the $ \kappa_n(\mu) $ are not strongly sensitive to some IR mass.
Since the $\mu$ integral in the definition of $\hat P(a)$ is over a finite range we may safely write this as a dispersion relation for $\hat P(a)$ as
\ba
\hat P(a) &=& \sum_{n=0}^{N_s-1} \int_{\Lambda^2}^{\Lambda_r^2} \d \mu \left[ \kappa_n(\mu) (a+a_*)^n \( \frac{2\mu-3a}{\pi\mu^4} \) \right]\\
&+&\frac{(a+a_*)^{N_s}}{\pi}  \int_{\Lambda^2}^{\Lambda_r^2} \d \mu \int_{\Lambda^2}^{\infty}  \d \tilde \mu \frac{\rho(\mu,\tilde \mu)}{(b(\mu ,\tilde \mu)+a_*)^{N_s}} \frac{(\mu+2 \tilde \mu) (\mu-b(\mu,\tilde \mu))}{\mu \tilde \mu (\mu+\tilde \mu)-a(\mu^2+\mu \tilde \mu + \tilde \mu^2)} \( \frac{2\mu-3a}{\pi\mu^4} \) \, , \nn
\ea
where for the loop contributions we have in mind $\Lambda^2 = 4m^2$. Our goal is to remove the low energy contribution from both the single and double discontinuities. The former is easy to do, but the latter is more subtle because there are contributions in the double integral where $\mu$ is small but $\tilde \mu$ is large and vice versa.\\

The formation of double discontinuities in perturbation theory is determined by the Landau curves \cite{Correia:2021etg}. For instance, a double discontinuity from a massive loop is summarised by writing the spectral density as a sum over two `wings'
\be
\rho(\mu,\tilde \mu) = \theta(\mu \tilde \mu - 4 \Lambda^2 \mu -\Lambda^2 \tilde \mu) \rho_1(\mu,\tilde \mu)+  \theta(\mu \tilde \mu - 4 \Lambda^2 \tilde \mu -\Lambda^2 \mu) \rho_1(\mu,\tilde \mu) \, ,
\ee
where $\Lambda^2 = 4m^2$ and $ \rho_1(\mu,\tilde \mu)$ itself is not required to be symmetric. When working in a low energy effective theory with cutoff $\Lambda_c$, we can calculate the double discontinuity $\rho(s,t)$, however this calculation is clearly only valid in a finite region. The precise shape of this region is difficult to determine a priori and does not have to correspond to a particular Landau curve. We shall be conservative and suppose that we can meaningfully trust the calculation for $\mu+\tilde \mu< \Lambda_c^2$, which bearing in mind that both spectral parameters are positive and have lower bound $\Lambda^2$, defines a triangular region bounded on one side by the lowest Landau-curve. We could equivalently choose a quarter-circular region $\mu^2+\tilde \mu^2<\Lambda_c^4$.

With the former choice, the low-energy contribution to the Regge function from the double discontinuity is then
\be
 \frac{(a+a_*)^{N_s}}{\pi}   \int_{\Lambda^2}^{\Lambda_r^2} \d \mu \int_{\Lambda^2}^{\infty}  \d \tilde \mu \, \theta(\Lambda_c^2-\mu-\tilde \mu) \, \[  \frac{\rho(\mu,\tilde \mu)}{(b(\mu ,\tilde \mu)+a_*)^{N_s}} \frac{(\mu+2 \tilde \mu) (\mu-b(\mu,\tilde \mu))}{\mu \tilde \mu (\mu+\tilde \mu)-a(\mu^2+\mu \tilde \mu + \tilde \mu^2)}\]\( \frac{2\mu-3a}{\pi\mu^4} \) \, .\nn
\ee
We can now split the Regge function cleanly into an IR contribution ($\mu<\Lambda_c^2$) and a sub-Regge UV contribution $\Lambda_c^2<\mu<\Lambda_r^2$
\ba
\hat P(a)=\hat P^{\rm (IR)}(a)+\hat P^{\rm (UV)}(a)\,,
\ea
similar to our previous high and low energy split, but we do so in such a way that $\hat P^{\rm (UV)}(a)$ contains no low energy contributions either from the double discontinuity, or the single discontinuity. The choice of IR contribution which achieves this is
\ba\label{PIR}
&& \hspace{-1cm}\hat P^{\rm (IR)}(a) =\int_{\Lambda^2}^{\Lambda_c^2} \d \mu \sum_{n=0}^{N_s-1}\[  \frac{1}{n!} \partial^n_{a}\disc_s\cA\(\mu,\tau(\mu,-a_*)\)  (a+a_*)^n\( \frac{2 \mu-3a}{\pi \mu^4}\) \]  \\ \nn
&& \hspace{-1cm} + \frac{(a+a_*)^{N_s}}{\pi}   \int_{\Lambda^2}^{\Lambda_r^2} \d \mu \int_{\Lambda^2}^{\infty}   \d \tilde \mu \,   \theta(\Lambda_c^2-\mu-\tilde \mu) \, \[  \frac{\rho(\mu,\tilde \mu)}{(b(\mu ,\tilde \mu)+a_*)^{N_s}} \frac{(\mu+2 \tilde \mu) (\mu-b(\mu,\tilde \mu))}{\mu \tilde \mu (\mu+\tilde \mu)-a(\mu^2+\mu \tilde \mu + \tilde \mu^2)} \]\( \frac{2\mu-3a}{\pi\mu^4} \) \, .\hspace{-1cm}
\ea
It is important to stress that $\hat P^{\rm (IR)}(a) $ contains only terms which are calculable within a given low-energy EFT which describes the scattering below the cutoff scale $\Lambda_c$. The only in principle unknown is the number of subtractions $N_s$. However, within a given low-energy EFT, there is a pragmatic choice which is to take $N_s$ to be the value needed to write a dispersion relation for the given low energy amplitude computed to the desired order in loops etc. In practice this means taking $N_s$ to be such that using the IR calculation of $\rho(\mu,\tilde \mu)$ the double integral in \eqref{PIR} converges if the theta function is removed so that the upper limit is taken to $\infty$. \\

The sub-Regge UV contribution is similarly
\ba\label{PUV}
&& \hspace{-1cm} \hat P^{\rm (UV)}(a) =\int_{\Lambda_c^2}^{\Lambda_r^2} \d \mu \sum_{n=0}^{N_s-1}\[  \frac{1}{n!} \partial^n_{a}\disc_s\cA\(\mu,\tau(\mu,-a_*)\)  (a+a_*)^n\( \frac{2 \mu-3a}{\pi \mu^4}\) \] \\ \nn
&& \hspace{-1cm}+ \frac{(a+a_*)^{N_s}}{\pi}   \int_{\Lambda^2}^{\Lambda_r^2} \d \mu \int_{\Lambda^2}^{\infty}  \d \tilde \mu \,    \theta(\mu+\tilde \mu-\Lambda_c^2) \, \[  \frac{\rho(\mu,\tilde \mu)}{(b(\mu ,\tilde \mu)+a_*)^{N_s}} \frac{(\mu+2 \tilde \mu) (\mu-b(\mu,\tilde \mu))}{\mu \tilde \mu (\mu+\tilde \mu)-a(\mu^2+\mu \tilde \mu + \tilde \mu^2)} \]\( \frac{2\mu-3a}{\pi\mu^4} \) \, ,\hspace{-1cm}
\ea
although it is of course undetermined unless at least a partial UV completion which describes the region $\Lambda_c^2<\mu<\Lambda_r^2$ of the EFT is already known. What is understood and relevant here is its analytic structure. Since the remaining double discontinuity  in $\hat P^{\rm (UV)}(a)$ has only support from $\mu+\tilde \mu >\Lambda_c^2$ then $\hat P^{\rm (UV)}(a)$ is an analytic function of $a$ up to a right hand branch cut which begins at $a=2\Lambda_c^2/3$. Unlike our previous definition $\hat P^{\rm high}(a)$, $\hat P^{\rm (UV)}(a)$ is now devoid of any contribution from low-energy physics and so we have succeeded in disentangling the two contributions. The historical application of finite sum rules \cite{Igi:1962zz,Logunov:1967dy,Igi:1967zzb,Igi:1967zza,Dolen:1967jr} was to assume that $\hat P(a)$ can be well approximated entirely by its IR contribution $\hat P^{\rm (IR)}(a)$. This is equivalent to assuming $\Lambda_c \sim \Lambda_r$ or that at least $|\hat P^{\rm (UV)}(a)| \ll |\hat P^{\rm (IR)}(a)|$. This led to the bootstrap hypothesis that the UV could be calculated via the IR. A more conservative modern point of view would be that the one taken here that we can meaningfully compute the IR contribution $\mu<\Lambda_c^2$ to UV quantities $\mu>\Lambda_r^2$.

\subsection{Continuous Moment Sum Rules}

The finite energy sum rule we have derived so far \eqref{sumrule} determines the Regge function $\hat P(a) \approx P(a)$ which is a particular function of the Regge residue $r(a)$ and the Regge slope $\alpha(a)$ \eqref{PRalpha}. However ideally we would like to disentangle this information so as to better understand each function separately. One approach following \cite{Dolen:1967jr} would be to derive higher order sum rules from the analyticity of $x^n \cA(x,a)$ for integer $n>0$. Whilst in principle possible, in practice increasing powers of $n$ requires increasing knowledge about the UV properties of the amplitude for which we can no longer trust the simple ansatz \eqref{CSRegge}. One solution is to improve the ansatz to assume the asymptotics are given by a sum over Regge poles so that there are subleading contributions each with their own trajectory
\be\label{CSRegge2}
	\lim_{\mu\xr\infty}\disc_s\cA\(\mu,\tau(\mu,a) \)=\sum_I r_I(a)\Lambda_r^4\left(\frac{\mu}{\Lambda_r^2}\right)^{\alpha_I(a)}\,.
\ee
As well as being a rather strong assumption, for instance it neglects the possibility of Regge cuts, it introduces a significant number of free functions.
A better approach would be to allow $n$ to be small and non-integer to avoid increasing dependence on subleading Regge trajectories.

Fortunately this can be achieved. To see how this works, let us first define an exact function which has the correct Regge asymptotics  \eqref{CSRegge} and prescribed analytic structure
\be
\cA_{\rm Regge}\(x,a\)=-\frac{r(a)}{\sin \(\frac{\pi}{2} \alpha(a)\) }  \left[ \( x+\frac{\Lambda^6}{\Lambda^2-a}\) ^{\alpha(a)/2} -\( \frac{\Lambda^6}{\Lambda^2-a}\) ^{\alpha(a)/2}  \right]\, .
\ee
Consider now the combination \cite{Liu:1967xaf,Olsson:1968fuu}
 \be
 \cB(x,a)=\(\frac{\Lambda^6}{\Lambda^2-a}+x\)^{\sigma}\( \cA(x,a) -\cA_{\rm Regge}\(x,a\) \)\, .
 \ee
The assumption that $\cA_{\rm Regge}$ correctly captures the asymptotic form of $\cA$ can be written as that
\be
\lim_{|x| \rightarrow \infty} \frac{|\cA(x,a) -\cA_{\rm Regge}(x,a)|}{|x|} =0 \, ,
 \ee
at least in the vicinity of $a=0$. We can now consider small enough moment $\sigma$ so that the following relation remains satisfied,
\be
\lim_{|x| \rightarrow \infty} \frac{|\cB(x,a)|}{|x|} =0 \, .
 \ee
It then follows that we can write a dispersion relation with a single subtraction similar to \eqref{crossamp10}
\be
\cB(x,a)= \cB(0,a)-\frac{x}{\pi}\int_{\omega_{0}(a)}^{\infty}\d \omega \,  \disc_s\cB\(\mu,\tau(\mu,a) \) \frac{1}{\omega(\omega+x)}\,,
\ee
and in the integrand $\omega$ is related to $\mu$ by $\omega = \frac{\mu^3}{\mu-a}$ and we have defined $\omega_0(a)= \frac{\Lambda^6}{(\Lambda^2-a)}$. For the graviton scattering amplitudes \eqref{triplecrossingamp} we always have $\cA(0,a)=0$ and we have conveniently chosen $\cA_{\rm Regge}\(0,a\)=0$ whence $\cB(0,a)=0$.
Differentiating and evaluating at $x=0$ gives
\be
\omega_0(a)^{\sigma} \partial_x \cA(x,a)|_{x=0} =\omega_0(a)^{\sigma} \partial_x \cA_{\rm Regge}(x,a)|_{x=0} -\frac{1}{\pi}\int_{\omega_0(a)}^{\infty}\d \omega \,  \disc_s\cB\(\mu,\tau(\mu,a) \) \frac{1}{\omega^2} \, .
\ee
Remembering that $a_{1,0}=0$ for graviton scattering amplitudes this gives
\be
\frac{1}{\mpl^{D-2} a}+a_{0,1} a =-\frac{\alpha(a) r(a) \omega(a)^{\alpha(a)/2-1}}{\sin \(\frac{\pi}{2} \alpha(a)\)}- \frac{1}{\omega_0(a)^{\sigma}} \frac{1}{\pi}\int_{\omega_{0}(a)}^{\Lambda_r^4}\d \omega \,  \disc_s\cB\(\mu,\tau(\mu,a) \) \frac{1}{\omega^2} \, ,
\ee
where we have assumed that $\cA_{\rm Regge}\(x,a\) $ is a sufficiently good approximation to the form of the amplitude for $\omega>\Lambda_r^4$ that we may neglect the second integral for $\omega>\Lambda_r^4$.
The $x/a$ pole in the amplitude $\cA$ propagates along the branch cut in $\cB$ and so it is helpful to separate it out and focus on the discontinuity that comes from $\hat \cA$. This results in the {\it continuous moment sum rule}  \cite{Liu:1967xaf,Olsson:1968fuu}  which relates the Regge behaviour with the IR or more precisely sub-Regge physics\footnote{See \cite{Jackson:1970aw} for a review of this approach.}
\be\label{contsumrule}
\hspace{-0.5cm}\boxed{
P_\sigma(a)= a_{0,1} a  +\frac{1}{\pi  \omega_0(a)^{\sigma}}\int_{\omega_0(a)}^{\Lambda_r^4}\d \omega \,   \frac{\(\omega-\omega_0(a)\)^{\sigma}  }{\omega^2}\( \cos (\pi \sigma) \disc_s\hat \cA\(\mu,\tau(\mu,a)\)-\sin(\pi \sigma) {\rm Re}\hat \cA\(\mu,\tau(\mu,a) \)\)\, .}
\ee
Here we have defined the new Regge function $P_\sigma(a)$ as
\be\label{PS1}
 P_\sigma(a)=-\frac{1}{\mpl^{D-2} a} - \frac{1}{\pi \omega_0^{\sigma}}\int_{\omega_0(a)}^{\Lambda_r^4} \d \omega \frac{\sin(\pi \sigma) (\omega-\omega_0(a))^{\sigma}}{\mpl^{D-2} a \omega } +r(a) \frac{K_{\sigma}(a)}{\sin (\frac{\pi}{2} \alpha(a)) } \, ,
\ee
with
\ba
 K_\sigma(a)&=&- \frac{\alpha(a)  \omega_0(a)^{\alpha(a)/2-1}}{2 }  \\ \nn
 &+&\frac{1}{\pi \omega_0^{\sigma}}\int_{\omega_0(a)}^{\Lambda_r^4} \d \omega \frac{(\omega-\omega_0(a))^{\sigma}}{\omega^2 } \left[ \sin \(\frac{\pi}{2} \alpha(a)+\pi \sigma\) (\omega-\omega_0(a))^{\alpha(a)/2} -\sin \(\pi \sigma\)  \omega_0(a)^{\alpha(a)/2}\right]\,.
\ea
Although not immediately apparent $ P_\sigma(a)$ has no pole at $a=0$ given \eqref{Reggecond1} and a similar cancellation amongst the integrals. Furthermore it is an analytic function of $a$ with a branch cut beginning at $a=2\Lambda^2/3$.
For the specific choice $\sigma=0$ we have
\be
 P_0(a) = -\frac{1}{\mpl^{D-2} a}-\frac{1}{\pi }\int_{\Lambda_r^4}^{\infty} \d \omega \frac{r(a)(\omega-\omega_0(a))^{\alpha(a)/2} }{\omega^2 }\,,
\ee
which for $\Lambda_r^4 \gg \omega_0(a)$ gives \eqref{PRalpha} so that \eqref{contsumrule} reduces to our previous sum rule \eqref{sumrule}. The virtue of the continuous moment sum rules is that we can now separate the information on the Regge residue and the Regge slope. For instance we can solve \eqref{contsumrule} for the residue
\be
r(a) = \frac{\sin \(\frac{\pi}{2} \alpha(a)\)}{K_{\sigma}(a)}\(P_{\sigma}(a) +\frac{1}{\mpl^{D-2} a}+\frac{1}{\pi \omega_0^{\sigma}}\int_{\omega_0(a)}^{\Lambda_r^4} \d \omega \frac{\sin(\pi \sigma) (\omega-\omega_0(a))^{\sigma}}{\mpl^{D-2} a \omega } \)  \, ,
\ee
and since the equation must be true for any value of $\sigma$ we can identify it at two different values
\be
\frac{K_{\tilde \sigma}(a)}{K_{\sigma}(a)}=\frac{\(P_{\tilde \sigma}(a) +\frac{1}{\mpl^{D-2} a}+\frac{1}{\pi \omega_0^{\tilde \sigma}}\int_{\omega_0(a)}^{\Lambda_r^4} \d \omega \frac{\sin(\pi \tilde \sigma) (\omega-\omega_0(a))^{\tilde \sigma}}{\mpl^{D-2} a \omega } \) }{\(P_{\sigma}(a) +\frac{1}{\mpl^{D-2} a}+\frac{1}{\pi \omega_0^{\sigma}}\int_{\omega_0(a)}^{\Lambda_r^4} \d \omega \frac{\sin(\pi \sigma) (\omega-\omega_0(a))^{\sigma}}{\mpl^{D-2} a \omega } \) } \, ,
\ee
which gives us an equation for the Regge trajectory $\alpha(a)$ in terms of the integrals over the discontinuities in \eqref{contsumrule}.

An alternative way to write this that removes explicit mention of the Planck scale is from \eqref{PS1}
\be
r(a) = \sin \(\frac{\pi}{2} \alpha(a)\) \frac{\partial_{\sigma}(P_{\sigma} \Sigma_{\sigma})}{\partial_{\sigma}(K_{\sigma} \Sigma_{\sigma})}\, ,
\ee
where
\be
\Sigma_{\sigma}^{-1}=1+\frac{1}{\pi \omega_0^{\sigma}}\int_{\omega_0(a)}^{\Lambda_r^4} \d \omega \frac{\sin(\pi \sigma) (\omega-\omega_0(a))^{\sigma}}{ \omega } \, ,
\ee
from which we obtain
\be
\partial_{\sigma}\left[ \frac{\partial_{\sigma}(K_{\sigma} \Sigma_{\sigma})}{\partial_{\sigma}(P_{\sigma} \Sigma_{\sigma})} \right]=0 \, ,
\ee
or equivalently
\be\label{slope}
 \frac{\partial^2_{\sigma}(K_{\sigma} \Sigma_{\sigma})}{ \partial_{\sigma}(K_{\sigma} \Sigma_{\sigma})}= \frac{\partial^2_{\sigma}(P_{\sigma} \Sigma_{\sigma})}{ \partial_{\sigma}(P_{\sigma} \Sigma_{\sigma})} \, .
\ee
It is sufficient to evaluate this at any $0 \le \sigma<1$ to an obtain an implicit equation for the Regge trajectory. \\

\subsection{Approximate expression for Regge trajectory}

To illustrate this, consider the limit where there is a large hierarchy between the beginning of the branch cut $\Lambda$ and the Regge scale $\Lambda_r$ so that $\Lambda_r^4 \gg \omega_0$. In this limit we have
\be
\Sigma_{\sigma}^{-1} \approx \frac{\sin (\pi \sigma)}{\pi \sigma} \(\frac{\Lambda_r^4}{\omega_0} \)^{\sigma} \, ,
\ee
and so
\be
K_{\sigma} \Sigma_{\sigma} \approx \frac{\pi \sigma}{\sin (\pi \sigma)} \frac{\sin (\pi \alpha/2+\pi \sigma)}{\pi (\sigma+\alpha/2-1)} \Lambda_r^{2\alpha-4} \, .
\ee
Evaluating the LHS of \eqref{slope} at $\sigma=1/2$ gives a monotonically decreasing function of $\alpha$ over the relevant range $0<\alpha<4$
\be\label{LHS}
 \frac{\partial^2_{\sigma}(K_{\sigma} \Sigma_{\sigma})}{ \partial_{\sigma}(K_{\sigma} \Sigma_{\sigma})}\Big|_{\sigma=1/2}=-\frac{4 (\alpha(a) -2) \left(\pi  (\alpha(a) -1) \sin \left(\frac{\pi  \alpha(a) }{2}\right)+2
   \cos \left(\frac{\pi  \alpha(a) }{2}\right)\right)}{(\alpha(a) -1) \left(2 (\alpha(a) -2) \cos
   \left(\frac{\pi  \alpha(a) }{2}\right)-\pi  (\alpha(a) -1) \sin \left(\frac{\pi  \alpha(a) }{2}\right)\right)} \, .
\ee
Choosing for simplicity the special polarization $\phi=\phi_0$ for which $a_{0,1}$ vanishes  then in the same limit we have
\be
P_{\sigma} \Sigma_{\sigma} \approx \frac{\sigma}{\sin (\pi \sigma)  \Lambda_r^{\sigma}}\int_{0}^{\Lambda_r^4}\d \omega \,   \omega^{\sigma-2}  \( \cos (\pi \sigma) \disc_s\hat \cA\(\mu,\tau(\mu,a)\)-\sin(\pi \sigma) {\rm Re}\hat \cA\(\mu,\tau(\mu,a) \)\)\,,
\ee
which on evaluating the RHS of \eqref{slope} at $\sigma=1/2$ gives
\be\label{RHS}
 \frac{\partial^2_{\sigma}(P_{\sigma} \Sigma_{\sigma})}{ \partial_{\sigma}(P_{\sigma} \Sigma_{\sigma})}\Big|_{\sigma=1/2}\hspace{-0.4cm}=\frac{\int_{0}^{\Lambda_r^4}\d \omega \,   \omega^{-3/2} \left[ 2 \pi \disc_s\hat \cA\(\mu,\tau(\mu,a)\) (2-y)-{\rm Re}\hat \cA\(\mu,\tau(\mu,a)\)  y(4-y)\right]}{\int_{0}^{\Lambda_r^4}\d \omega \,   \omega^{-3/2}  \left[  \pi \disc_s\hat \cA\(\mu,\tau(\mu,a)\) + {\rm Re}\hat \cA\(\mu,\tau(\mu,a) \) (2-y)\right]} \, ,
\ee
with $y=\ln(\Lambda_r^4/\omega)$. For instance one may check that with the Regge ansatz $ \disc_s\hat \cA\(\mu,\tau(\mu,a)\) =r(a)\omega^{\alpha(a)/2}$, ${\rm Re}\hat \cA\(\mu,\tau(\mu,a)\)  = -r(a) \cot(\pi \alpha/2) \omega^{\alpha(a)/2}+\omega/{\mpl^{D-2}a}$ taken to be valid at all scales then \eqref{LHS} equals \eqref{RHS} where the (crossing symmetric) $t$-channel pole cancels because of the integral identities
\be
\int_{0}^{\Lambda_r^4} \d \omega \, \omega^{-1/2} (2-y) = \int_{0}^{\Lambda_r^4} \d \omega \, \omega^{-1/2} y(4-y)=0 \, ,
\ee
which means the hat may be dropped from the amplitude in \eqref{RHS}. Thus equation \eqref{slope} serves to uniquely determine the Regge trajectory in terms of sub-Regge physics. \\

Using the knowledge that at $a=0$ the Regge trajectory satisfies $\alpha(0)=2$ in order to reproduce the $t$-channel pole we infer a non-trivial constraint
\be
\frac{16}{\pi^2-4} = \frac{\int_{0}^{\Lambda_r^4}\d \omega \,   \omega^{-3/2} \left[ 2 \pi \disc_s \cA\(\mu,\tau(\mu,0)\) (2-y)-{\rm Re} \cA\(\mu,\tau(\mu,0)\)  y(4-y)\right]}{\int_{0}^{\Lambda_r^4}\d \omega \,   \omega^{-3/2}  \left[  \pi \disc_s \cA\(\mu,\tau(\mu,0)\) + {\rm Re}\cA\(\mu,\tau(\mu,0) \) (2-y)\right]} \, .
\ee
In practice if $\Lambda_r$ is taken too large, well into the Regge region then the integrals in \eqref{RHS} will be dominated by the contribution from the Regge region in which case we do not learn anything new. Thus the optimal situation is where $\Lambda_r$ is taken to be the smallest value at which the transition to Regge behaviour kicks in and ideally for which $\Lambda_r \sim \Lambda_c$ so that data from the EFT region can be used to anticipate the Regge trajectory. The historical use of such sum rules was to precisely predict the Regge parameters from known low energy phenomenology. \\

\subsection{Physical Implications}

The above sum rules show that there is a non-trivial IR-UV mixing whereby the high-energy $\mu>\Lambda_r^2$ Regge behaviour is determined by the sub-Regge physics $\mu<\Lambda_r^2$. Furthermore, positivity bounds constrain the high energy sub-Regge contributions $\Lambda_c^2<\mu<\Lambda_r^2$. The coefficients in the low-energy expansion of the Regge functions are potentially dominated by contributions from light loops in a way which cannot be compensated by high energy sub-Regge physics. The dependence of the Regge functions on light loops is arguably not surprising since the traditional Regge limit is stated as large $s$ and small $t$ and small $t$ physics is expected to be sensitive to the IR through light $t$-channel exchanges. In our crossing symmetric dispersion relation $a$ plays the role of $t$ in the traditional fixed $t$ dispersion relation and so a similar feature holds. What is remarkable though is that this is determined in a calculable way via the non-trivial finite energy sum-rule \eqref{sumrule}. This very much differs from the traditional approach where ad hoc Feynman diagram resummation methods are used to anticipate Regge behaviour from the bottom up in particular theories with questionable success\footnote{See for example the beautiful textbook \cite{Gribov:2003nw} for a review of this approach.}. Rather our approach follows closely the duality bootstrap approach via finite and continuous energy sum rules which were the origin for the Dolen-Horn-Schmid duality \cite{Dolen:1967jr}. Indeed the need for a Regge trajectory with $\alpha(0)=2$ given the exchange of a massive spin-2 field is a direct consequence of this duality mindset. This duality historically gave rise to string theory by means of the Veneziano's explicit amplitude \cite{Veneziano:1968yb}. The motivation of Dolen, Horn and Schmid and others was to use the equivalent of \eqref{sumrule} as a bootstrap from which Regge parameters could be predicted from low-energy data alone. \\

Our modern perspective is arguably slightly more conservative. From a bottom approach we typically assume we are working with a given EFT with cutoff $\Lambda_c$ for which the first non-analyticity arises at $\Lambda\le \Lambda_c$. Our sum rules \eqref{sumrule} and \eqref{contsumrule} can be used to unambiguously determine the IR contribution to the Regge functions. In order to realize the historical bootstrap idea we  also have to include some notion of the sub-Regge UV contribution. The question is to what extent can we anticipate this from the low-energy EFT alone? A plausible approach is to follow the spirit of the modern S-matrix bootstrap methods \cite{Paulos:2017fhb,Guerrieri:2020bto,Tourkine:2021fqh,Figueroa:2022onw,Huang:2022mdb}. We can make an ansatz for the discontinuity for $\mu<\Lambda_r^2$ by considering a large but finite number of multipoles and number of mass states, similar to what is assumed in numerical optimization procedures for low energy positivity bounds. Rather than using this to put bounds on low energy coefficients we can rather fix a subset of the free parameters in this ansatz by known low-energy expansion coefficients. Further coefficients can be constrained by means of the null constraints which have already been shown to be powerful at restricting the range of the low-energy coefficients. Then the continuous moment sum rules can be used to fix the leading Regge trajectory and residue. It should then be possible at least in principle to perform an optimization procedure to find the possible range of the Regge functions. \\

This statement that the Regge behaviour can be strongly dependent on IR physics has important consequences for potential proofs of the Weak Gravity Conjecture \cite{Hamada:2018dde,Tokuda:2020mlf} which have attempted to bypass the difficulties of gravitational positivity bounds by making specific assumptions about UV completions. For example, one could in principle have imagined that all of the scales in the Regge physics encoded in $P(a)$ are set by the string or Regge scale which is essentially the scale of the massive higher spin states $M_*$. This is effect what is implied by assuming \eqref{tree} alone.
If this were true, then one can derive positivity bounds on low energy effective fields theories which allow a mild negativity set only by $1/({\mpl^{D-2} M_*^2})$. Assume the string scale is much larger than the scale of the low energy EFT then it would appear that traditional non-gravitational positivity bounds would continue to hold (approximately) even with gravity despite the $t$-channel pole. The finite energy sum rule \eqref{sumrule} shows that assumption is incorrect, by crossing symmetry, whatever happens in the IR is embedded in the UV Regge behaviour. \\

Beyond applications to the Weak Gravity Conjecture, we note that already within known physics of the Standard Model, this behaviour could have far-reaching consequences. Considering the contribution from Standard Model loops, starting from neutrino loops at a scale $m_\nu\sim 10^{-3}$eV, and involving contributions from all known standard matter loops (including notably electrons and W and Z bosons as was performed in \cite{Noumi:2021uuv,Alberte:2021dnj} in the context of graviton-photon amplitude), all the way up to the $\Lambda_c$ which may be taken as the TeV scale. The calculable effects contribute to the various moment sum rules $P_{\sigma}^{\rm low}$. By contrast the energy scale which determines the sub-Regge high energy contribution $P^{\rm high}$ or more precisely $P^{\rm UV}$ comes in is at least 15 orders of magnitude higher than the neutrino scale. Thus (ignoring massless loops) we expect neutrino loops to significantly dominate the expansion coefficients of the Regge functions at small non-zero $a$. Similarly, contemplating (string-inspired) axion models with masses potentially as low as $m_{\rm axion}\sim 10^{-20}$eV$\sim 10^{-47}\mpl$ would directly imply that the Regge functions must carry a scale some 47 orders of magnitude below the Planck scale and contributions from these axions loops would entirely dominate the Regge behaviour of any gravitational amplitude. These results are consistent with what was already argued in \cite{Alberte:2021dnj} and also observed in \cite{Herrero-Valea:2022lfd} however we note that our results rely on no charged particles under any gauge symmetry (other than gravity itself), making this IR/UV mixing entirely generic to any gravitational theory and involving a mixing scale carried by the lightest massive particle (rather than the lightest charged particle). It this important to both include and consistently remove along the lines outline in section \eqref{separating} these light loop contributions to get meaningful constraints on the UV behaviour.

\section{Conclusions}
\label{sec:conclusion}


It is well known that the standard Einstein-Hilbert term of GR should be considered as the leading operator in an infinite EFT expansion that may resum to string theory or other quantum gravity completion at high-energy. In any concrete realization, the next order operators are always expected to be highly suppressed so as to remain virtually unobservable in any concrete realization. Nevertheless with the tremendous leap forward brought to us by the direct detection of gravitational waves and ever more precise cosmological observations, the possibility of constraining higher operators in the EFT of gravity has gained traction. Furthermore it is important to what types of EFTs can be consistently UV completed.
With this in mind, folding in information from their potential high-energy completion provides powerful theoretical constraints on the higher order operators, carving out the appropriate region of parameter space, while potentially allowing us to establish contact between gravitational observables at low-energy (even if in the strong gravitational field regime), and the UV.  \\

This work complements and pushes further the previous results obtained in that direction by providing an explicitly triple crossing symmetric gravitational amplitude on which non-linear compact bounds following from unitarity, locality, causality and Lorentz-invariance may be implemented fully analytically. Up to dimension-10 operators, our results agree with previously derived bounds, although smeared bounds optimized numerically allow for better constraints in some cases. The analytical bounds obtained here have however the advantage of being directly generalizable to arbitrary orders, and easily implemented in any dimension, with new bounds on dimension-12 operators being derived and highlighting features at higher order. \\

Besides the derivation of positivity bounds on low energy physics, our framework highlights the existence of exact finite energy sum rules and continuous moment sum rules whereby the characteristics of the UV Regge behaviour is directly related to the low-energy dispersion relation implying a IR/UV mixing or equivalently a bootstrap relation which may be read either way. The existence of these nontrivial sum rules is connected with the presence of the spin-2 $t$-channel pole. Higher order subtracted dispersion relations for which this pole drops out are largely insensitive to the Regge behaviour as the denominator in the integrand will contain increasing powers of energy (squared). It is the usual twice subtracted (in $s$, once in $x$) relation which is strongly sensitive to the Regge trajectory and residue. The continuous moment sum rules we have defined are similarly sensitive provided the moment $\sigma$ is kept close to zero. Using the continuous moment sum rules we can in principle determine the scattering amplitude for energies above $\Lambda_r$ when the leading Regge trajectory is assumed to dominate, in terms of a dispersive integral of the amplitude at low energies, potentially fulfilling the spirit of the bootstrap program. \\

This framework confirms that the perturbative expansion of the Regge functions away from the forward limit, loop corrections of the lightest massive particles are shown to always dominate over tree-level contributions. This is expected from their relative contributions to the dispersion relation. While such conclusions could have been expected by a resummation of a perturbative low-energy expansion, the validity of such an expansion is not always well-justified and it is only by developing a dispersion relation valid at all scales that we can correctly identify how low energy physics shows up in the Regge behaviour. The relations obtained here do not rely on any assumptions related to the validity of a perturbative resummation of subsets of Feynman diagrams which are likely insufficient to see the emergence of Regge behaviour. The IR/UV mixing presented in the context of the pure gravitational amplitude resembles that already derived in scattering involving the photon \cite{Alberte:2020jsk,Alberte:2020bdz,Alberte:2021dnj,Herrero-Valea:2022lfd}, however unlike these previous cases our result is generic to any gravitational theory.  As an illustrative example, our dispersion relation/sum rules can determine how loops of standard model fields such as the neutrino, ultra-light axions or other dark matter particles considered directly contribute to the Regge trajectory and residue, mixing with the UV tree level physics.


\bigskip
\noindent{\textbf{Acknowledgments:}}
We would like to thank Arshia Momeni, Aninda Sinha, Piotr Tourkine  and Daniel Waldram for useful discussions. We would also like to thank the organizers and attendees of the IAS workshop ``Possible and Impossible in Effective Field Theory: From the S-Matrix to the Swampland" for useful discussions. The work of AJT and CdR is supported by STFC grant ST/T000791/1.  CdR is supported  by a Simons Investigator award 690508.  SJ is supported by an STFC studentship.

\newpage

\appendix


\section{Mandelstam's Double Spectral Representation}
\label{app:mandel}

In order to get a better understanding of the crossing symmetric dispersion relation considered in the text it is helpful to compare it with the more familiar Mandelstam double spectral representation for a manifestly crossing symmetric amplitude \cite{Mandelstam:1958xc}. Ignoring the issue of subtractions \cite{Froissart:1961ux} this is given by
\be
\cA(s,t,u)=\frac{1}{\pi^2} \int_{\Lambda^2}^{\infty} \d \mu \int_{\Lambda^2}^{\infty}  \d \tilde \mu \left[  \frac{\rho(\mu,\tilde \mu)}{(\mu-s)(\tilde \mu-t)}+ \frac{\rho(\mu,\tilde \mu)}{(\mu-s)(\tilde \mu-u)}+\frac{\rho(\mu,\tilde \mu)}{(\mu-t)(\tilde \mu-u)} \right] \, ,
\ee
where the spectral density itself is required to be symmetric by crossing symmetry $\rho(\mu,\tilde \mu)=\rho(\tilde \mu,\mu)$. Writing this in terms of $x$ and $a$ variables, we have
\be
\cA(s,t,u)=\frac{1}{\pi^2} \int_{\Lambda^2}^{\infty}  \d \mu \int_{\Lambda^2}^{\infty}  \d \tilde \mu  \rho(\mu,\tilde \mu) \frac{(2\mu-3a)\mu^2(\mu+2 \tilde \mu)}{(\mu^3-ax+\mu x)(\mu \tilde \mu(\mu+\tilde \mu)-a(\mu^2+\mu \tilde \mu+\tilde \mu^2)) } \, .\
\ee
As expected, for fixed $a$ this gives a dispersion relation in $x$ of the form \eqref{crossamp1} modulo the precise number of subtractions. By comparing the discontinuities in $x$ for fixed $a$ we infer
\be
\disc_s\cA\(\mu,\tau(\mu,a)\) = \frac{1}{\pi} \int_{\Lambda^2}^{\infty}  \d \tilde \mu \rho(\mu,\tilde \mu) \frac{(\mu+2 \tilde \mu) (\mu-a)}{\mu \tilde \mu (\mu+\tilde \mu)-a (\mu^2+\mu \tilde \mu + \tilde \mu^2)} \, ,
\ee
modulo a subtraction polynomial in $a$. This is in turn a dispersion relation in $a$ meaning that the single discontinuity is an analytic function of $a$ for fixed $\mu$ with a right hand branch cut on the real axis beginning at the minimum value of $\mu \tilde \mu (\mu+\tilde \mu)/(\mu^2+\mu \tilde \mu + \tilde \mu^2)$ which is in general $\mu$ dependent but is at least $2\Lambda^2/3$. This justifies our use of the dispersion relation  \eqref{crossamp1} up to $a=2\Lambda^2/3$. Furthermore for fixed $\mu$ there is a maximum value of the branch cut which is given by $a=\mu^2$.
The double discontinuity is related to Mandelstam's double discontinuity by
\be
\disc_a \disc_s\cA\(\mu,\tau(\mu,a)\) \Big|_{a=\mu \tilde \mu (\mu+\tilde \mu)/(\mu^2+\mu \tilde \mu + \tilde \mu^2)}= \frac{ \rho(\mu,\tilde \mu) \mu^3 (\mu+2 \tilde \mu)}{ (\mu^2+\mu \tilde \mu + \tilde \mu^2)^2} \,.
\ee

\section{(Partial) UV completions}
\label{app:coeff}

\subsection{One-loop four graviton scattering discontinuity}
\label{oneloopscalar}
Consider the theory of a minimally coupled massive scalar field in $D=4$,
\begin{dmath}
	S = \int\rd^4 x\,\sqrt{-g}\left(\frac{\mpl^2}{2}R-\frac{1}{2}(\del\phi)^2-\frac12 m^2\phi^2\right)\,.
\end{dmath}
Our goal is to compute the discontinuity across the branch cut for the crossing symmetric combination of helicity amplitudes, and use this to compute the low energy portion of the branch cut integral appearing in Eq.~\eqref{phatdef}. Even though this discontinuity is for illustration purposes uniquely, it will give insight into a concrete realisation of how massive loops can appear in the Regge-subtracted dispersion relation. 

\begin{figure}
\begin{center}
	\includegraphics[scale=0.35]{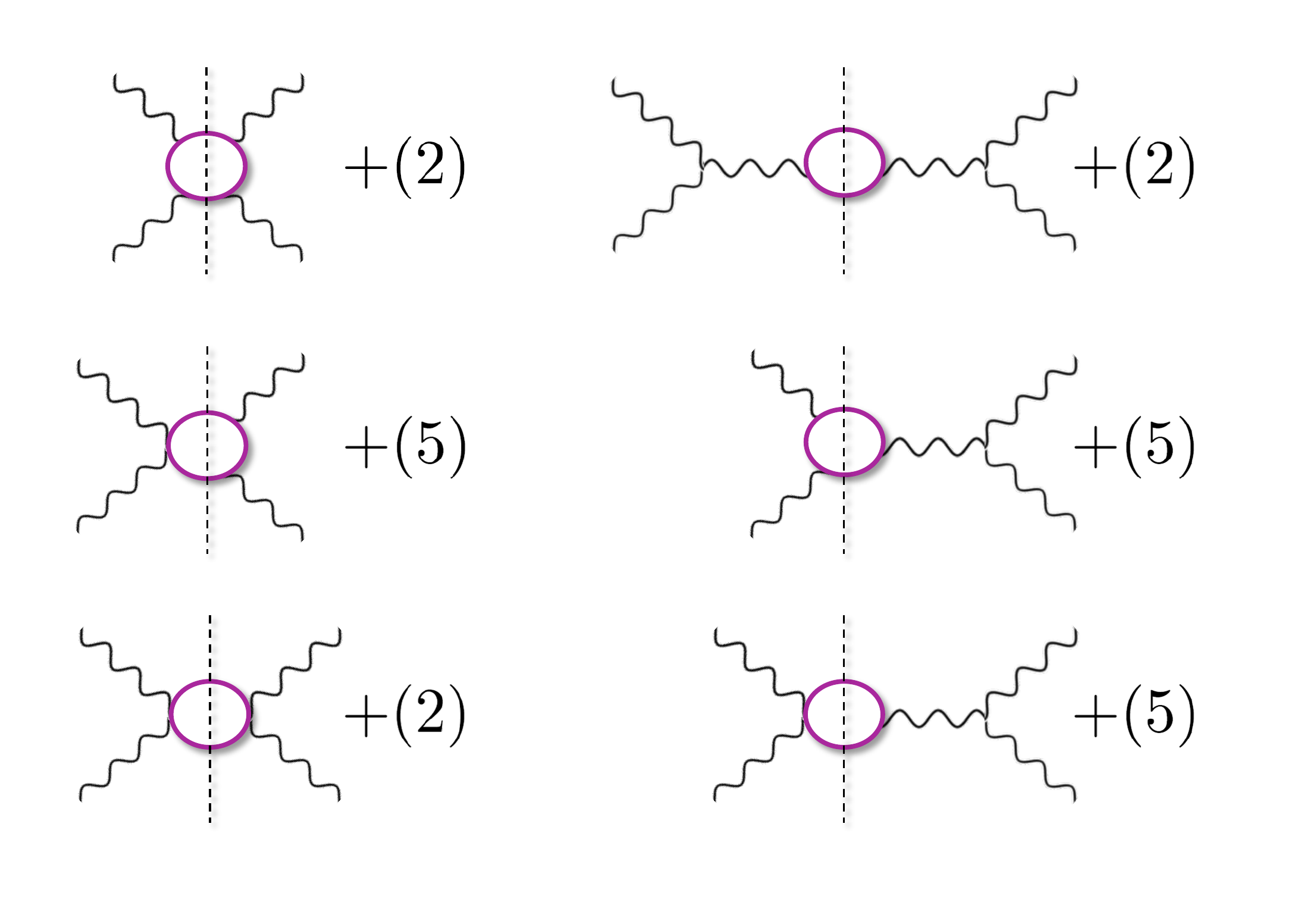}
\caption{\label{Fig:discon}
Feynman graphs that produce a discontinuity in one of the three four graviton scattering channels. The number in brackets indicates how many additional diagrams of that topology there are (including crossed versions etc\ldots).
}
\end{center}
\end{figure}

In the four graviton amplitude there are a subset of diagrams that can produce discontinuities in the $s$-channel as depicted in Fig.~\ref{Fig:discon}. Since we are concerned with triple crossing symmetric amplitudes, we need only compute the discontinuity across one of the three energy channels, e.g. the $s$-channel.
For illustration we take $a=0$ in order to compute the coefficient $d_{0}^{\rm low}$
\ba
	\disc_s \cA(s,0,\phi) &=& \frac{1}{1920 \pi  s^2 \mpl^4}\Big(\sqrt{s \left(s-4 m^2\right)} \left(60 m^6 \cos (2 \phi )+120 m^6+146 m^4 s-18 m^2 s^2+s^3\right) \nn \\
&-&120 m^6 \log \left(\frac{-\sqrt{s \left(s-4 m^2\right)}-2 m^2+s}{2 m^2}\right) \left(m^2 \cos (2 \phi )+2 m^2-2 s\right)\Big)\,.
\ea
Inserting into the forward limit sum rule above, gives,
\ba
	\hat P(0)&=& \int_{\Lambda_c^2}^{\Lambda_r^2}\d \mu \,  \disc_s\cA\(\mu,0,\phi_0 \)\( \frac{2}{\pi\mu^3} \) \\
&+& \frac{1}{960 \pi ^2\mpl^4}\Bigg\{\log \left(  \frac{\Lambda^2_c}{2m^2} +\frac{\Lambda_c}{2m}\sqrt{\frac{\Lambda_c ^2}{m^2}-4 }-1\right)\(1-80 \frac{m^6}{\Lambda_c^6} +30 \frac{m^8}{\Lambda_c^8}\(2+ \cos \phi_0 \)\)\nn\\
&-&\frac{\sqrt{\Lambda_c ^2-4 m^2}}{20 \Lambda_c}
\left((68-5 \cos \phi_0 )
- \frac{2m^2}{\Lambda_c^2} (5 \cos \phi_0 +152)
+ \frac{2m^4}{\Lambda_c^4} (634-15 \cos \phi_0 )
+300 \frac{m^6}{\Lambda_c^6} (\cos \phi_0 +2)\right)\Bigg\}\,,\nn\hspace{-2cm}
\ea
which can be expanded in powers of $m/\Lambda_c \ll 1$,
\begin{dmath}
\label{eq:P0scalar}
	\hat P(0)= \int_{\Lambda_c^2}^{\Lambda_r^2}\d \mu \,  \disc_s\cA\(\mu,0,\phi_0 \)\( \frac{2}{\pi\mu^3} \) + \frac{20 \log \frac{\Lambda_c^2}{m^2}+5 \cos (2\phi_0) -68}{19200 \pi ^2\mpl^4}+\frac{m^2}{48 \pi ^2 \Lambda_c^2 \mpl^4}-\frac{3 m^4}{32 \pi ^2 \Lambda_c^4 \mpl^4}+\mathcal{O}\left(\frac{m^6}{\Lambda_c^6 }\right)\,.
\end{dmath}
The higher order coefficients for $\hat P(a)^{\rm low}$ can be obtained similarly by computing the higher derivatives of the discontinuity.

\subsection{Coefficients from loops of lower-spin fields}

Assuming the Wilsonian coefficients entering our low-energy EFT are dominated by contributions from loops of (heavier) massive lower spin-$S\le 2$ particles, we can provide explicit examples of (partial) UV completions and the resulting amplitude coefficients.
In the cases of the single-helicity-flip amplitudes (e.g. $++\xr +-$)  and the `all-plus' amplitude ($++\xr++$) this leads to a simple relation between certain operator coefficients as detailed below. Note that in the main text we identify the scale $\Lambda^2$ (appearing in the EFT action) as the start of the branch cut, and so in this section one should take $\Lambda^2 = 4m^2$ to maintain consistency with what is written above. We have kept the two scales separate for illustrative purposes. Defining,
\begin{equation}
	n\coloneqq\frac{1}{(4\pi)^2}\frac{1}{30240}\frac{\Lambda^4}{\mpl^2 m^2}\,,
\end{equation}
and using the explicit one-loop amplitudes gives,
	\begin{equation}
\hspace{-0.5cm}		\nn \begin{array}[t]{cccccccccccc}
			S&\quad& c_3& c_+&  c_- & e_+& e_-& f_+& g_+& f_-& g_-& j_1\\[5pt]
			0&\quad& n & \frac{3}{5}\frac{\Lambda^2}{ m^2}n &  \frac12\frac{\Lambda^2}{ m^2}n  & \frac{2}{11}\frac{\Lambda^4}{m^4}n & \frac{21}{110}\frac{\Lambda^4}{m^4}n &  \frac{155}{2002}\frac{\Lambda^6}{m^6}n &  \frac{1}{143}\frac{\Lambda^6}{m^6}n &  \frac{3037 \Lambda ^6}{40040 m^6 }n &  \frac{113 \Lambda ^6}{20020 m^6 }n & -\frac{128 \Lambda ^6}{5005 m^6}n \\[5pt]
			\frac12&\quad& -2n & \frac{87}{40}\frac{\Lambda^2}{ m^2}n &  --  & \frac{59}{110}\frac{\Lambda^4}{m^4}n & -- &\frac{1136 \Lambda ^6}{5005 m^6} n  &\frac{29 \Lambda ^6}{715 m^6}n &-- &-- &-- \\[5pt]
			1&\quad& 3n & \frac{93}{10}\frac{\Lambda^2}{ m^2}n &  --  & \frac{159}{110}\frac{\Lambda^4}{m^4}n & -- &\frac{13113 \Lambda ^6}{20020 m^6}n   &\frac{433 \Lambda ^6}{1430 m^6} n &-- &-- &--  \\[5pt]
			\frac32&\quad& -4n & \frac{1257}{20}\frac{\Lambda^2}{ m^2}n &  --  & \frac{25}{11}\frac{\Lambda^4}{m^4}n & -- &\frac{11554 \Lambda ^6}{5005 m^6}n   &\frac{503 \Lambda ^6}{143 m^6} n &-- &-- &-- \\[5pt]
			2&\quad& 5n & \frac{2013}{2}\frac{\Lambda^2}{ m^2}n &  --  & -\frac{1993}{22}\frac{\Lambda^4}{m^4}n & -- &\frac{254269 \Lambda ^6}{20020 m^6}n   &\frac{25009 \Lambda ^6}{286 m^6}n &-- &-- &--   \,.
		\end{array}
	\end{equation}	
To obtain the blank entries (`$--$') for spin $S$ simply multiply the spin-0 result by $(-1)^{2S}(2S+1)$. We note that $c_3=(2S+1) (-1)^{2S} n$ and so unbroken massive supermultiplets make no contribution to $c_3$.
\subsection{String theory completions}
Following \cite{Bern:2021ppb}, the independent four-graviton tree-level amplitudes for different string theories are given by \cite{Kawai:1985xq} (with $\alpha'$ dependence made explicit),
\begin{equation}
	\begin{aligned}
		\mpl^{D-2}\cA_{22}^{\text{ss}}&=-\frac{u^4 \left(\alpha '\right)^3 \Gamma \left(-\frac{1}{4} \left(s \alpha '\right)\right) \Gamma \left(-\frac{1}{4} \left(t \alpha '\right)\right) \Gamma \left(-\frac{1}{4} \left(u \alpha '\right)\right)}{64 \Gamma \left(\frac{s \alpha '}{4}+1\right) \Gamma \left(\frac{t \alpha '}{4}+1\right) \Gamma \left(\frac{u \alpha '}{4}+1\right)}\\
		\mpl^{D-2}\cA_{22}^{\text{hs}}&=-\frac{u^4 \left(\alpha '\right)^3 \Gamma \left(-\frac{1}{4} \left(s \alpha '\right)\right) \Gamma \left(-\frac{1}{4} \left(t \alpha '\right)\right) \Gamma \left(-\frac{1}{4} \left(u \alpha '\right)\right) \left(1-\frac{s t \left(\alpha '\right)^2}{4 u \alpha '+16}\right)}{64 \Gamma \left(\frac{s \alpha '}{4}+1\right) \Gamma \left(\frac{t \alpha '}{4}+1\right) \Gamma \left(\frac{u \alpha '}{4}+1\right)}\\
		\mpl^{D-2}\cA_{22}^{\text{bs}}&=-\frac{u^4 \left(\alpha '\right)^3 \Gamma \left(-\frac{1}{4} \left(s \alpha '\right)\right) \Gamma \left(-\frac{1}{4} \left(t \alpha '\right)\right) \Gamma \left(-\frac{1}{4} \left(u \alpha '\right)\right) \left(\frac{s t \left(\alpha '\right)^2}{4 u \alpha '+16}-1\right)^2}{64 \Gamma \left(\frac{s \alpha '}{4}+1\right) \Gamma \left(\frac{t \alpha '}{4}+1\right) \Gamma \left(\frac{u \alpha '}{4}+1\right)}\\
		\mpl^{D-2}\cA_{12}^{\text{bs}}&=-\frac{s^2 t^2 u^2 \left(\alpha '\right)^5 \Gamma \left(-\frac{1}{4} \left(s \alpha '\right)\right) \Gamma \left(-\frac{1}{4} \left(t \alpha '\right)\right) \Gamma \left(-\frac{1}{4} \left(u \alpha '\right)\right)}{1024 \Gamma \left(\frac{s \alpha '}{4}+1\right) \Gamma \left(\frac{t \alpha '}{4}+1\right) \Gamma \left(\frac{u \alpha '}{4}+1\right)}\\
		\mpl^{D-2}\cA_{14}^{\text{bs}}&=-\frac{s^2 t^2 u^2 \left(\alpha '\right)^5 \Gamma \left(-\frac{1}{4} \left(s \alpha '\right)\right) \Gamma \left(-\frac{1}{4} \left(t \alpha '\right)\right) \Gamma \left(-\frac{1}{4} \left(u \alpha '\right)\right) \left(s t u \left(\alpha '\right)^3-128\right)^2}{1024 \left(s \alpha '+4\right)^2 \left(t \alpha '+4\right)^2 \left(u \alpha '+4\right)^2 \Gamma \left(\frac{s \alpha '}{4}+1\right) \Gamma \left(\frac{t \alpha '}{4}+1\right) \Gamma \left(\frac{u \alpha '}{4}+1\right)}\,.
	\end{aligned}
\end{equation}
The mass $M$ of the first spin-4 field exchanged between gravitons for these string theory completions is $M^2 = 4/\alpha'$.

To match the string amplitudes with the EFT, we should first bear in mind that both in bosonic and heterotic string theory, the presence of the Gauss-Bonnet term in higher dimensions and the massless dilaton $\varphi$ affects the low-energy amplitude. In any viable vacuum, the massless dilaton should be stabilized, and working below its mass would lead to the same EFT as that considered in \eqref{eq:EFT1}, however for completeness, we provide here the matching with non-stabilized massless dilaton, where the following operator is included (along with a kinetic term for the massless scalar $\Psi$),
\begin{dmath}
	\Delta S = \mpl^{(D-2)/2}\int\rd^D x\,\sqrt{-g}\left(\frac{c}{\Lambda^2}\varphi \cG + \frac{\tilde c}{\Lambda^2}\Psi \tilde\cC\right)\,,
\end{dmath}
with $\cG$ being the Gauss-Bonnet term. The second operator added here is parity violating and is needed specifically to match to the heterotic string amplitude. Its origin can be understood as the cross term between the field strength 3-form of the 2-form $B$ field, and the Lorentz Chern-Simons term (see e.g. \cite{Metsaev:1986yb}) which gives an interaction of the schematic form $(\del B) \omega \text{Riem}$ with $\omega$ the spin-connection. In four dimensions, the 3-form field strength is dual to a 1-form field strength for a massless shift symmetric scalar gauge field ($\Psi$); writing the interaction in terms of $\Psi$ and integrating by parts leads to an interaction as above.
This leads to the following contributions to the amplitude,
\be
\begin{aligned}
	\mpl^{D-2}\Delta\cA_{11}&=-\frac{4s^3}{\Lambda^4}\(\frac{2(D-4)}{D-2}c_{\rm GB}^2+c^2-\tilde c^2 \)\,,\\
	\mpl^{D-2}\Delta\cA_{14}&=\frac{12}{\Lambda^4}\(5c_3-\frac{2 (D-4) }{(D-2)}c_{\rm GB}^2-c^2-\tilde c^2\)y\,,\\
	\mpl^{D-2}\Delta\cA_{13}&=\mpl^{D-2}\Delta\cA_{12}=\mpl^{D-2}\Delta\cA_{24}=\mpl^{D-2}\Delta\cA_{21}=0\,.
\end{aligned}
\ee
Including the contributions from the massless dilaton and working in $D=4$ dimensions, we can now match the string amplitudes to the EFT coefficients. The string amplitudes expanded in powers of the Mandelstam invariants are,
\ba
	\mpl^{{2}}\cA_{11}(\text{ss})&=& \frac{s^3}{tu}-\frac{s^4 \psi ^{(2)}}{M ^6}+\frac{s^4 x \psi ^{(4)} }{12 M ^{10}}+\ldots
	\,,\\
	\mpl^{{2}}\cA_{11}(\text{hs})&=& \frac{s^3}{t u}-\frac{s^3}{M ^4}-\frac{s^4 (\psi ^{(2)}-1)}{M ^6}-\frac{s^5}{M ^8}-\frac{s^4 x (12 \psi ^{(2)}+\psi ^{(4)}) }{12 M ^{10}}\\
&-&\frac{s^6 (1+\psi ^{(2)})}{M ^{10}}+\ldots
	\,,\nn \\
	\mpl^{{2}}\cA_{11}(\text{bs})&=& \frac{s^3}{t u}-\frac{2 s^3}{M ^4}-\frac{s^4 (\psi ^{(2)}-2)}{M ^6}+\frac{s^3 x}{M ^8}-\frac{s^5 }{M ^8}+\frac{2 s^6 \psi ^{(2)}}{M ^{10}}\\
&+&\frac{s^4 (-24+24 \psi ^{(2)}
+\psi ^{(4)}) x}{12 M ^{10}}+\ldots \nn
	\,,\\
	\mpl^{{2}}\cA_{14}(\text{bs})&=&\frac{4 y}{M ^4}-\frac{8 y x}{M ^8}-\frac{4y^2  (3+\psi ^{(2)})}{M ^{10}}+\ldots
	\,,\\
	\mpl^{{2}}\cA_{13}(\text{bs})&=&\frac{y}{M ^4}-\frac{y^2 \psi ^{(2)}}{M ^{10}}+\ldots\,,
\ea
where $M^2 \equiv 4/\alpha'$ and $\psi$ denotes the polygamma function evaluated at $1$, $\psi^{(n)}\equiv \psi^{(n)}(1)$.

Given that the $1\xr2$ and $1\xr4$ configurations are zero for the heterotic and superstring theories, this implies that the coefficients $c_3=c_-=e_-=f_-=g_-=j_1=c^2+\tilde c^2=0$ for these theories. The non-zero EFT coefficients are:

\ba
{\textbf{Superstring:}}\quad && \frac{c_+}{\Lambda^6}=-\frac{ \psi ^{(2)}}{8 M ^6}\,,\ \frac{f_+}{\Lambda^{10}} = -\frac{\psi ^{(4)}}{48 M ^{10}}\,,\
\frac{g_+}{\Lambda^{10}} = -\frac{ \psi ^{(4)}}{24 M ^{10}}\,.\\
{\textbf{Heterotic string:}} \quad  &&
 \frac{c^2-\tilde c^2}{\Lambda^4}=\frac{1}{4  M ^4}\,,\ \frac{c_+}{\Lambda ^6}=-\frac{ \psi ^{(2)}-1}{8 M ^6}\,,\ \frac{e_+}{\Lambda ^8}=-\frac{1}{4 M ^8}\,,\\
  && \frac{f_+}{\Lambda ^{10}} = \frac{ 24+12 \psi ^{(2)}-\psi ^{(4)}}{48 M ^{10}}\,,\ \frac{g_+}{\Lambda ^{10}} =-\frac{ 12 \psi ^{(2)}+\psi ^{(4)}}{24 M ^{10}}.\nn \\
 {\textbf{Bosonic string:}}\quad  &&  \frac{c_3}{\Lambda ^4} = \frac{1}{6 M ^4}\,,\ \frac{c^2-\tilde c^2}{\Lambda ^4}=\frac{1}{2  M ^4}\,,\ \frac{c^2+\tilde c^2}{\Lambda ^4} = \frac{1}{2  M ^4}\,,\ \frac{c_+}{\Lambda ^6}=-\frac{ \psi ^{(2)}-2}{8 M ^6}\,,\qquad \\
  && \frac{e_-}{\Lambda ^8}=\frac{2 }{5 M ^8}\,,\ \frac{e_+}{\Lambda ^8}=-\frac{1}{2 M ^8}\,,\ \frac{f_+}{\Lambda ^{10}} = \frac{24+24 \psi ^{(2)}-\psi ^{(4)}}{48 M ^{10}}\,,\
   \frac{g_-}{\Lambda ^{10}} = \frac{2 }{M ^{10}}
  \nn \\
  && \frac{g_+}{\Lambda ^{10}} =\frac{24-24 \psi ^{(2)}-\psi ^{(4)}}{24 M ^{10}}\,,\ \frac{j_1}{\Lambda ^{10}}= \frac{16  \psi ^{(2)}}{3 M ^{10}}\,, \frac{f_-}{\Lambda ^{10}} = -\frac{1}{M ^{10}}\,.\nn
\ea
%
%
%
If the lowest lying state outside of the regime of validity of the EFT is the spin-4 state exchanged by the two gravitons, we identify the scale appearing in the Lagrangian $\Lambda$ with the mass $\Lambda=M$.
\section{Hilbert series at mass dimension-12 for $D=4$}
\label{app:Hilbert}
Naturally from the EFT perspective, effective actions may include any operators that are consistent with the crucial symmetries of the physical system. Without further restriction, this principle can allow an unwieldy overabundance of redundant operators. The effective field theory of gravity is no stranger to this issue, where we encounter thousands of possible operators arising from the various index contractions between curvature tensors as we move to higher mass dimensions. Reducing these massive sets of operators to their minimal `non-redundant operator basis', i.e. the smallest set of operators where no two are related via field redefinition, algebraic/tensor identity or integration by parts, is a complicated task. Fortunately in recent years much work has been undertaken to simplify this process via the underlying group theoretic structure of the desired symmetries, and in the particular context of constructing EFT Lagrangians \cite{Henning:2015alf, Henning:2017fpj, Lehman:2015coa, Lehman:2015via}. Here we utilise the developments made by \cite{Ruhdorfer:2019qmk} to verify that we have included all mass dimension-12 operators that may contribute to the four graviton amplitude at tree level.

The most general action built out of the curvature tensors involves parity-odd terms, containing an odd number of factors of the Levi-Civita symbol. A non-redundant operator basis for gravity in $D=4$ was provided in \cite{Ruhdorfer:2019qmk} up to and including operators of mass dimension-10. In our analysis we have gone to the next order, requiring operators of mass dimension-12 to apply our crossing symmetric bounds. At this order one may have operators built out of six factors of the Riemann tensor, five factors of the Riemann tensor and two covariant derivatives, and so on.

The Hilbert Series computed in an expansion that arranges operators by mass dimension (indicated by $\epsilon$) gives,
\begin{dmath}
	\mathcal H (\mathcal D, C_L, C_R;\epsilon) = \ldots + \left(2 C_L^6+C_L^4 C_R^2+C_L^3 C_R^3+C_L^2 C_R^4+\mathcal D^2 \left(C_L^5+C_L^3 C_R^2+C_L^2 C_R^3+C_R^5\right)+\mathcal D^4 \left(2 C_L^4+C_L^3 C_R+2 C_L^2 C_R^2+C_L C_R^3+2 C_R^4\right)+2 C_R^6\right) \epsilon^{12} + \mathcal O (\epsilon^{14})\,.
\end{dmath}
Here the factors of $C_{L/R}$ represent the left and right handed parts of the Weyl tensor and $\mathcal D$ represents a covariant derivative. Each term corresponds to a different class of operator, built out of the above objects, but does not describe the precise structure of index contractions between them. The coefficient in front of the term is equal to the number of operators that are truly independent and cannot be related to others via tensor identities, equations of motion or integration by parts etc. The series tells us that at this mass dimension there are eight independent operators involving four factors of the Riemann tensor and/or its dual, either of odd or even parity.

As detailed above, in order to match the EFT of gravity to the one-loop or string theory amplitudes provided in Appendix~\ref{app:coeff}, we required five independent parity even operators at dimension-12. In addition to this we present the following parity-odd operators and their contribution to the amplitude:
\begin{dmath}
	\Delta S = \mpl^{2} \int \rd^4 x\sqrt{-g} \left( \frac{n_1}{\Lambda^{10}} [\mathcal F] [\tilde{\mathcal F}] + \frac{n_2}{\Lambda^{10}} \mathcal F_{\alpha\beta} \tilde{\mathcal F}^{\alpha\beta}\\+\frac{n_3}{\Lambda^{10}}\nabla^\mu\nabla^\nu R^{\alpha\beta\gamma\delta}\nabla^\varepsilon R^{\chi}_{\phantom{\chi}\zeta\delta \gamma}\nabla^{\zeta}R_{\varepsilon\alpha \iota\beta} \tilde R^{\iota}_{\phantom{\iota}\nu\mu\chi}\right)\,,
\end{dmath}
with,
\be
\begin{aligned}
	\mpl^{{2}}\Delta\cA_{11}&=0
	\,,\\
	\mpl^{{2}}\Delta\cA_{14}&=  - \frac{2 i}{\Lambda^{10}}n_1(2 x^3 -3 y^2) - \frac{i}{\Lambda^{10}}n_2(2 x^3 +3 y^2)-\frac{3 i}{4}n_3 y^2\,,\\
	\mpl^{{2}}\Delta\cA_{13}&=\frac{3 i}{16}n_3 y^2\,,
\end{aligned}
\ee
the usual crossing symmetry is obeyed and in addition these amplitude contributions satisfy $\cA_{\text{P}(i)\text{P}(j)}+\cA_{ij}=0$, where $\text{P}$ is the parity operator.

These three CP violating operators give independent contributions to the amplitude, meaning we have found eight independent operators involving four Riemann tensors at dimension-12, which matches the number predicted by the Hilbert series method. For the purposes of computing four-point amplitudes between gravitons at tree-level in the EFT, we have identified all the operators at dimension-12 that could possibly contribute.
\bibliographystyle{JHEP}
\bibliography{references}

\appendix

\end{document}